\documentstyle[preprint,prl,aps,epsf]{revtex}
\begin{document}
\newcommand\ie {{\it i.e. }}
\newcommand\eg {{\it e.g. }}
\newcommand\etc{{\it etc. }}
\newcommand\cf {{\it cf.  }}
\newcommand\etal {{\it et al. }}
\newcommand{\be}{\begin{eqnarray}}
\newcommand{\ee}{\end{eqnarray}}
\tightenlines
\def\lsim{\mathrel{\raise.3ex\hbox{$<$\kern-.75em\lower1ex\hbox{$\sim$}}}}
\def\gsim{\mathrel{\raise.3ex\hbox{$>$\kern-.75em\lower1ex\hbox{$\sim$}}}}
\vskip 2.5cm
\preprint{LBNL-42900}

\title{The Effect of Shadowing on Initial Conditions, Transverse Energy
and Hard Probes in Ultrarelativistic Heavy Ion Collisions}

\author{V. Emel'yanov$^1$, A. Khodinov$^1$, S. R. Klein$^2$ and R. Vogt$^{2,3}$}

\address{{
$^1$Moscow State Engineering Physics Institute (Technical
University), Kashirskoe Ave. 31, Moscow, 115409, Russia\break 
$^2$Nuclear Science Division, Lawrence Berkeley National Laboratory, 
Berkeley, CA
94720, USA\break 
$^3$Physics Department, University of California, Davis, CA 95616, USA}\break}

\vskip .25 in
\maketitle
\begin{abstract}

The effect of shadowing on the early state of ultrarelativistic heavy
ion collisions is investigated along with transverse energy and hard
process production, specifically Drell-Yan, $J/\psi$, and $\Upsilon$
production.  We choose several parton distributions and
parameterizations of nuclear shadowing, as well as the spatial
dependence of shadowing, to study the influence of shadowing on
relevant observables.  Results are presented for Au+Au collisions at
$\sqrt{s_{NN}} = 200$ GeV and Pb+Pb collisions at $\sqrt{s_{NN}} =
5.5$ TeV.

\end{abstract}

\section{Introduction}

Experiments \cite{Arn} have shown that the proton and neutron
structure functions are modified in the nuclear environment.  The
modification depends on the parton momentum fraction $x$.  For medium
$x$, $0.3<x<0.7$, the nuclear parton distributions are depleted
relative to those in isolated nucleons.  For intermediate $x$, $0.1 <
x < 0.3$, the distributions are enhanced, an effect known as
antishadowing.  Finally, for small $x$, $x<0.1$, the nuclear depletion
returns.  We refer to the entire characteristic modification as a
function of $x$ as shadowing.  To date, most measurements of shadowing
have studied charged partons, quarks and antiquarks, through
deep-inelastic scattering (DIS), $eA \rightarrow e' X$, while the
behavior of the nuclear gluon distribution has been inferred from the
modifications to the charged partons.

Almost all of these measurements were blind to the position
of the nucleons in the nucleus. However, most models of shadowing
predict that the structure function modifications should be correlated
with the local nuclear density.  For example, if shadowing is due to
gluon recombination, it should be proportional to the local nuclear
density.  The only experimental study of spatial dependence of parton
distributions relied on dark tracks in emulsion to tag more central
collisions \cite{E745}.  They found evidence of a spatial dependence
but could not determine the form.

This paper studies the effect of shadowing and its position dependence
in ultrarelativistic Au+Au collisions at a center of mass energy of
200 GeV per nucleon, as will be studied at the Brookhaven Relativistic
Heavy Ion Collider (RHIC) and in 5.5 TeV per nucleon Pb+Pb collisions
expected at the CERN Large Hadron Collider (LHC).  We determine the
initial quark, antiquark and gluon production and the first two $E_T$
moments of minijet production for two commonly used parton
distributions, with three shadowing parameterizations.  Following
previous calculations, we find the initial energy density and the
average energy per particle.  We critically examine the concept of fast
thermalization in these collisions.

The spatial dependence of shadowing is reflected in particle
production as a function of impact parameter, $b$, which may be
inferred from the total transverse energy, $E_T$, produced in a heavy
ion collision \cite{us,yad}.  We discuss the relationship between
$E_T$ and $b$ including both hard and soft contributions.  
We then consider the effect of shadowing on
the production of hard probes such as $J/\psi$, $\Upsilon$, 
and Drell-Yan dileptons 
as a function of $b$.  These latter calculations
complement studies of shadowing in open charm and bottom
production \cite{firstprl}.

Section 2 discusses the initial state nuclear parton distributions,
including shadowing and its spatial dependence.  Section 3 then
considers minijet production and the effect on initial conditions for
further evolution of the system.  Section 4 is devoted to the
relationship between transverse energy and impact parameter.  Section
5 discusses $J/\psi$ and Drell-Yan production and their sensitivity to
the nuclear parton distributions.  Finally, section 6 gives some
conclusions.

\section{Nuclear Parton Distributions}

The nuclear parton densities, $F_i^A$, are assumed be the product of
the nucleon density in the nucleus $\rho_A(s)$, the nucleon parton density
$f_i^N(x,Q^2)$, and a shadowing function $S^i(A,x,Q^2,\vec{r},z)$
where $A$ is the atomic mass number, $x$ is the parton momentum fraction,
$Q^2$ is the interaction scale, and $\vec{r}$ and $z$ are the transverse and
longitudinal location of the parton in position space with 
$s=\sqrt{|\vec{r}|^2+z^2}$ so that
\begin{eqnarray}
F_i^A(x,Q^2,\vec{r},z) & = & \rho_A(s) S^i(A,x,Q^2,\vec{r},z) 
f_i^N(x,Q^2) \, \, .  \label{fanuc} 
\end{eqnarray}
In the absence of nuclear modifications,
$S^i(A,x,Q^2,\vec{r},z)\equiv1$.  The density of nucleons in the nucleus
is given by the
Woods-Saxon distribution
\begin{equation}
\rho_A(s)= \rho_0{1 + \omega (s/R_A)^2 \over 
1 + \exp[(s-R_A)/d]} \, \, . \label{density}
\end{equation}
where the nuclear radius, $R_A$, skin thickness, $d$, and oblateness,
$\omega$, are determined from low energy electron-nuclear scattering 
\cite{Vvv}.  The central density is determined by the normalization $\int d^2r
dz \rho_A(s) = A$.  Results are given for Au+Au collisions at RHIC and
Pb+Pb collisions at the LHC with $R_{\rm Au} = 6.38$ fm and $R_{\rm Pb}
= 6.62$ fm respectively.

\begin{figure}[htb]
\setlength{\epsfxsize=0.5\textwidth}
\setlength{\epsfysize=0.7\textheight}
\centerline{\epsffile{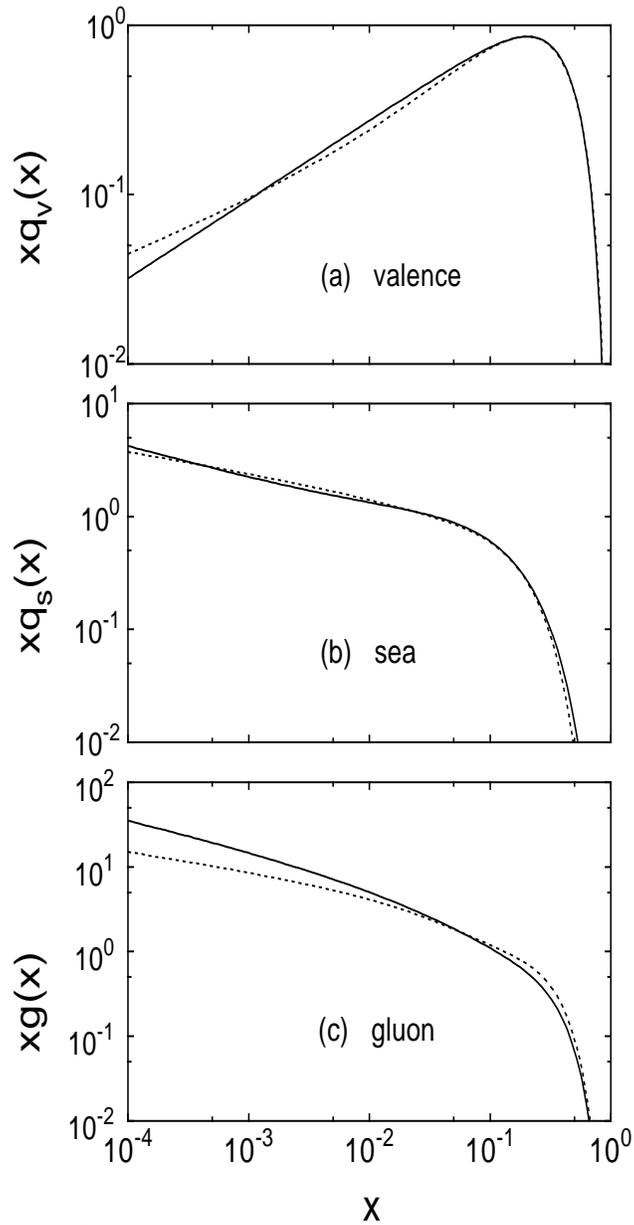}}
\caption[]{The (a) valence, (b) sea, and (c) gluon 
distributions in a proton are given at $Q^2 = p_0^2 = 4$ GeV$^2$ for
the GRV 94 LO (solid) and MRST LO (dashed) sets.}
\label{pdffig}
\end{figure}

The densities of parton $i$ in the nucleon are obtained from fits to
DIS data.  These fits are necessary because the distributions at the
initial scale $Q_0$ are nonperturbative.  However, the
parameterizations of $f_i^N$ are only reliable where measurements
exist.  The continually improving DIS data from HERA \cite{HERA}
shows that uncertainties still exist at small $x$.  Therefore, we
consider two different parton distribution sets.  Both are chosen because they
are leading order, LO, sets which are more consistent with a leading order
calculation.  They also have a relatively low initial scale.  The GRV 94 LO
\cite{grv94} distributions have a lower scale, $Q_0 =
0.63$ GeV, than the MRST LO \cite{mrsg} central gluon distribution with
$Q_0 = 1$ GeV.
Figure~\ref{pdffig} compares the valence, $q_V= u_V + d_V$, sea quark,
$q_S=2(\overline u + \overline d + \overline s)$, and the gluon, $g$
distributions at $Q^2 = 4$ GeV$^2$ of the two sets.  In the
low $x$ region probed by the LHC the valence and sea quark distributions in
both sets are similar.  However, the MRST LO gluon distribution is less
than half as large as the GRV 94 LO gluon distribution.  As we show in
the next section, the gluons dominate particle production 
at the LHC.  Thus the low $x$ gluon density will
significantly affect the initial conditions obtained for these high
energies.  At $x \sim 0.01$ the parton densities are well known so
that the two sets are similar and the choice of the proton parton
distributions do not strongly influence the initial conditions at
RHIC.

Shadowing is an area of intense study with numerous models available
in the literature\cite{Arn}.  However, none of the models can
satisfactorily explain the behavior of the nuclear parton
distributions over the entire $x$ and $Q^2$ range.  Therefore, we
choose to use parameterizations of shadowing based on data.  We use
three different fits, all based on nuclear DIS data.  As in DIS with
protons, the nuclear gluon distribution is not directly measured and
can only be inferred from conservation rules.  The first
parameterization, $S_1(A,x)$, treats quarks, antiquarks, and gluons
identically without $Q^2$ evolution \cite{EQC}. The other two evolve with
$Q^2$ and conserve baryon number and total momentum.  The
$S_2^i(A,x,Q^2)$ parameterization, 
starting from the Duke-Owens parton densities
\cite{DO}, modifies the valence quarks, sea quarks and gluons
separately and includes $Q^2$ evolution for $Q_0=2 < Q < 10$ GeV
\cite{KJE}.  The third parameterization, $S_3^i(A,x,Q^2)$, 
is based on the GRV LO
\cite{GRV} parton densities.  Each parton type is evolved separately
above $Q_0 = 1.5$ GeV \cite{EKRS3,EKRparam}.  The initial gluon
distribution in $S_3$ shows significant antishadowing for $0.1<x<0.3$
while the sea quark distributions are shadowed.  In contrast, $S_2$
has less gluon antishadowing and essentially no sea quark effect in
the same $x$ region.  Since $S_3$ includes the most recent nuclear DIS
data, it should perhaps be favored.  Figure~\ref{fshadow} compares
$S_1$, $S_2$ and $S_3$ for $Q=Q_0$ and $Q=10$ GeV.

The remaining ingredient is the spatial dependence of the shadowing.
Unfortunately, there is little relevant data.  Fermilab experiment
E745 studied the spatial distribution of nuclear structure functions
with $\nu N$ interactions in emulsion.  The presence of one or more
dark tracks from slow protons is used to infer a more central
interaction\cite{E745}.  For events with no dark tracks, no shadowing
is observed while for events with dark tracks, shadowing is enhanced
over spatially independent measurements from other experiments.
Unfortunately, this data is too limited to be used in a fit of the
spatial dependence.
 
Most models of shadowing predict that the nuclear parton densities 
should depend on
the interaction point within the nucleus.  In one model, at
high parton density gluons and sea quarks from one nucleon can
interact with partons in an adjacent nucleon\cite{hot} so that
shadowing is proportional to the local density, Eq.~(\ref{density})
\cite{us,firstprl}.  Then
\begin{eqnarray}
S^i_{\rm WS} = 
S^i(A,x,Q^2,\vec{r},z) & = & 1 + N_{\rm WS}
[S^i(A,x,Q^2) - 1] \frac{\rho_A(s)}{\rho_0} \label{wsparam} \, \, ,
\end{eqnarray}
where $N_{\rm WS}$ is chosen so that $(1/A) \int d^2r dz \rho_A(s)
S^i_{\rm WS} = S^i$. At large distances, $s \gg R_A$, the nucleons
behave as free particles while in the center of the nucleus, the
modifications are larger than the average value $S^i$.

In another approach, shadowing stems from multiple interactions
by an incident parton\cite{ayala}. Parton-parton interactions are
spread longitudinally over a distance known as the coherence
length, $l_c=1/2mx$, where $m$ is the nucleon mass\cite{ina}.  For
$x<0.016$, $l_c$ is greater than any nuclear radius and the
interaction of the incoming parton is delocalized over the entire
trajectory.  The incident parton interacts coherently with all of the
target partons along this interaction length. At large $x$,
$l_c\ll R_A$ and shadowing is proportional to the local density at the
interaction point, while for small $x$, it depends on the density
integrated over the incident parton trajectory.  Both formulations
reproduce the spatially independent shadowing data quite well.
Unfortunately, the available data\cite{E745} is inadequate to test
these theories.

Because of the difficulty of matching the shadowing at large and small
$x$ while maintaining baryon number and momentum conservation, 
we do not include the
multiple scattering model explicitly in our calculations.  However, we do
consider the small $x$ and large $x$ limits separately.
Equation~(\ref{wsparam}) corresponds to 
the large $x$ limit.  In the small $x$ regime, the
spatial dependence may be parameterized
\be S^i_\rho(A,x,Q^2,\vec{r},z) = 1 + N_\rho (S^i(A,x,Q^2) - 1) \int dz
\rho_A(\vec{r},z) \label{rhoparam} \, \, . \ee
The integral over $z$ includes the material traversed by
the incident nucleon.  
The normalization requires $(1/A) \int d^2r dz \rho_A(s)
S^i_{\rho} = S^i$. We find $N_\rho > N_{\rm WS}$.  

There are a number of difficulties with the coherent-interaction
picture.  While traversing the formation length, both the incident and
the produced partons will undergo multiple interactions, which will
reduce the effective coherence length, analogous to the
Landau-Pomeranchuk-Migdal effect\cite{LPM}.  Also, the picture of a
single incident parton interacting with a static nucleus is
inappropriate in heavy ion collisions since the parton density rises
rapidly as many interactions occur simultaneously.  A step-by-step
calculation cannot solve this problem because non-local depictions of
heavy ion collisions are inevitably Lorentz frame dependent
\cite{bnl96}.  Finally, in a model where the parton densities are
spread out over an $x$-dependent distance, baryon number is not
locally conserved.

We previously considered a variant, $S_{\rm R}^i$, where
shadowing is proportional to the thickness of a spherical nucleus at
the collision point\cite{us}, 
\be S^i_{\rm R}(A,x,Q^2,\vec{r},z) =
\left\{ \begin{array}{ll} 1 + N_{\rm R} (S^i(A,x,Q^2) - 1) \sqrt{1 -
(|\vec{r}|/R_A)^2} & \mbox{$r \leq R_A$} \label{rparam} \\ 1 &
\mbox{$r > R_A$} \end{array} \right.  \, \, .  
\ee 
The normalization,
$N_{\rm R}$, obtained after averaging over $\rho_A(s)$, is similar to
$N_{\rho}$.  This model suffers from a discontinuous
derivative at $r=R_A$ with no shadowing predicted for $r>R_A$, but is
otherwise fairly similar to $S_\rho$.

\begin{figure}[htb]
\setlength{\epsfxsize=0.8\textwidth}
\setlength{\epsfysize=0.45\textheight}
\centerline{\epsffile{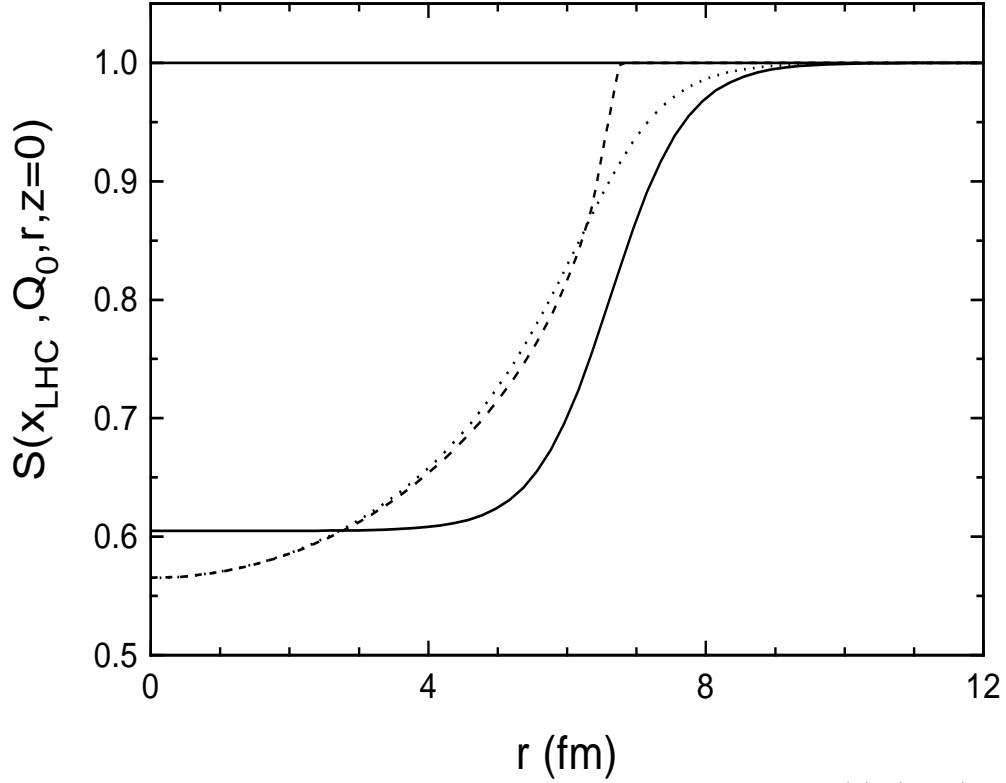}}
\caption[]{The radial distribution of shadowing for three models:
$S_{\rm WS}$, Eq.~(\ref{wsparam}), (solid), $S_{\rm R}$,
Eq.~(\ref{rparam}) (dashed) and $S_\rho$, Eq.~(\ref{rhoparam}).  All
curves are normalized to a homogeneous $S^i(A,x,Q^2)$ of 0.7.  Note that
$S_{\rm WS}$ is evaluated at $z=0$.}
\label{shadowspace}
\end{figure}

Figure~\ref{shadowspace} compares the radial dependence of $S_{\rm
WS}$, $S_\rho$, and $S_{\rm R}$ for $S^i(A,x,Q^2)=0.7$.  For the
comparison, $S_{\rm WS}$ is evaluated at $z=0$.  The $S_\rho$ and
$S_{\rm R}$ results are very similar except near the nuclear surface
where they differ by $\approx 10$\%.  Later we compare calculations of
the first $E_T$ moment with $S_{\rm WS}$ and $S_\rho$ and show that
the two results are very similar.  Calculations using $S_\rho$ and
$S_{\rm R}$ would be in closer agreement, effectively
indistinguishable.

Other mechanisms such as 
nuclear binding have also been suggested as
possible explanations of shadowing\cite{close}.  These calculations can 
explain only a small fraction of the
observed effect\cite{li}, at least for
$x>0.1$. However, many of these models would also
predict some spatial dependence.

Given the difficulties of matching spatial dependencies for different
$x$ and $A$ while preserving baryon and momentum conservation in the
multiple interaction model, we focus our calculations on the local
density model, and perform most of our calculations using $S_{\rm
WS}$. However, as we will show, the calculations are relatively
insensitive to the exact parameterization suggesting that heavy ion
collision studies will not distinguish between different models.

For simplicity, we will refer to homogeneous (without spatial
dependence) and inhomogeneous (position dependent) shadowing.

\section{Initial Conditions in $A+A$ Collisions}

At RHIC and LHC perturbative QCD processes are expected to be an
important component of the total particle production.  At early times,
$\tau_i \sim 1/p_T\leq 1/p_0\sim 0.1$ fm for $p_0 \sim 2$ GeV,
semihard production of low $p_T$ minijets will set the stage for
further evolution \cite{EG}.  Copious minijet production, especially
gluonic minijets, in the initial $NN$ collisions has been suggested
as a mechanism for rapid thermalization, particularly at the LHC.  We
critically examine this idea, with special attention to the effects of
shadowing on these expectations.

Minijet production is calculated from the jet cross
section for $p_T > p_0$.  At leading order the rapidity, $y$,
distribution of a parton flavor $f$ produced in the parton
subprocess $ij\rightarrow kl$ in $AB$ collisions is \cite{EKinit}
\begin{eqnarray}
\frac{d\sigma^f(p_0)}{d^2b d^2r dy} & = & K_{\rm jet}
\int dp_T^2 dy_2 dz dz'
\sum_{{ij=}\atop{\langle kl \rangle}}
x_1F_i^A(x_1,p_T^2,\vec{r},z) x_2 F_j^B(x_2,p_T^2,\vec{b} - \vec{r},z')
 \nonumber \\ 
&  & \mbox{} \times  \frac{1}{1 + \delta_{kl}}
\left[\delta_{fk} {d\hat\sigma\over d\hat t}^{ij\rightarrow kl}(\hat t, \hat u)
+ \delta_{fl} {d\hat\sigma\over d\hat t}^{ij\rightarrow kl}(\hat u, \hat t)
\right] \, \, ,
\label{dsdy} 
\end{eqnarray}
where $\hat t = -p_T^2(1 + e^{-(y-y_2)})$ and $\hat u = -p_T^2(1 +
e^{(y-y_2)})$. 
The limits of integration on $p_T^2$ and $y_2$ are $p_0^2 < p_T^2 <
s_{NN}/(4 \cosh^2 y)$ and $\ln(r_{p_T} - e^{-y}) \leq y_2 \leq
\ln(r_{p_T} - e^y)$ where $|y|\leq \ln(r_{p_0} + \sqrt{ r_{p_0}^2 - 1
})$, $r_{p_T} = \sqrt{s_{NN}}/p_T$ and $r_{p_0} = \sqrt{s_{NN}}/2p_0$.
The sum over initial states includes all combinations of two parton
species with three flavors while the final state includes all pairs
without a mutual exchange and four flavors (including charm) so that
$\alpha_s(p_T)$ is calculated at one loop with four flavors.  The
factor $1/(1 + \delta_{kl})$ accounts for the identical particles in
the final state.  The factor $K_{\rm jet}$ in Eq.~(\ref{dsdy}) is the
ratio of the NLO to LO jet cross sections and indicates the size of
the NLO corrections.  Previous analysis of high $p_T$ jets predicted
$K_{\rm jet} \approx 1.5$ at LHC energies \cite{hpcjet}.  A more
recent NLO calculation of minijet production found $K_{\rm jet}
\approx 2$ at both RHIC and LHC \cite{LO}.  Assuming $K_{\rm jet}=1$,
as in Ref.~\cite{EKinit}, gives a conservative lower limit on minijet
production.  The cutoff $p_0$, represents the lowest $p_T$ scale at
which perturbative QCD is valid.  There is some uncertainty in the
exact value of $p_0$ which can be constrained by soft physics
\cite{EMW}.  However, 2 GeV should be a safe value for heavy ion
collisions, especially at the LHC\cite{EKinit}.  The effects of
different choices for $p_0$ will be discussed later. 

The parton densities are evaluated at scale $p_T$, with $x$ values at
as low as $x_{1,2} \sim 2p_0/\sqrt{s_{NN}}\sim 7 \times 10^{-4}$ at
$y=y_2=0$ in Pb+Pb collisions at 5.5 TeV/nucleon.  At higher
rapidities, $x_1$ or $x_2$ can be even smaller.  Thus the small $x$
behavior of the parton densities strongly influences the initial
conditions of the minijet system.

\begin{figure}[htb]
\setlength{\epsfxsize=0.7\textwidth}
\setlength{\epsfysize=0.6\textheight}
\leftline{\epsffile{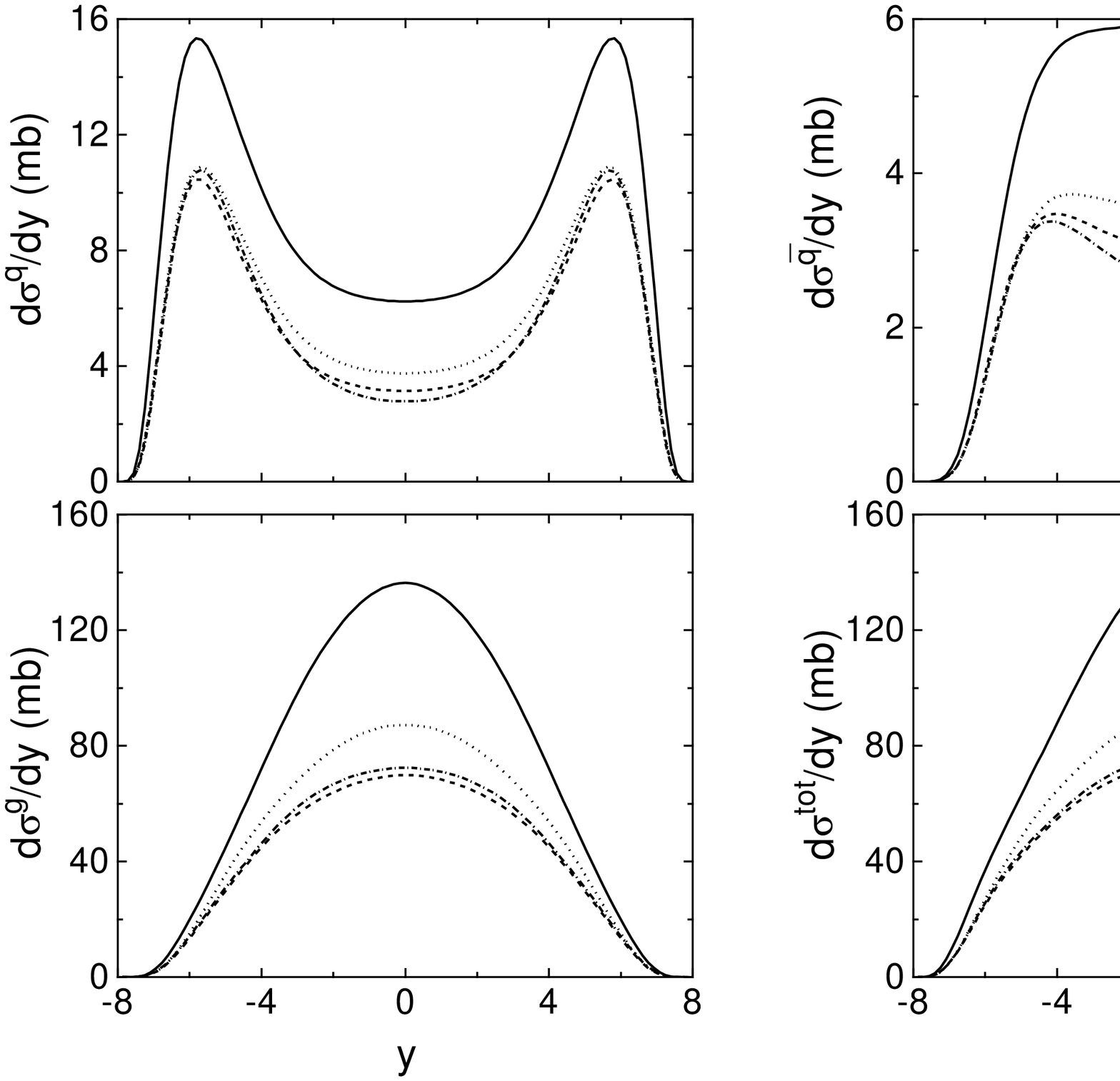}}
\caption[]{The rapidity distributions of quarks, antiquarks, gluons
and the sum of all contributions in Pb+Pb collisions at $\sqrt{s_{NN}}=5.5$
TeV integrated over $b$ and divided by $AB$ calculated with the GRV 94 LO
parton distributions for $p_0=2$ GeV.  The solid curve is without shadowing, 
the dashed is with shadowing parameterization $S_1$, the dot-dashed is
with $S_2$ and the dotted uses $S_3$.  }
\label{dsdygrv}
\end{figure}

\begin{figure}[htb]
\setlength{\epsfxsize=0.7\textwidth}
\setlength{\epsfysize=0.6\textheight}
\leftline{\epsffile{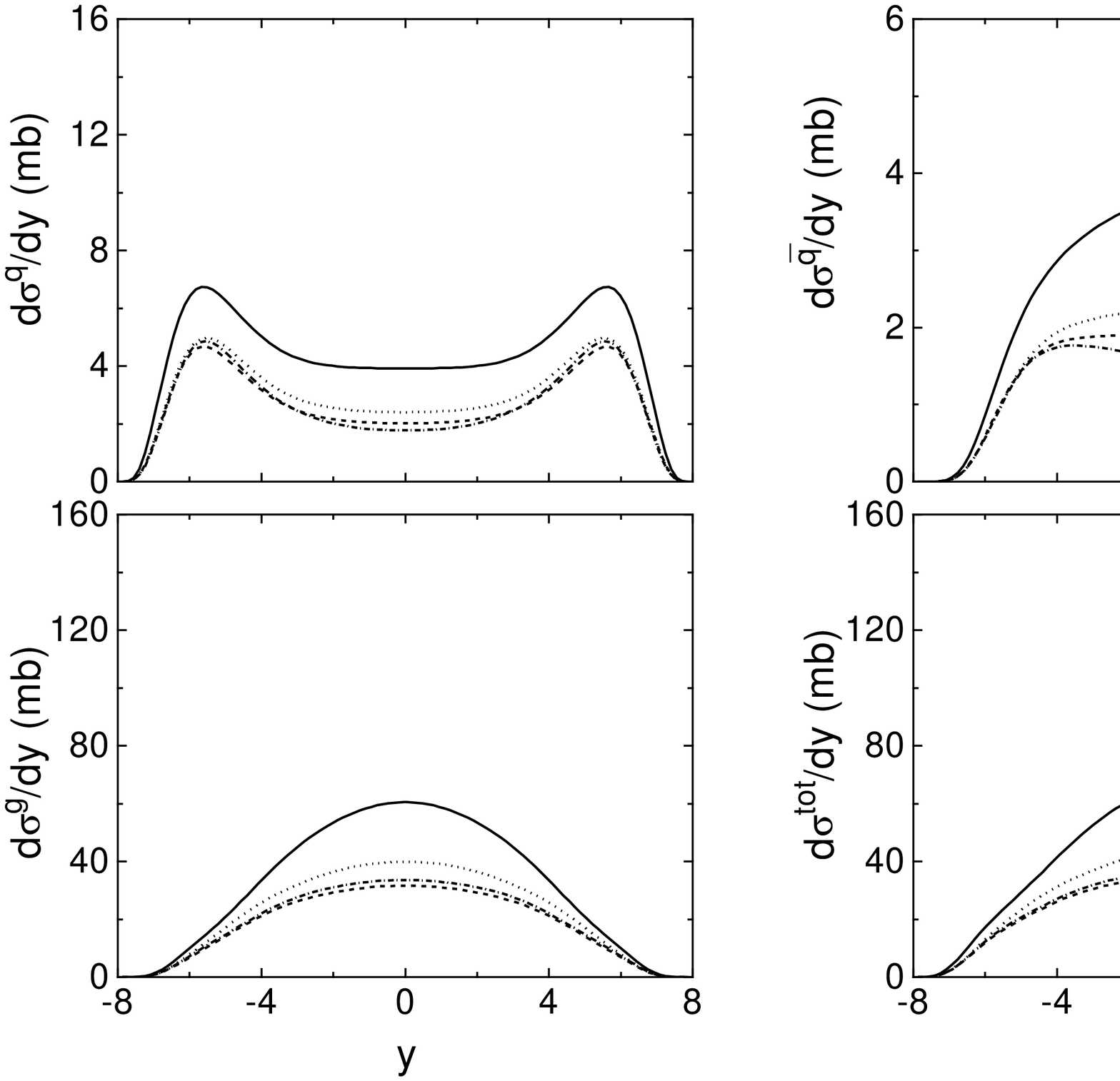}}
\caption[]{The rapidity distributions of quarks, antiquarks, gluons
and the sum of all contributions in Pb+Pb collisions at $\sqrt{s_{NN}}=5.5$
TeV integrated over $b$ and divided by $AB$ calculated with the MRST LO parton
distributions for $p_0 = 2$ GeV.  The solid curve is without 
shadowing, the dashed is
with shadowing parameterization $S_1$, the dot-dashed is with
$S_2$ and the dotted uses $S_3$.  }
\label{dsdymrsg}
\end{figure}

\begin{figure}[htb]
\setlength{\epsfxsize=0.7\textwidth}
\setlength{\epsfysize=0.6\textheight}
\leftline{\epsffile{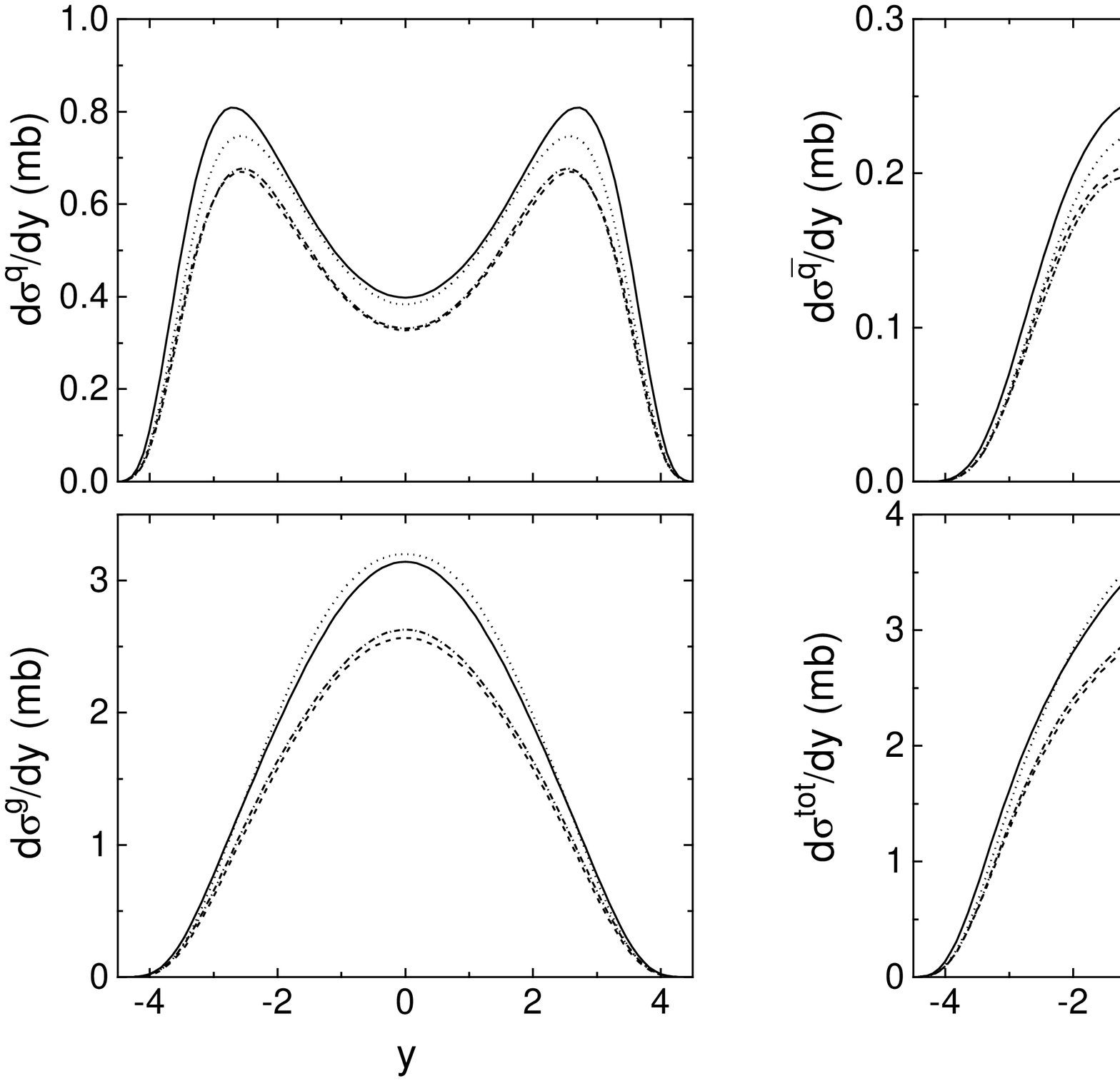}}
\caption[]{The rapidity distributions of quarks, antiquarks, gluons
and the sum of all contributions in Au+Au collisions at $\sqrt{s_{NN}}=200$
GeV integrated over $b$ and divided by $AB$ calculated with the GRV 94 LO
parton distributions for $p_0 = 2$ GeV.  The solid curve is without 
shadowing, the
dashed is with shadowing parameterization $S_1$, the dot-dashed is
with $S_2$ and the dotted uses $S_3$.  }
\label{dsdyrgrv}
\end{figure}

\begin{figure}[htb]
\setlength{\epsfxsize=0.7\textwidth}
\setlength{\epsfysize=0.6\textheight}
\leftline{\epsffile{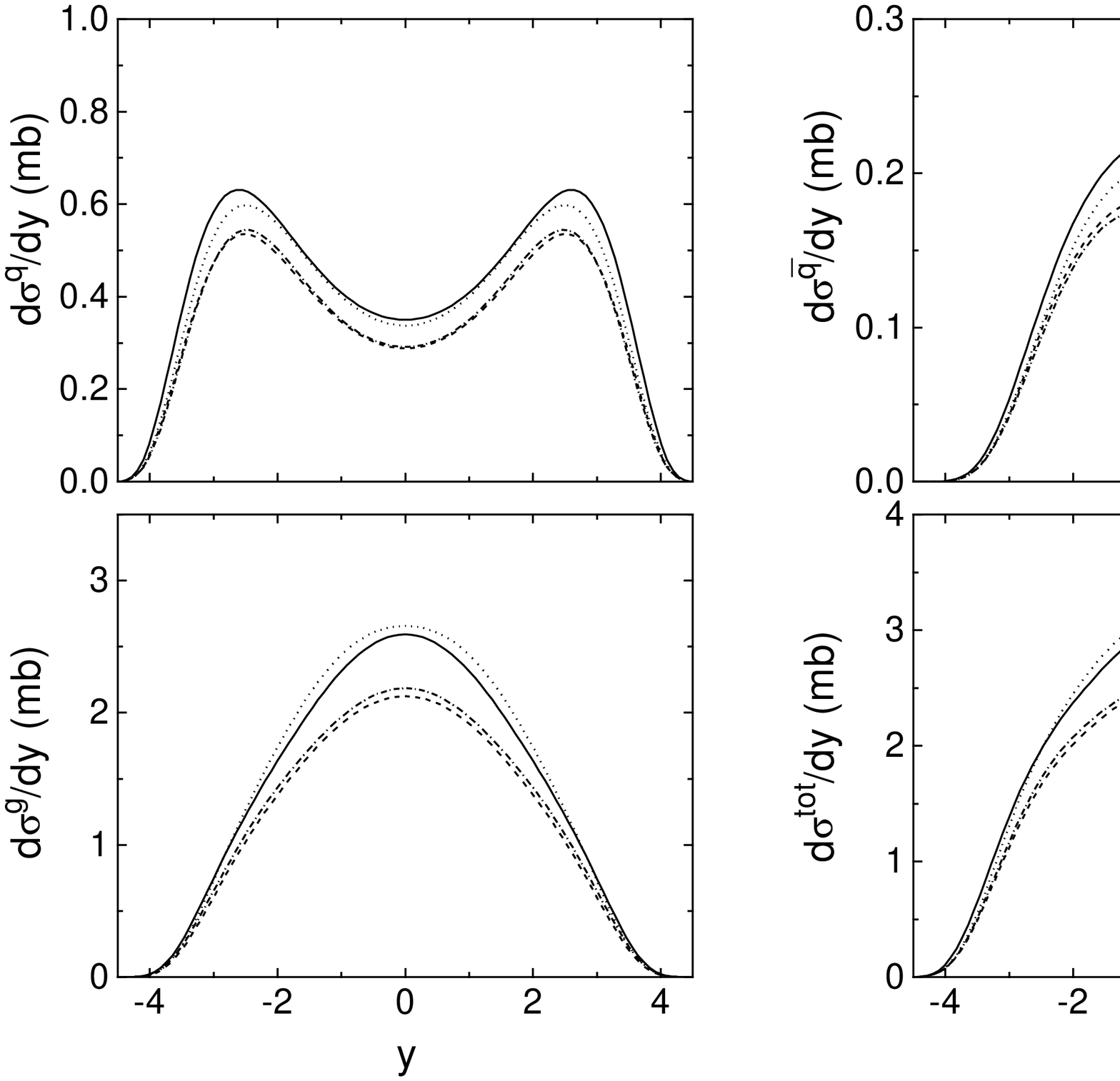}}
\caption[]{The rapidity distributions of quarks, antiquarks, gluons
and the sum of all contributions in Au+Au collisions at $\sqrt{s_{NN}}=200$
GeV integrated over $b$ and divided by $AB$ calculated with the MRST LO parton
distribution for $p_0 = 2$ GeV.  The solid curve is without 
shadowing, the dashed is
with shadowing parameterization $S_1$, the dot-dashed is with
$S_2$ and the dotted uses $S_3$.  }
\label{dsdyrmrsg}
\end{figure}

The resulting minijet rapidity distributions are shown in
Figs.~\ref{dsdygrv}-\ref{dsdyrmrsg} for the two sets of parton
distributions at the LHC and RHIC both without shadowing and with
homogeneous shadowing.  Shadowing can reduce the number of produced
partons by up to a factor of two at the LHC, depending on the
parameterization and the parton type.  The smallest effect is observed
with the newer $S_3$ parameterization.  At the lower RHIC energy,
$x_{1,2} \sim 0.02$, and shadowing is smaller, as is shown in
Fig.~\ref{fshadow}.  Due to the strong antishadowing, gluons are actually 
enhanced with $S_3$.

Since each collision has two final state partons, the total number of
partons of flavor $f$ at impact parameter $b$ is
\begin{eqnarray}
\overline N^f(b,p_0) =
2 \, \frac{d\sigma^{f}(p_0)}{d^2b} \, \,  \label{npartaa}
\end{eqnarray}
where $d\sigma^{f}(p_0)/d^2b$ is the integral of 
Eq.~(\ref{dsdy}) over $d^2r$ and $dy$ normalized so that
\begin{eqnarray}
\int \frac{d \sigma^f(p_0)}{d^2b dy} dy = 
2 \frac{d\sigma^{f}(p_0)}{d^2b} \label{signorm}
\end{eqnarray}
because there are two final-state partons in each collision.  
The total hard scattering cross section as a function of impact parameter
is the sum over all parton flavors so that
\begin{eqnarray}
2 \sum_f \frac{d \sigma^f(p_0)}{d^2b} = 
2 \frac{d\sigma (p_0)}{d^2b} \equiv \sigma^H(b,p_0) \, \, .
\label{sighnorm}
\end{eqnarray}
When
$S=1$ or $S \equiv S^i(A,x,Q^2)$ the spatial dependence factorizes,
the per nucleon cross section is independent of $b$, and the total
cross section scales with the nuclear overlap function, $T_{AB}(b)$,
\cite{overlap}. The overlap function is the convolution of the
nuclear density distributions \cite{Vvv}
\begin{equation}
T_{AB} (b) = \int d^2\vec{r} T_A(\vec{r}) T_B(\vec{b}-\vec{r})
\end{equation}
with the nuclear thickness function $T_{A}(\vec{r}) = \int dz
\rho_A(z, \vec r)$.  For $AA$ collisions,
$T_{AA}(0)\approx A^2/(\pi R_A^2)\propto A^{4/3}$.  
The transverse area of the system and the initial
volume at $b=0$ are
\begin{eqnarray}
 A_T & = & \pi R_A^2 \\
 V_i & = & A_T \Delta y \tau_i = A_T \Delta y / p_0 \label{vol} 
\, \, , \end{eqnarray}
where $\tau_i = 1/p_0$ and $\Delta y$ is the rapidity range.

Parton production saturates when the transverse area occupied by
the partons is larger than the total transverse area available.  The total
number of partons produced in the collision is the sum over flavors,
\begin{eqnarray}
\overline N^H (b) = \sum_f \overline N^f(b,p_0)
\label{ntot}
\end{eqnarray}
In a $b=0$ collision, the partons occupy a transverse area $\pi
\overline N^H(0)/p_0^2$.
Saturation occurs when the area occupied by partons is equivalent to the
transverse area of the target in a symmetric heavy ion collision at $b=0$, 
$\overline N^H(0) > R_A^2 p_0^2$.  In Pb + Pb
collisions $T_{AA}(0)=30.4$/mb and saturation occurs if the hard cross
section is greater than 74 ($p_0$/2 GeV)$^2$ mb.  At the LHC,
gluons alone are sufficient to saturate the transverse area, 
even with shadowing.
For Au + Au collisions at RHIC, the hard cross section must be
more than 71 ($p_0$/2 GeV)$^2$ mb.  This condition is not satisfied at
RHIC, unless $p_0$ is lowered to $\sim 1$ GeV\cite{EMW}.  However,
1 GeV is close to if not within the nonperturbative regime,
suggesting that soft physics still dominates particle production at
RHIC.

These conclusions depend on the small $x$ behavior of the gluon
distribution, the factor $K_{\rm jet}$, the cutoff $p_0$, and the
shadowing parameterization.  Transverse saturation does not occur at
the LHC when the MRST LO set is used if $K_{\rm jet} = 1$.  An
empirical $K_{\rm jet}$ may be obtained by comparing model
calculations to data, giving some freedom in the value of $K_{\rm
jet}$ for different parton distributions.  However, less
variation is allowed in the theoretical values of $K_{\rm jet}$ obtained from
the ratio of the NLO and LO cross sections.  The theoretical $K_{\rm jet}$ 
does, however, tends
to rise as $p_T$ decreases, rendering calculations with $p_0 < 2$ GeV
less reliable.

Transverse saturation at $p_0 = 2$ GeV implies that the minijet cross
section exceeds the inelastic $pp$ cross section, violating unitarity.
This is especially a problem for the GRV 94 LO distributions because
of the high gluon density at low $x$.  At very low $x$ then, the
proton is like a black disk and instead of further splitting to
increase the density of partons, the partons begin to recombine,
acting to lower the density below that without recombination.
Therefore at very low $x$, the density of partons should not increase
without bound but begin to saturate.  This recombination corresponds
to one picture of shadowing in the proton \cite{hot}.  A recent HERA
measurement of the derivative of the structure function $F_2$ found
that at low $x$ and $Q^2$, $dF_2/d\ln Q^2$ no longer increases, in
contrast to the GRV 94 parton densities which continue to increase
over the range of their validity \cite{ZEUS}.  The newer MRST
distributions have been tuned to fit this behavior for $Q^2 > 1$
GeV$^2$.  This data implies that the unitarity violation in $pp$
interactions is likely an artifact of the free proton parton
densities.  

\begin{figure}[htb]
\setlength{\epsfxsize=0.7\textwidth}
\setlength{\epsfysize=0.6\textheight}
\leftline{\epsffile{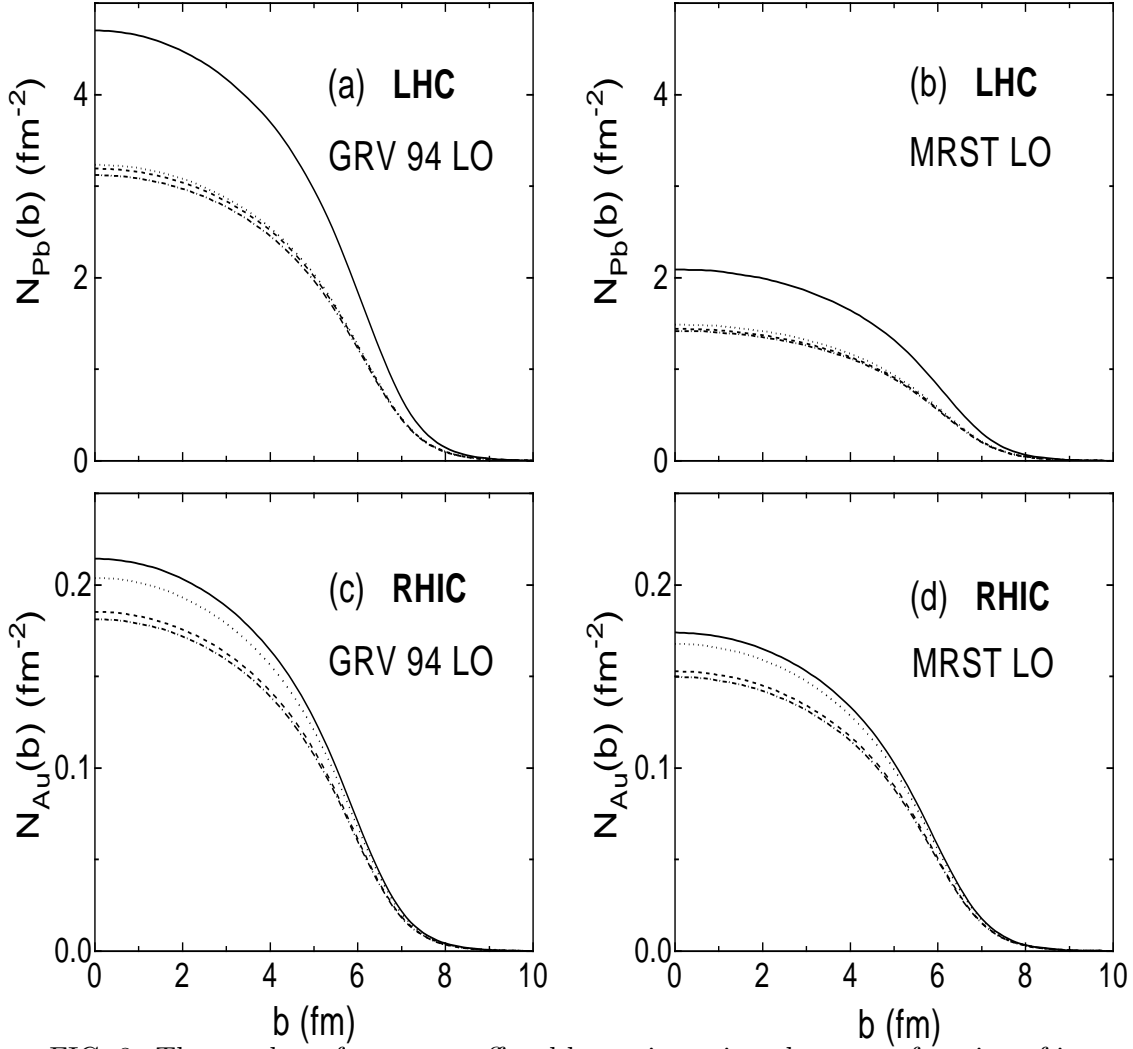}}
\caption[]{The number of scatters suffered by an incoming gluon as a function
of impact parameter.  The LHC results at $x_1 = 10^{-2}$
are shown in (a) and (b) for GRV 94 LO
and MRST LO parton densities respectively while the RHIC calculations at $x_1 =
10^{-1}$ are shown
in (c) and (d) for the GRV 94 LO and MRST LO sets.  The solid curve is without 
shadowing, the dashed is
with shadowing parameterization $S_1$, the dot-dashed is with
$S_2$ and the dotted uses $S_3$.  }
\label{partscat}
\end{figure}

The magnitude of the problem can be gauged by calculating the number of
collisions suffered by incoming partons.  If, on average, a parton collides
more than once while crossing the nucleus, unitarity violation is a serious
problem.  The higher the incoming parton $x_1$, the more low $x_2$ target
partons are available for it to interact with, the larger the interaction cross
section, and the subsequent number of collisions.  The minimum $x_2$ depends on
$p_0$ and $\sqrt{s_{NN}}$. Since the gluon interaction cross sections are
larger than those of quarks, we focus on
incoming gluons with $x=0.1$.  The average number of collisions
experienced by such an initial gluon 
at the LHC is shown in Fig.~\ref{partscat}(a) and (b) for
GRV 94 LO and MRST LO distributions respectively.  The scattering
cross section has been multiplied by the nuclear profile function
$T_A(b)$ to give the number of collisions.  
A gluon can suffer up to an average of 5 hard scatterings
in central collisions with GRV 94 LO and $S=1$. It experiences less
than one collision in the target when $b>5-6$ fm.  Shadowing reduces
the severity of the problem by decreasing 
the number of scatterings by $\approx 30$\%.
On the other hand, $u$ and $\overline u$ quarks with $x=0.1$ typically
scatter once or less in the target, even without shadowing. With the
MRST LO distributions, the unitarity violation is less severe, with $1.4-2$
scatterings per central collision for gluons and 0.5 $u$ or $\overline
u$ collisions per central event.

Therefore we might expect that to satisfy unitarity, transverse
saturation cannot be used as a criteria for determining $p_0$ and
early equilibration by minijet production is unlikely in reality.  At
the lower RHIC energy, unitarity is always satisfied with incoming
partons experiencing an average of less than one
collision. Figure~\ref{partscat}(c) and (d) shows this for the gluon.
The $q$ and $\overline q$ results are considerably smaller.

The quark rapidity distribution, $d\sigma^q/dy$ is indicative of
baryon stopping due to hard processes.  As Tables~\ref{table1}-\ref{table4} 
and Figs.~\ref{dsdygrv}-\ref{dsdyrmrsg} show, at
LHC energies, the GRV 94 LO parton distributions predict considerably
larger stopping than MRST LO.  These homogeneous shadowing cross
sections can be converted to $dN/dy$ at any impact parameter by
multiplying by $T_{AB}(b)$.  Although both parton distributions
predict similar baryon densities at mid-rapidity, GRV 94 LO predicts
about twice as many baryons at large rapidity than MRST
LO.  Because of the unitarity problems and the high gluon
density at low $x$, at LHC the final state baryon
number, $\int dy (dq/dy - d\overline q/dy)/3$, exceeds the baryon
number of the two incoming nuclei.  This is a clear result of the
unitarity violation.  Previous works\cite{EKinit} noted this
but suggested that the problem is reduced if only central
rapities are considered, typically $|y| < 0.5$.  
A better solution would include a more complete treatment of multiple
scattering.  However, such a calculation involves even more uncertainties. 
At RHIC energies, the
cross sections are lower, and baryon conservation is not an issue.
The two sets of parton distributions make similar predictions, with
MRST LO finding a somewhat higher baryon density at mid-rapidity.

We present calculations covering the entire range of rapidities, even
though at the LHC, at large rapidity, $|y|>5$, either $x_1$ or $x_2$
is outside the stated validity range of the parton distributions.
This range problem could affect calculations at all $y$ since a parton
density that satisfies the unitarity bound at $p_0$ will be different
at all rapidities since more of the low $x$ rise will be subsumed into
higher $x$ values to maintain momentum conservation.

We would like to determine the effects of shadowing on the quantities
such as the energy density which are important for our understanding
of the initial conditions.  The initial energy density is directly
related to the cross section times first $E_T$ moment of each flavor,
$\sigma^f(p_0) \langle E_T^f \rangle$, which is calculated within a
specific acceptance.  A crude approximation of the acceptance is \be
\epsilon(y) = \left\{ \begin{array}{ll} 1 \,\, \, \, \, \, \, &
\mbox{if $|y| \leq y_{\rm max}$} \\ 0 & \mbox{otherwise}
\end{array} \right.  \ee where $y_{\rm max}$ is the highest measurable
rapidity.  At leading order, the parton pairs are produced
back-to-back. The $E_T$ distribution of each flavor as a function of
impact parameter is \cite{EKinit}
\begin{eqnarray}
\frac{d\sigma^f(p_0)}{dE^f_T d^2b d^2r} & = & 
\frac{K_{\rm jet}}{2} \int dp_T^2 dy_2 dy dz dz'
\sum_{{ij=}\atop{\langle kl \rangle}}
x_1F_i^A(x_1,p_T^2,\vec{r},z) x_2 F_j^B(x_2,p_T^2,\vec{b} - \vec{r},z') 
\nonumber \\ 
&  & \mbox{} \times \frac{1}{1 + \delta_{kl}} 
\left\{ {d\hat\sigma\over d\hat t}^{ij\rightarrow kl}(\hat t, \hat u)
\delta(E^f_T - \left[\delta_{fk} \epsilon(y) + \delta_{fl} \epsilon(y_2)
\right] p_T ) \right. \nonumber \\
&  & \mbox{} \left. + {d\hat\sigma\over d\hat t}^{ij\rightarrow kl}(\hat u, 
\hat t)\delta(E^f_T - \left[\delta_{fl} \epsilon(y) + \delta_{fk} \epsilon(y_2)
\right] p_T )
\right\} \, \, . \label{dsdet} \end{eqnarray}
Equation (\ref{dsdet}) is valid for $E_T > E_{T \, {\rm min}}$ where 
the $E_{T\, {\rm min}}$ required in $pp$ collisions\footnote{A 
comparison of the LO
and NLO jet $E_T$ distributions with UA2 data \cite{UA2} suggests that
below $E_T = 55$ GeV, the discrepancy between the calculations and
data can be attributed to further higher-order corrections or
higher-twist effects such as initial and final-state radiation
\cite{Leonidov}.} is such that
$\sigma^H \leq \sigma_{\rm inelastic}$ for $E_{T \, {\rm min}}
\approx 2$ GeV at 5.5 TeV and 1 GeV at 200 GeV \cite{LO}.  
Therefore, integration over $d^2r$ and $E_T > E_{T\, {\rm min}}$
reduces Eq.~(\ref{dsdet}) to the total hard
cross section as a function of impact parameter
\begin{eqnarray}
\sum_f \int d^2r \int_{E_T \, {\rm min}}^\infty \frac{d 
\sigma^f(p_0)}{d^2b d^2r dE_T^f}
dE_T^f = 2 \frac{d\sigma(p_0)}{d^2b} 
\equiv \sigma^H(b,p_0) \,\,\, . \label{setnorm}
\end{eqnarray}
The last definition in Eq.~(\ref{setnorm}) holds for $E_{T\, {\rm min}} = p_0$,
as in Eq.~(\ref{sighnorm}).

\begin{figure}[htb]
\setlength{\epsfxsize=0.7\textwidth}
\setlength{\epsfysize=0.6\textheight}
\leftline{\epsffile{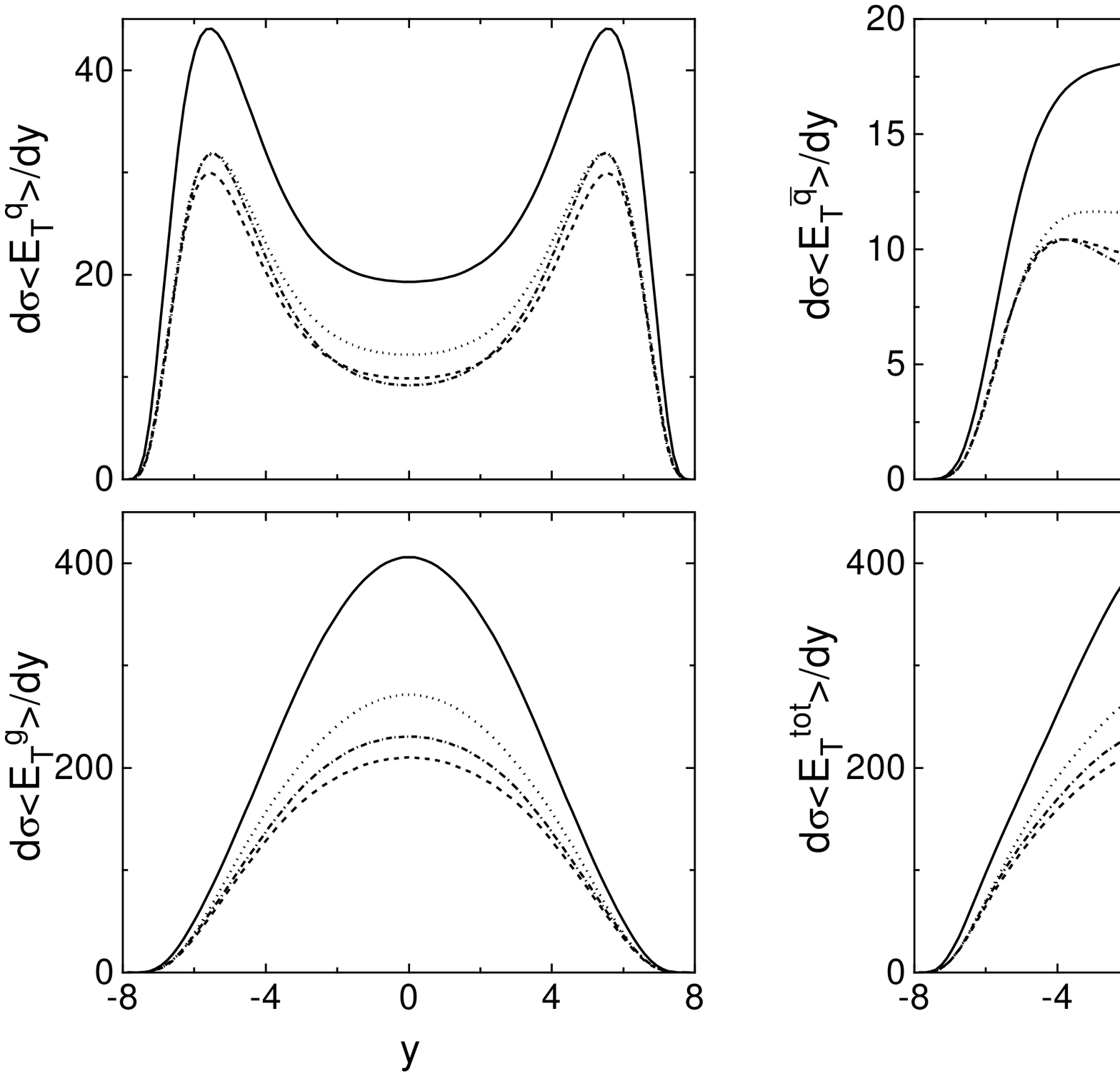}}
\caption[]{The first $E_T$ moment, $\sigma(p_0) \langle
E_T^f \rangle$, as a function of rapidity for quarks, antiquarks,
gluons and the sum of all contributions in Pb+Pb collisions at
$\sqrt{s_{NN}}=5.5$ TeV integrated over $b$ and divided by $AB$ calculated with
the GRV 94 LO parton distributions for $p_0=2$ GeV.  The solid curve is without
shadowing, the dashed is with shadowing parameterization $S_1$, the
dot-dashed is with $S_2$ and the dotted uses $S_3$.}
\label{dsetdygrv}
\end{figure}

\begin{figure}[htb]
\setlength{\epsfxsize=0.7\textwidth}
\setlength{\epsfysize=0.6\textheight}
\leftline{\epsffile{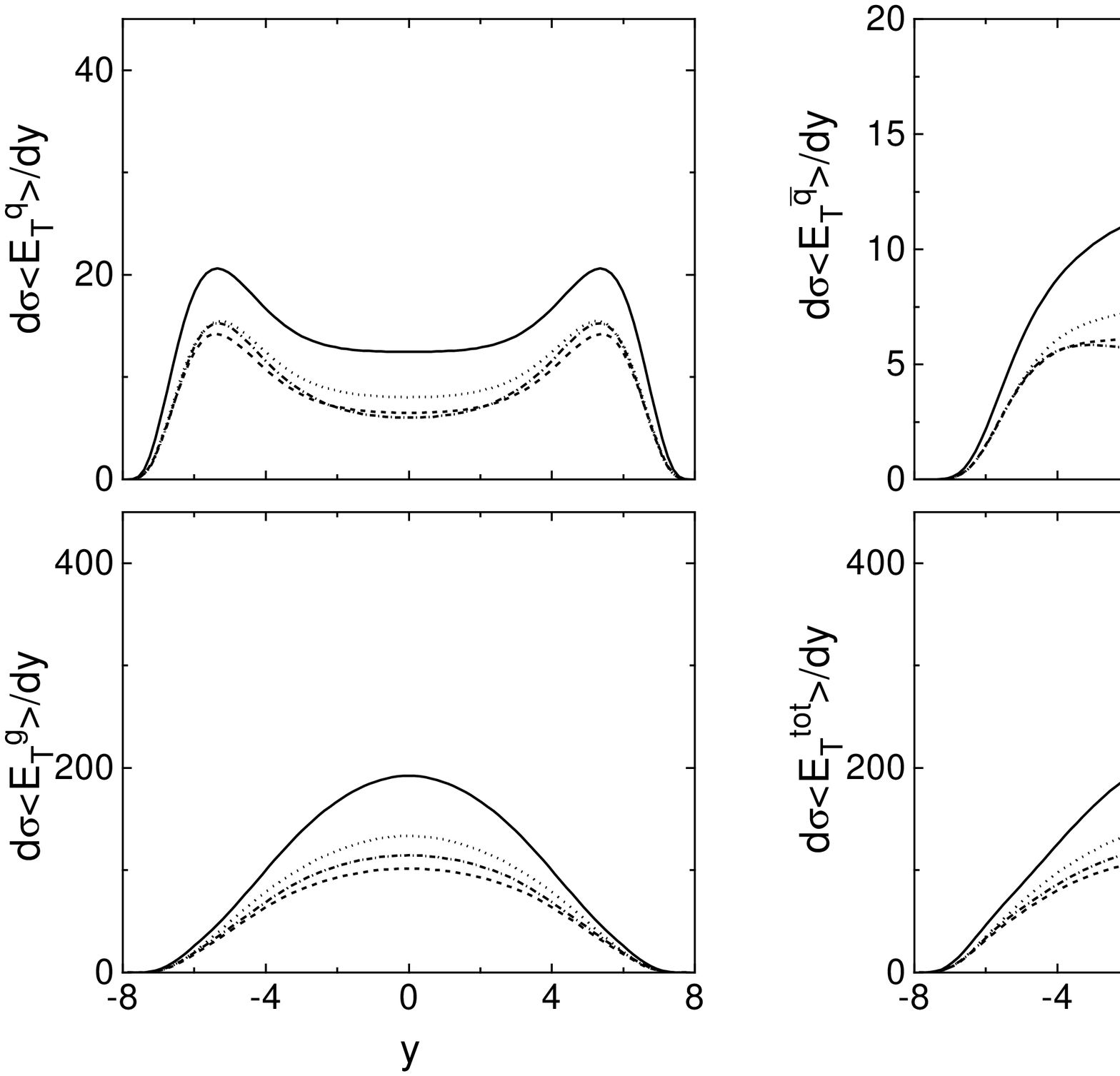}}
\caption[]{The first $E_T$ moment, $\sigma(p_0) \langle
E_T^f \rangle$, as a function of rapidity for quarks, antiquarks,
gluons and the sum of all contributions in Pb+Pb collisions at
$\sqrt{s_{NN}}=5.5$ TeV integrated over $b$ and divided by $AB$ calculated with
the MRST LO parton distributions for $p_0 = 2$ GeV.  The solid curve is without
shadowing, the dashed is with shadowing parameterization $S_1$, the
dot-dashed is with $S_2$ and the dotted uses $S_3$.}
\label{dsetdymrsg}
\end{figure}

\begin{figure}[htb]
\setlength{\epsfxsize=0.7\textwidth}
\setlength{\epsfysize=0.6\textheight}
\leftline{\epsffile{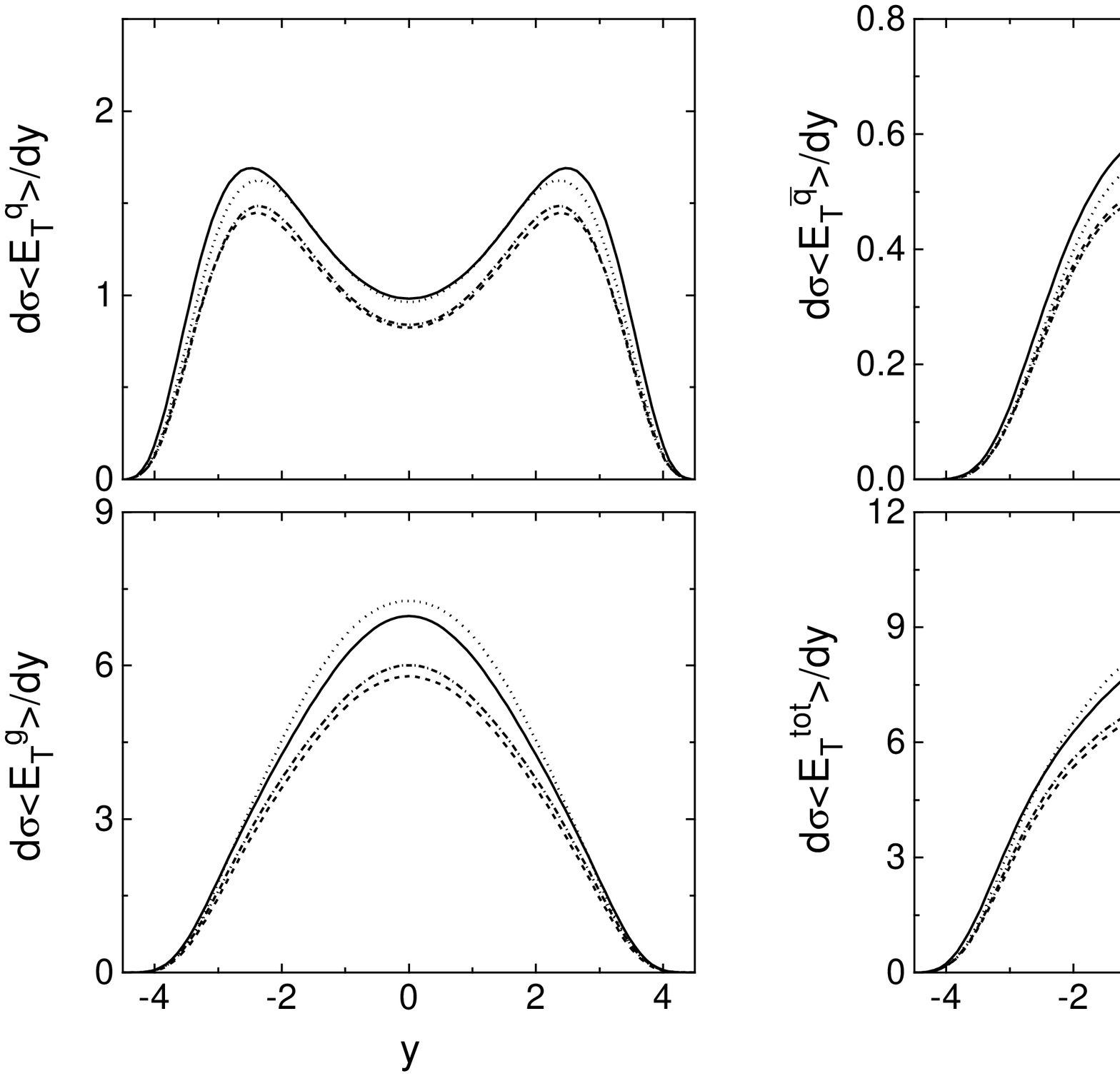}}
\caption[]{The first $E_T$ moment, $\sigma(p_0) \langle
E_T^f \rangle$, as a function of rapidity for quarks, antiquarks,
gluons and the sum of all contributions in Au+Au collisions at
$\sqrt{s_{NN}}=200$ GeV integrated over $b$ and divided by $AB$ calculated with
the GRV 94 LO parton distributions for $p_0 = 2$ GeV.  
The solid curve is without
shadowing, the dashed is with shadowing parameterization $S_1$, the
dot-dashed is with $S_2$ and the dotted uses $S_3$.}
\label{dsetdyrgrv}
\end{figure}

\begin{figure}[htb]
\setlength{\epsfxsize=0.7\textwidth}
\setlength{\epsfysize=0.6\textheight}
\leftline{\epsffile{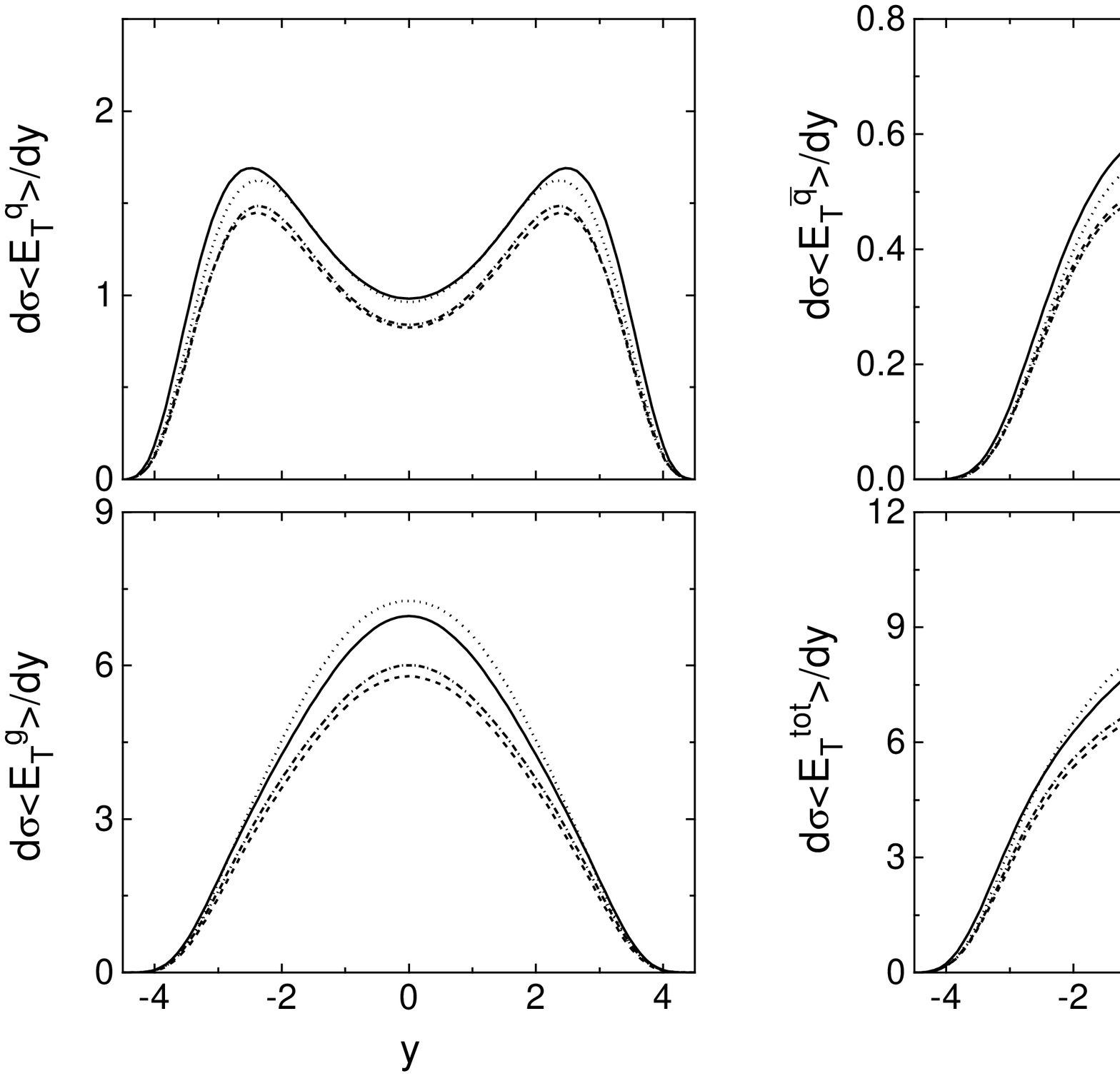}}
\caption[]{The first $E_T$ moment, $\sigma(p_0) \langle
E_T^f \rangle$, as a function of rapidity for quarks, antiquarks,
gluons and the sum of all contributions in Au+Au collisions at
$\sqrt{s_{NN}}=200$ GeV integrated over $b$ and divided by $AB$ calculated with
the MRST LO parton distributions for $p_0 =2$ GeV.  The solid curve is without
shadowing, the dashed is with shadowing parameterization $S_1$, the
dot-dashed is with $S_2$ and the dotted uses $S_3$.}
\label{dsetdyrmrsg}
\end{figure}

The first
$E_T$ moment is obtained by weighting Eq.~(\ref{dsdet}) with $E^f_T$ and
integrating over $E^f_T$; we neglect particle masses so that
$E^f_T=p_T$,
\begin{eqnarray}
\frac{d\sigma^f(p_0)\langle E_T^f \rangle}{d^2b d^2r} & = & 
K_{\rm jet} \int dp_T^2 dy_2 dy dz dz'
\sum_{{ij=}\atop{\langle kl \rangle}}
x_1F_i^A(x_1,p_T^2,\vec{r},z) x_2 F_j^B(x_2,p_T^2,\vec{b} - \vec{r},z')
\nonumber \\ 
&  & \mbox{} \times \frac{ \epsilon(y) p_T}{1 + \delta_{kl}} 
\left[\delta_{fk} {d\hat\sigma\over d\hat t}^{ij\rightarrow kl}(\hat t, \hat u)
+ \delta_{fl} {d\hat\sigma\over d\hat t}^{ij\rightarrow kl}(\hat u, \hat t)
\right] \, \, . \label{etmom} \end{eqnarray}
The first $E_T$ moment is given as a function of rapidity in 
Figs.~\ref{dsetdygrv}-\ref{dsetdyrmrsg} both 
with and without impact parameter averaged
shadowing for the GRV 94 LO and MRST LO parton densities at LHC and RHIC.
The average transverse energy given to a particular parton species 
in a central $AB$ collision is then 
\begin{eqnarray}
\overline E_T^f(b,p_0) = 
\frac{d\sigma^f(p_0) \langle E_T^f \rangle}{d^2b} 
\label{aveetaa}
\end{eqnarray}
where $d\sigma^f(p_0) \langle E_T^f \rangle/d^2b$ is the integral of
Eq.~(\ref{etmom}) over $d^2r$. If the nuclear structure functions are
homogeneous, then the spatial effects factorize and $\overline E_T^f(b,
p_0)$ is proportional to $T_{AB}(b)$.  The first $E_T$ moment is
proportional to the energy density, as we discuss shortly.

The second moment of each flavor is calculated similarly,
\begin{eqnarray}
\frac{d\sigma^f (p_0)\langle E_T^{2\,f} \rangle}{d^2b d^2r} 
& = & K_{\rm jet} \int dp_T^2 dy_2 dy dz dz'
\sum_{{ij=}\atop{\langle kl \rangle}}
x_1F_i^A(x_1,p_T^2,\vec{r},z) x_2 F_j^B(x_2,p_T^2,\vec{b} - \vec{r},z')
\nonumber \\ 
&  & \mbox{} \times \frac{p_T^2}{1 + \delta_{kl}} 
\left\{ \left[\delta_{fk} {d\hat\sigma\over d\hat t}^{ij\rightarrow kl}(\hat 
t, \hat u)
+ \delta_{fl} {d\hat\sigma\over d\hat t}^{ij\rightarrow kl}(\hat u, \hat t)
\right] \epsilon(y) \right. \nonumber \\
&  & \mbox{} + \left. \left[{d\hat\sigma\over d\hat t}^{ij\rightarrow ff}(\hat 
t, \hat u)
+ {d\hat\sigma\over d\hat t}^{ij\rightarrow ff}(\hat u, \hat t)
\right] \epsilon(y) \epsilon(y_2) \right\} \, \, . \label{etmom2} 
\end{eqnarray}
The terms proportional to $\epsilon(y) \epsilon(y_2)$ in Eq.~(\ref{etmom2}) 
correspond to only those processes that contain
identical particles in the final state:  $qq \rightarrow qq$,
$\overline q \overline q \rightarrow \overline q \overline q$, $q
\overline q \rightarrow gg$ and $gg \rightarrow gg$.  These
terms are negligible for $f = q$ and $\overline q$ but large for $f = g$.  
Indeed, $ff = gg$ in Eq.~(\ref{etmom2})
contributes $\approx 30$\% of the total second $E_T$ moment of the gluon.  
The second moment is
\begin{eqnarray}
\overline{E^{2\ f}_T}(b,p_0) =
\frac{d\sigma^f(p_0) \langle E_T^{2 \,f} 
\rangle}{d^2b} \,\, 
\label{stddevaa}
\end{eqnarray}
where $d\sigma^f(p_0) \langle E_T^{2 \,f} \rangle/d^2b$ is the
integral of Eq.~(\ref{etmom2}) over $d^2r$.  For homogeneous structure
functions, factorization again occurs and $\overline{E^{2\ f}_T}(b,p_0)$
scales with $T_{AB}(b)$.

We now discuss the results characteristic to specific detectors.  We
will concentrate on the coverage around midrapidity, thereby excluding
some detector subsystems from our consideration.  In all cases we assume full
azimuthal coverage.  At the LHC, there
will be two detectors taking data with heavy ion beams, CMS
\cite{CMS}, optimized for $pp$ studies but with a broad rapidity
coverage, $y_{\rm max}= 2.4$, and ALICE \cite{Alice}, a dedicated
heavy ion experiment with central rapidity coverage up to $y_{\rm
max}= 1$.  We do not include the ALICE forward muon spectrometer.  The
heaviest ions accelerated will be lead.  STAR \cite{STAR} and PHENIX
\cite{cdr} are the two large heavy ion experiments at RHIC, a
dedicated heavy ion collider that will accelerate ions through gold.
STAR has the larger acceptance at central rapidities, $y_{\rm max}=
0.9$ for the electromagnetic calorimeter, while the central electron
arms of PHENIX only cover up to $y_{\rm max}= 0.35$\footnote{Since we quote
results over full azimuthal coverage, the actual PHENIX cross sections would
be lower because the central electron arms only cover a fraction of the total
azimuth.}.  The PHENIX muon
arms will cover more forward rapidities but will not increase the
coverage at midrapidity except for high mass lepton pairs such as
those from $\Upsilon$ decay.  The cross sections per nucleon pair and
the first and second $E_T$ moments with and without
homogeneous shadowing are given
in Tables~\ref{table1} and \ref{table2} in the given CMS and ALICE
rapidity ranges respectively.  At
this energy, shadowing can reduce the parton yield and the
$E_T$ moments by up to
a factor of two.  The corresponding results from RHIC are presented in
Tables~\ref{table3} and \ref{table4} for STAR and PHENIX.  The effect
of shadowing is much smaller at RHIC than at the LHC.  In fact, with
$S_3$, gluon antishadowing can increase the yield relative to $S=1$.
Recall that the cross sections and moments are all calculated with $K_{\rm
jet}=1$ and another choice would scale the results correspondingly.

\begin{figure}[htb]
\setlength{\epsfxsize=0.5\textwidth}
\setlength{\epsfysize=0.7\textheight}
\centerline{\epsffile{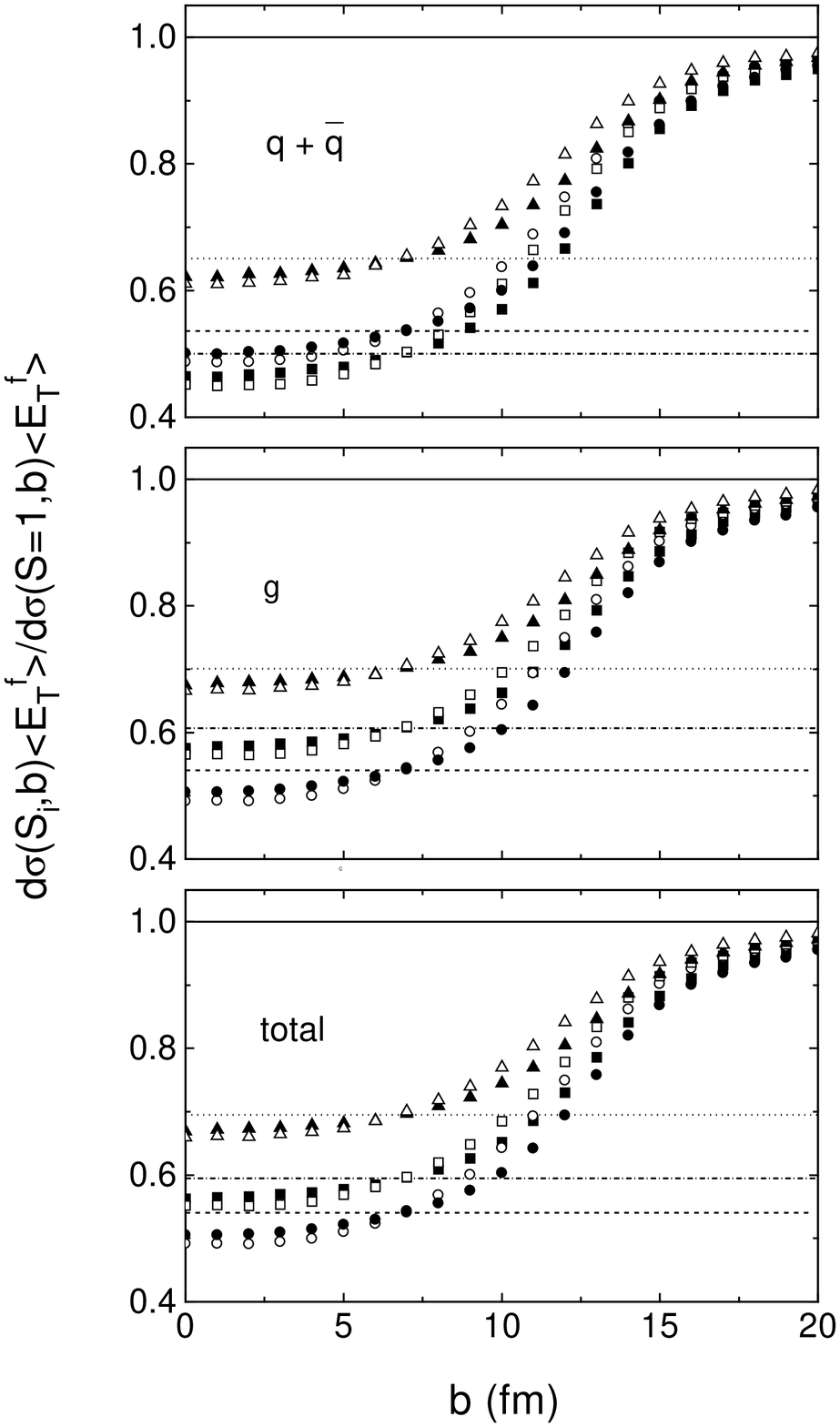}}
\caption[]{The impact parameter dependence of the first $E_T$ moment,
$\sigma(p_0) \langle E_T^f \rangle$ relative to the $E_T$ moment with
$S=1$ in CMS, $|y| \leq 2.4$, calculated with the MRST LO
distributions with $p_0 =2$ GeV.  The upper plot shows the ratio for
quarks and antiquarks, the middle plot is the gluon ratio and the
lower plot is for the total.  The horizontal lines show the homogeneous
shadowing results: dashed for $S_1$, dot-dashed for $S_2$, and dotted
for $S_3$.  The inhomogeneous shadowing results for $S_1$, circles,
$S_2$, squares, and $S_3$, diamonds are shown for $S_{\rm WS}$ (filled
symbols) and $S_{\rho}$ (open symbols).  }
\label{cmsmom1}
\end{figure}

\begin{figure}[htb]
\setlength{\epsfxsize=0.5\textwidth}
\setlength{\epsfysize=0.7\textheight}
\centerline{\epsffile{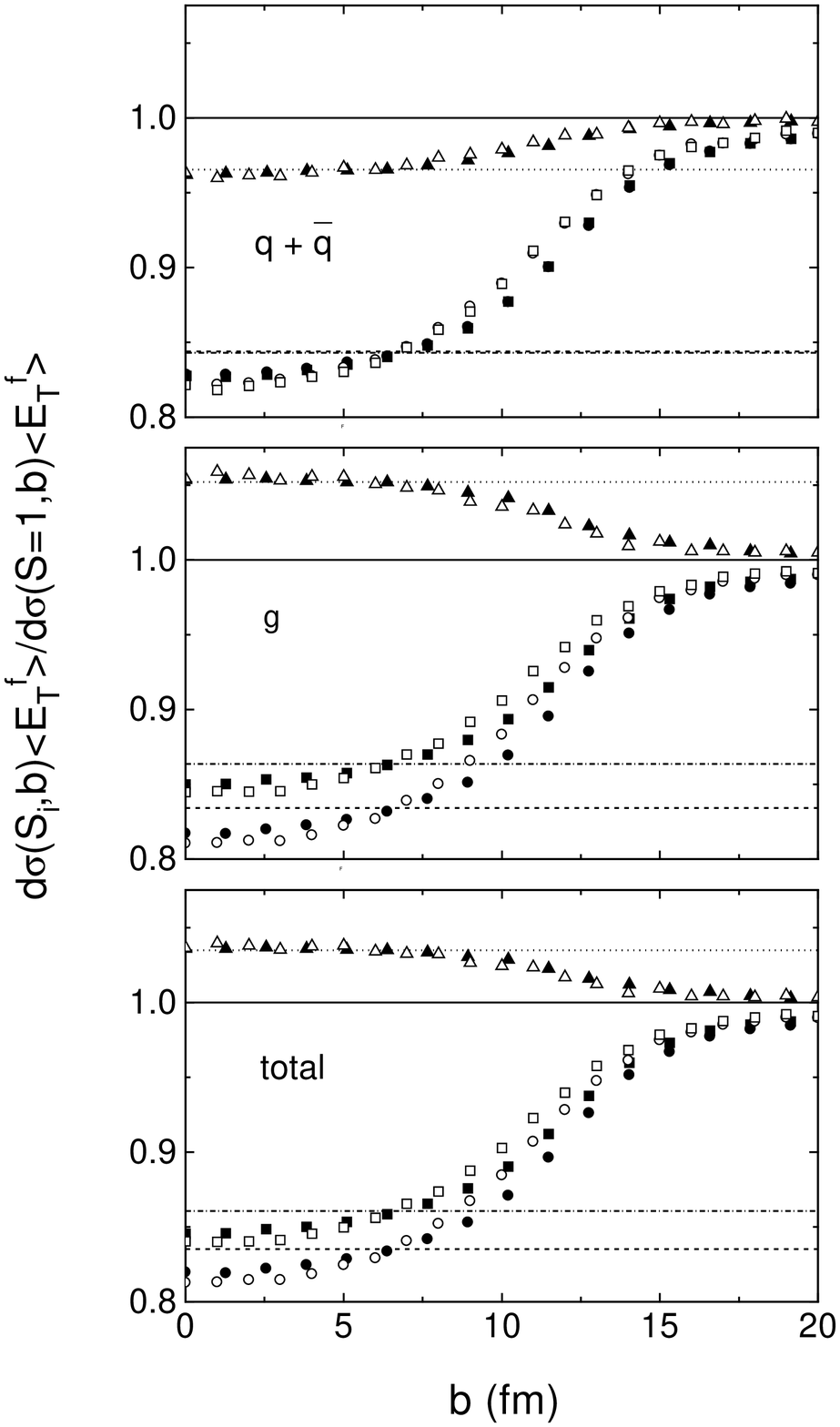}}
\caption[]{The impact parameter dependence of the first $E_T$ moment,
$\sigma(p_0) \langle E_T^f \rangle$ relative to the
$E_T$ moment with $S=1$ in STAR, $|y| \leq 0.9$, calculated with the
MRST LO distributions with $p_0 =2$ GeV.  
The upper plot shows the ratio for quarks
and antiquarks, the middle plot is the gluon ratio and the lower plot
is for the total.  
The horizontal lines show the homogeneous shadowing results: dashed for
$S_1$, dot-dashed for $S_2$, and dotted for $S_3$.  
The inhomogeneous
shadowing results for $S_1$, circles, $S_2$, squares, and $S_3$,
diamonds are shown for $S_{\rm WS}$ (filled symbols) and $S_{\rho}$
(open symbols).  }
\label{starmom1}
\end{figure}

The effect of the inhomogeneous shadowing is shown for the first $E_T$
moment calculated with the GRV 94 LO parton densities in
Figs.~\ref{cmsmom1} and \ref{starmom1} for CMS and STAR.  The ALICE
and PHENIX ratios are similar to those shown here.  The ratios of the
other moments do not differ greatly from the first moment.  The impact
parameter dependence is calculated using Eqs.~(\ref{wsparam}) and
(\ref{rhoparam}). When $x$ lies in the shadowing region, central
collisions are more shadowed than the average.  In the antishadowing
region, central collisions are more antishadowed than the average.
When $b \sim R_A$, the homogeneous and inhomogeneous shadowing are
approximately equal, as might be expected from an inspection of
Eqs.~(\ref{wsparam}) and (\ref{rhoparam}).  When $b \sim 2R_A$, the
shadowing or antishadowing is significantly reduced.  As $b$ further
increases, the approach to $S=1$ is asymptotic.  With $S_\rho$,
Eq.~(\ref{rhoparam}), the central shadowing is somewhat stronger than
with $S_{\rm WS}$ and the strength of the shadowing decreases more
rapidly when $b>R_A$.  At $b \sim 2R_A$, the ratio with $S_\rho$ is
5\% higher than with $S_{\rm WS}$.  A calculation with $S_{\rm R}$,
Eq.~(\ref{rparam}), would have somewhat stronger shadowing than
$S_\rho$.  Since the inhomogeneous calculations agree within a few
percent, the exact dependence cannot be experimentally distinguished;
only the presence of inhomogeneity can be detected.  This conclusion
applies to a wide variety of observables\cite{us}.  Because the
differences are small, we use only the $S_{\rm WS}$ parameterization
in the remainder of this work.

The figures show the ratio of the first $E_T$ moment with shadowing
included relative to $S=1$ as a function of impact parameter for $q +
\overline q$, $g$, and the total, $q + \overline q +g$.  At the LHC,
quarks and antiquarks are $\approx 10$\% of the total minijet
production for the GRV 94 LO parton densities and $\approx 17$\% with
the MRST LO densities when $S=1$.  The overall $q + \overline q$
contribution decreases 1-2\% when shadowing is included.  At RHIC, the
$q + \overline q$ fraction is $\approx 19$\% with the GRV 94 LO
densities and $\approx 23$\% with the MRST LO set.  There is again
only an $\approx 1$\% variation in the fraction with shadowing
included.  The ratios in the given rapidity intervals with homogeneous
shadowing are given by the horizontal lines on Figs.~\ref{cmsmom1} and
\ref{starmom1} and correspond to the ratio of the moments in
Figs.~\ref{dsetdygrv} and \ref{dsetdyrgrv} integrated over the same
rapidity intervals.  At the LHC, the ratios are nearly the same for
the gluon fraction and the total because gluons dominate minijet
production.  At RHIC energies, since the $q + \overline q$
contribution is a larger fraction of the total, the difference between
the $g$ ratio and the total, shown in Figs.~\ref{cmsmom1} and
\ref{starmom1}, is visible, particularly for $S_3$ which is shadowed
for $q + \overline q$ and antishadowed for the gluons.  The total
remains antishadowed, but less than for gluons alone.

The homogeneous shadowing ratios can also be determined for the zeroth
moment, particle number, and the second moment of of the $E_T$
distribution, from Tables~\ref{table1}-\ref{table4}.   The second $E_T$
moment has a slightly smaller $q + \overline q$ fraction due to the
term in Eq.~(\ref{etmom2}) proportional to $\epsilon(y) \epsilon(y_2)$
which arises from identical particles in the final state.  The
dominant contribution to this term is $gg \rightarrow gg$, enhancing
the overall gluon contribution.

We now show how the initial conditions for further evolution of the system are
impacted by shadowing.  The initial energy density of each parton species is 
the ratio of the first
$E_T$ moment to the initial volume
\be \epsilon_i^f(b, p_0) 
= \frac{\overline E_T^f(b, p_0)}{V_i} 
\, \, . \label{edens} 
\ee 
The total initial energy density is the sum
over all species, $\epsilon_i = \sum_f \epsilon_i^f$.
The initial number densities are likewise
\be n_i^f(b, p_0) = \frac{\overline 
N^f(b, p_0)}{V_i} 
\label{ndens} \ee 
and $n_i = \sum_f n_i^f$.  The energy and number densities are given
in Tables \ref{table5}-\ref{table8}, both for gluons only and the
total minijet yield. Results are shown for both homogeneous and
inhomogeneous shadowing at $b=0$ where the volume is most clearly
defined.  Since shadowing is
stronger in central collisions, the energy density and multiplicity
are reduced with $S^i_{\rm WS}$ relative to the homogeneous case.  (See
Figs.~\ref{cmsmom1} and \ref{starmom1} for the impact parameter
dependence of $\overline E_T^f$.)  At the LHC, inhomogeneous shadowing
reduces the energy density by $\approx 3$-8\% at $b=0$.   At RHIC, the
difference is smaller and the energy density may even rise marginally
at $b=0$ with the $S_3$ parameterization.  
The average energy per particle for a given species is
$\epsilon_i^f/n_i^f \sim 3$ GeV, somewhat larger than $p_0$, as can be
expected since $\overline E_T^f$ reflects the average $p_T$ within the
rapidity range.

These densities can be compared to those obtained
for an ideal gas in thermal equilibrium.  An ideal gas has energy density
$\epsilon_{\rm th} = 3a T_{\rm th}^4$ and entropy density $s_{\rm th}
= 4a T_{\rm th}^3 = 3.6 n_{\rm th}$ where $a = 16 \pi^2/90$ for a
gluon gas and $a = 47.5 \pi^2 /90$ for a three flavor quark-gluon
plasma\footnote{N. Hammon {\it et al.}\ also calculated $\overline N$ and 
$\overline E_T$ using spatially
homogeneous nuclear structure functions at RHIC and LHC \cite{noteadd}.  
Since they take $K_{\rm jet} > 1$, they find larger energy densities and
effective temperatures than we do here.  They also neglected the unitarity
problem in their LHC estimates.}.  
The initial equilibrium temperatures of such gases are then
$T_{\rm th} = (\epsilon_{\rm th}/3a)^{1/4}$ and the ideal energy per
particle is \be \frac{\epsilon_{\rm th}}{n_{\rm th}} = 2.7 \, T_{\rm
th} \, \, .
\label{idealrat} \ee  We use the results of Tables \ref{table5}-\ref{table8}
with the assumption that $\epsilon_i= \epsilon_{\rm th}$.  When $S=1$,
$T_{\rm th} \approx 1.07$ GeV for gluons only and $\approx 840$ MeV
for a quark-gluon plasma at the LHC with the GRV 94 LO distributions
and $p_0=2$ GeV.  Using the MRST LO results with $p_0=2$ GeV yields
$T_{\rm th} \approx 860$ MeV and 680 MeV respectively.  The calculated
initial quark-gluon plasma temperature is lower than that for gluons
because, even though $\epsilon_i$ is larger for the sum of all
species, the larger number of available degrees of freedom reduces the
temperature.  Shadowing reduces $T_{\rm th}$ by 10-17\% for the gluons
and 10-15\% for the total with the largest effect due to $S_1$ and the
smallest from $S_3$ with its antishadowing.  At RHIC, the equivalent
temperatures extracted with $p_0=2$ GeV are smaller, 410 MeV
for gluons and 330 MeV for a quark-gluon plasma with $S=1$.  The
reduction due to shadowing is 5\% or less---in fact a slight
enhancement is possible because of the antishadowing in $S_3$.  The
temperatures are virtually independent of the parton distributions at
this energy since the two sets are very similar in the $x$ range of
RHIC.  The temperatures estimated for RHIC are
lower than those obtained elsewhere.  This difference will be
discussed in the next section.

These equivalent equilibrium temperatures are only approximate because
they depend on the rapidity range over which $T_{\rm th}$ is
calculated.  The extracted temperature rises as the rapidity range
decreases because the antiquark and gluon distributions are maximal at
$y=0$.  The fact that the width of the slice affects $T_{\rm th}$
shows that thermalization in the collision is incomplete.  To study
this further, we can compare these results with expectations from the
ideal gas.  The GRV 94 LO gluon temperature, $T_{\rm th} \approx 1.07$
GeV, satisfies Eq.~(\ref{idealrat}) when $S=1$, {\it i.e.}  \be
\frac{\epsilon_i^g(b=0, p_0)}{n_i^g(b=0, p_0)}\bigg|_{S=1} \approx
\frac{\epsilon_{\rm th}^g}{n_{\rm th}^g} = 2.7 \, T_{\rm th} \, \,
. \ee This equation suggests that, even if a quark-gluon plasma is far
from equilibrium, the gluons might equilibrate quickly, around $\tau_i
\sim 0.1$ fm, even without the secondary collisions required for
isotropization.  However, even this suggestion only holds at LHC
energies without shadowing. Shadowing drives the result away from
equilibrium so that \be \frac{\epsilon_i^g(b, p_0)}{n_i^g(b, p_0)}
\bigg|_{S \neq 1} > \frac{\epsilon_{\rm th}^g}{n_{\rm th}^g} = 2.7 \,
T_{\rm th} \, \, . \ee Note however that taking $K_{\rm jet}=1.5$
increases all the extracted temperatures by $\approx 10$\%, bringing
the shadowed results closer to the ideal in Eq.~(\ref{idealrat}).
Reducing $p_0$ for $S \neq 1$ at the LHC or for any scenario
at RHIC would also increase the
extracted temperatures so that the gluon result would again appear to
equilibrate.  The quarks alone or the $q + \overline q + g$ total will
not come to equilibrium, even when $p_0$ is reduced, due to the lower
equivalent temperature.  In any case, $p_0$ cannot be set arbitrarily
low for perturbative QCD to be valid.  In addition, $p_0$ should not be a
strong function of energy and should be independent of the shadowing
parameterization.
Shadowing thus reduces the likelihood of fast thermalization, even at the
LHC where the conditions are most favorable.

Given these uncertainties, one can nevertheless obtain an approximate lower
bound on the produced particle multiplicity.
In an ideal longitudinally expanding plasma, the energy 
density evolves following\cite{Bjorken}
\begin{eqnarray}
\frac{d\epsilon}{ d\tau} + \frac{\epsilon+P}{\tau} = 0
\end{eqnarray}
where $P$ is the pressure and $\tau$ is the proper time.  There are
two extreme solutions: free streaming, with $P=0$, leading to
$\epsilon \sim \tau^{-1}$, and ideal hydrodynamics, $P=\epsilon/3$,
where $\epsilon \sim \tau^{-4/3}$.  The lower limit of multiplicity is
obtained from ideal hydrodynamics where the system is treated as it it
were in thermal equilibrium at $\tau_i=1/p_0=0.1$ fm and expands
adiabatically with $\tau$. Then the initial entropy determines the
final-state multiplicity, neglecting final-state interactions, fragmentation,
and hadronization.  If only minijet production contributes to
the final-state multiplicity, the total number of particles in a
specific detector's central acceptance is then \cite{eskola}
\begin{eqnarray}
\frac{dN}{dy} \approx \frac{4}{3.6}\biggl[\frac{\tau_i \pi R_A^2a}{27}
\left\{\frac{\overline E_T (|y| \leq y_{\rm max})}{2y_{\rm max}}
\right\}^3 \biggr]^{\frac{1}{4}}. \, \, \label{mult} \end{eqnarray}
Equation~(\ref{mult}) suffers from some uncertainty due to the
$\tau_i^{1/4} \sim p_0^{1/4}$ dependence in the volume besides the dependence
on $p_0$ in $ \overline E_T$. 
In a complete calculation, the variation with $p_0$ would be
compensated by a corresponding variation in the soft component, as discussed
later.  With the GRV 94 LO
distributions at the LHC the total $dN/dy$ at $y=0$ from minijets is $\approx
4000-6000$ or about $2700-4000$ charged particles, $\approx 2/3$ the total 
$dN/dy$.  Shadowing reduces
the number of charged particles to $\approx 1800-2600$.  With the MRST LO
distributions, the total $dN/dy|_{y=0} \approx 2000-3500$ without shadowing
and $1400-2600$ with shadowing.  With inhomogeneous shadowing, the
LHC multiplicity drops 2-5\% for collisions at $b=0$.  The gluon $E_T$
moment dominates the total and drives the rapidity distribution, as
can be inferred from Figs.~\ref{dsetdygrv} and \ref{dsetdymrsg}.
We find total minijet multiplicities of
$220-350$ without shadowing and $200-360$ with shadowing.  The larger
$dN/dy$ with shadowing is a result of the antishadowing in $S_3$.
Since soft production is large at RHIC, the total $dN/dy$
found here is considerably lower than predicted by some event
generators \cite{HIJING}.  

\section{Correlation between E$_T$ and impact parameter}

Thus far, we have discussed the dependence of shadowing on the impact
parameter, a quantity which cannot be directly determined in a heavy ion
collision.  However, although the impact parameter is not measurable
it can be related to direct observables such as the transverse
energy, $E_T$ \cite{us,EKinit}.  The transverse energy 
is summed over all detected particles in the
event with masses $m_k$ and transverse momenta $p_{Tk}$ so that $E_T=\Sigma_k
\sqrt{m_k^2+p_{Tk}^2}$.  Besides an
$E_T$ measurement, it is also possible to infer the impact parameter
by a measurement of the nuclear breakup since the beam remnants
deposited in a zero degree calorimeter are also correlated with the
impact parameter \cite{ZDC}.  A measure of the total charged particle
multiplicity, proportional to $E_T$ \cite{DNDY}, could also refine the
impact parameter determination.

The transverse energy contains ``soft'' and ``hard'' components. The
``hard'' components, calculated in the previous section, arise from
quark and gluon interactions above the cutoff $p_0 = 2$ GeV.  
``Soft'' processes with $p_T<p_0$ are not perturbatively
calculable yet they can contribute a substantial fraction of the
measured $E_T$ at high energies (and essentially the entire $E_T$ at
CERN SPS energies).  These processes must be modeled
phenomenologically.  Our calculation of the total $E_T$
distribution follows Ref.\ \cite{EKinit}.
We assume that the soft cross section,
$\sigma^S_{pp}$, is equal to $\sigma^{\rm inelastic}_{pp}$, the inelastic $pp$
scattering cross section.  The hard part of the $E_T$
distribution can be expressed as
\begin{equation}
{d\sigma^H \over dE_T} = \int d^2b {\large \Sigma}_{N=1}^\infty 
{[\overline N^H(b)]^N \over N!} \exp{[-\overline N^H(b)]}
\int \prod_{i=1}^N dE_{Ti} {1 \over \sigma_{pp}^H} {d\sigma_{pp}^H \over 
dE_{Ti}} \delta(E_T - {\large \Sigma}_{i=1}^N E_{Ti}) \, \, . \label{fullet}
\end{equation}
The average number of hard parton-parton collisions is defined in
Eq.~(\ref{ntot}).  For most
$b<2R_A$, $\overline N^H$ is large and $d\sigma^H/dE_T$ can be
approximated by the Gaussian \cite{EKinit}
\begin{equation}
{d\sigma^H \over dE_T} = \int  {d^2b \over \sqrt{2\pi\sigma^{2\ H}_E(b)}}
\exp{\bigg( \,- \, { [E_T-\overline E_T^H(b)]^2 
\over 2\sigma^{2\ H}_E(b)} \bigg)}, \label{etgauss} \end{equation}
where the mean $E_T$, $\overline E_T^H(b)$,
is proportional to the first moment of the
hard cross section,
\begin{eqnarray}
\overline E_T^H(b)  = \Sigma_f \overline E_T^f (b,p_0).
\end{eqnarray}
The standard deviation, $\sigma^H_E(b)$ is computed from the
first and second moments,
\begin{equation}
\sigma^{2\ H}_E(b) = \sum_f \overline{E_T^{2\ f}} (b,p_0) - 
\frac{\overline E_T^{H \, 2}(b)}{\sigma^H(b,p_0)}
\, \, ,
\end{equation}
see Eqs.~(\ref{aveetaa}) and (\ref{stddevaa}).  The impact parameter
averaged values of the hard cross section and its first and second $E_T$
moments correspond to the ``total'' values in
Tables~\ref{table1}-\ref{table4} for the specified rapidity coverages
of the four detectors.  Note that these moments are a lower bound on particle
production from hard processes because hadronization has not been included.

The soft component is usually taken to be proportional to the number
of nucleon-nucleon collisions,
\begin{equation}
\overline N^S (b) = T_{AB}(b)\sigma_{pp}^S \, \, ,
\end{equation}
where $\sigma_{pp}^S \sim 40$ mb at RHIC and may increase to 60 mb at
the LHC \cite{pdg}.  Since $\sigma_{pp}^S$ depends only weakly on the
collision energy, the hard and soft components
are assumed to be separable on the $pp$ level 
and thus independent of each other at
fixed $b$\cite{EKinit}.  The soft component may be computed using the
first moment, $\overline E_T^S$, and second moment, $\overline{E_T^2}^S$, of the soft
$E_T$ distributions, obtainable from lower energy data
\cite{etdata}.  At the SPS, $\sqrt{s_{NN}}=19.4$ GeV,
$\sigma_{pp}^S=32$mb, $\overline E_T^S=15$ mb GeV and $\overline{E_T^2}^S=50$
mb GeV$^2$\cite{EKinit} for $|y| \leq 0.5$.  We assume that
$\overline E_T^S$ and $\overline{E_T^2}^S$ are independent of impact parameter
and scale with energy as $\sigma_{pp}^S$ and linearly
with rapidity acceptance.  The resulting
first and second moments for the four detectors
are given in Table~\ref{softet}. Alternatively, $\overline E_T^S$ and 
$\overline{E_T^2}^S$ could be scaled by the
charged particle production rate in the selected rapidity
interval\cite{EKinit}.  However, at these higher energies, the 
charged particle distributions will have a strong contribution from
hard production which could lead to double counting of the total rate.  
If the SPS multiplicity distribution is used,
then the effect of the rising cross section will dominate.

The total $E_T$ distribution is a convolution of the hard and soft
components with mean and standard deviation
\begin{eqnarray}
\overline E_T(b) & = &\overline E_T^H(b) 
+  T_{AB}(b) \overline E_T^S \label{sumave} \\
\sigma^2_E(b) & = & \sigma^{2\ H}_E(b) 
+ T_{AB}(b) \sigma_E^{2\ S} \label{stdave} \, \, .
\end{eqnarray}
The standard deviation for the soft component,
$\sigma_E^{2 \, S}$, is
\begin{equation}
\sigma_E^{2\ S}=\overline{E_T^2}^S-{\overline E_T^{S\ 2} \over \sigma^S_{pp}}
\, \, .
\end{equation}
We do not assume that the second moment $\overline{E_T^2}^S$ is equivalent to 
the standard deviation as in some previous calculations \cite{us,EKinit}.

Some caveats related to the soft $E_T$ contribution should be mentioned.
The second moment is system dependent\footnote{See $\omega$ in Table 8 of
Ref.~\cite{etdata}. In lighter targets $\omega$ is significantly
different than in heavy targets.}, perhaps because the fluctuations are
concentrated in the central region, making $\overline{E_T^2}^S$ sensitive
to the acceptance\cite{etproblem}.  There also may be some
contamination from hard processes.  Additionally, soft processes may also be
subject to a form of shadowing due to large mass
diffraction\cite{capella},
analogous to the multiple-scattering picture of shadowing except that
it affects soft interactions.  If correct,
the soft component would also be reduced 
and the soft and hard interactions would have a
similar impact parameter dependence.   Thus the
soft component is only accurate to the
20\% level at best.

At the LHC, the hard component is an order of magnitude larger than
the soft part. This can be seen from a comparison of the homogeneous
shadowing first and second $E_T$ moments in
Tables~\ref{table1}-\ref{table4} with $\overline E_T^S$ and
$\overline{E_T^2}^S$ in Table~\ref{softet}.  The results are directly
comparable because the first and second $E_T$ moments in all the
tables are given per nucleon pair.  At the LHC, the moments from
minijet production are $6-10$ times larger than $\overline E_T^S$ with
the GRV 94 LO parton densities and $3-7$ times larger than $\overline
E_T^S$ when calculated with the MRST LO parton densities. Total
particle production is then dominated by minijet production.  With
soft production included, the estimated $dN/dy$ in Sec. III would be
increased by $5-22$\%, less than the change due to shadowing.

At RHIC however, $\overline E_T^S$ is $1.3-2.2$ times larger than the first
$E_T$ moment, depending on the parton densities and shadowing
parameterization.  Thus the soft contribution to the total $E_T$ is
still somewhat larger than the hard contribution.  When soft
production is included in the estimated $dN/dy$ by adding $T_{AB}(b)
\overline E_T^S$ to $\overline E_T^H(b,p_0)$ in Eq.~(\ref{mult}), $dN/dy$ 
at $b=0$ could
increase by a factor of $1.9-2.4$, up to $680-750$ particles. 
Likewise, the extracted initial
temperature assuming thermal equilibrium would be $\approx 20$\%
higher when the soft contribution is included and could reach $\approx
500$ MeV with $S=1$, consistent with previous predictions \cite{XNW}.
If soft production is also affected by shadowing \cite{capella}, 
then the soft contribution to RHIC
central collisions would be reduced and the hard and soft components would
be more in balance.

\begin{figure}[htb]
\setlength{\epsfxsize=0.5\textwidth}
\setlength{\epsfysize=0.7\textheight}
\centerline{\epsffile{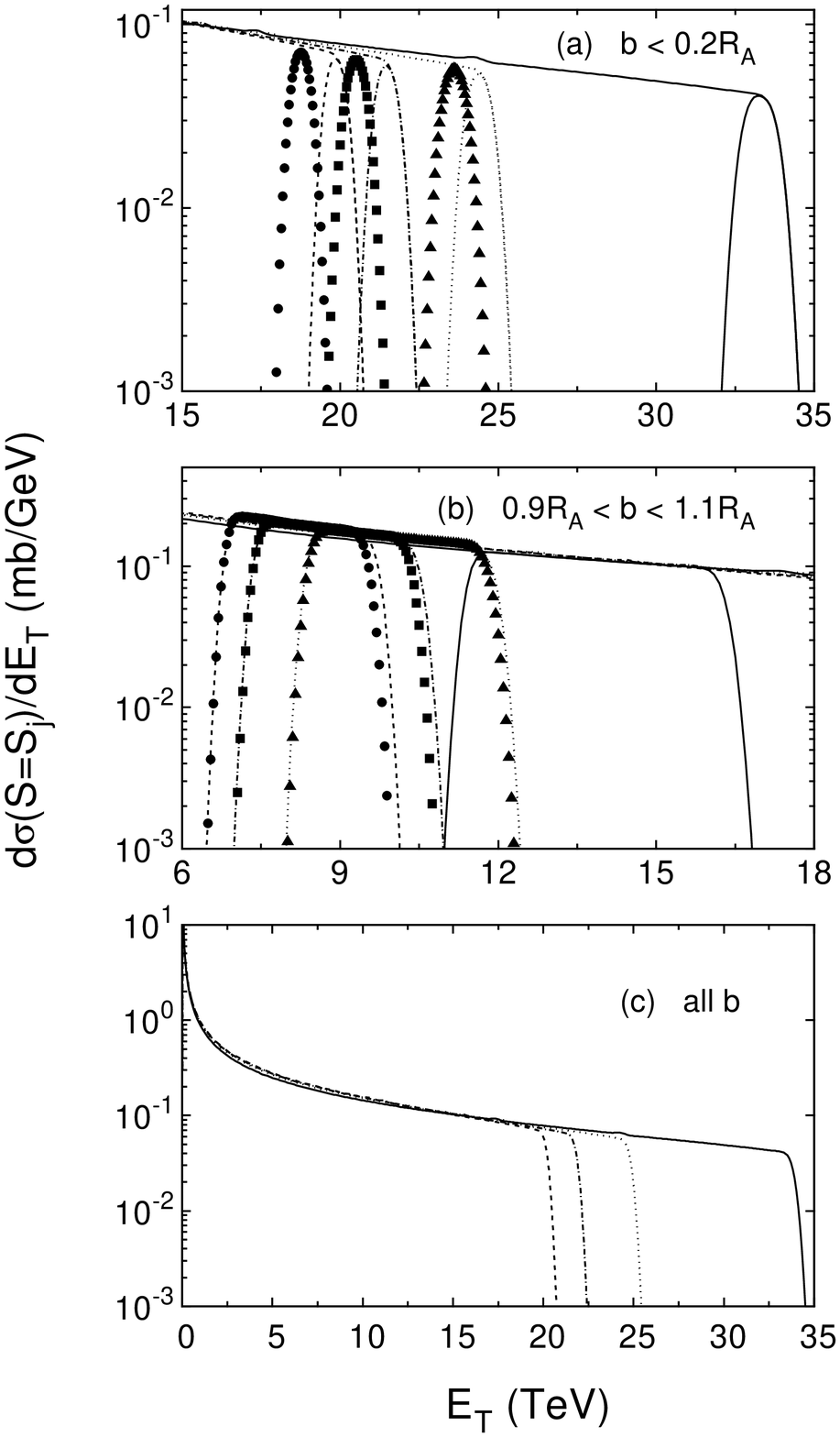}}
\caption[]{The $E_T$ distribution predicted for the CMS detector in
the interval $|y| \leq 2.4$, calculated with the MRST LO
distributions and $p_0=2$ GeV.  
The upper plot is for central collisions with
$b<0.2 R_A$, the middle plot shows the region $0.9R_A < b < 1.1R_A$,
and the lower plot shows the entire $E_T$ distribution.  
The lines indicate the homogeneous shadowing results: solid for no
shadowing, dashed line for $S_1$, dot-dashed for $S_2$, and dotted for
$S_3$.  The inhomogeneous shadowing results for $S_1$, circles, $S_2$,
squares, and $S_3$, diamonds are shown for $S_{\rm WS}$.}
\label{cmset}
\end{figure}

\begin{figure}[htb]
\setlength{\epsfxsize=0.5\textwidth}
\setlength{\epsfysize=0.7\textheight}
\centerline{\epsffile{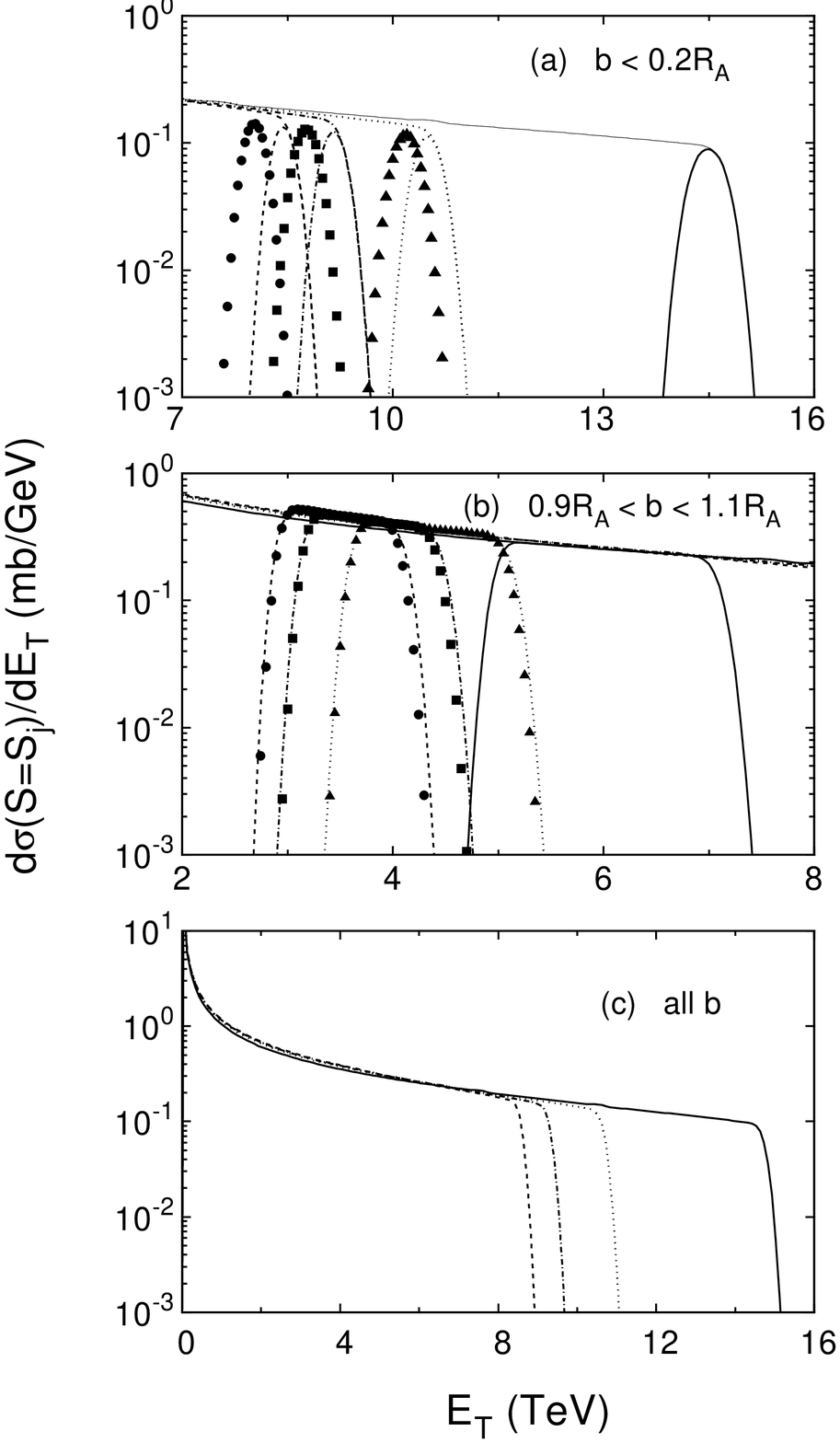}}
\caption[]{The $E_T$ distribution predicted for the ALICE detector in
the interval $|y| \leq 1$, calculated with the MRST LO
distributions and $p_0=2$ GeV.  
The upper plot shows central collisions with
$b<0.2 R_A$, the middle plot shows the region $0.9R_A < b < 1.1R_A$,
and the lower plot shows the entire $E_T$ distribution.  
The lines indicate the homogeneous shadowing results: solid for no
shadowing, dashed line for $S_1$, dot-dashed for $S_2$, and dotted for
$S_3$.  The inhomogeneous shadowing results for $S_1$, circles, $S_2$,
squares, and $S_3$, diamonds are shown for $S_{\rm WS}$.}
\label{aliceet}
\end{figure}

\begin{figure}[htb]
\setlength{\epsfxsize=0.5\textwidth}
\setlength{\epsfysize=0.7\textheight}
\centerline{\epsffile{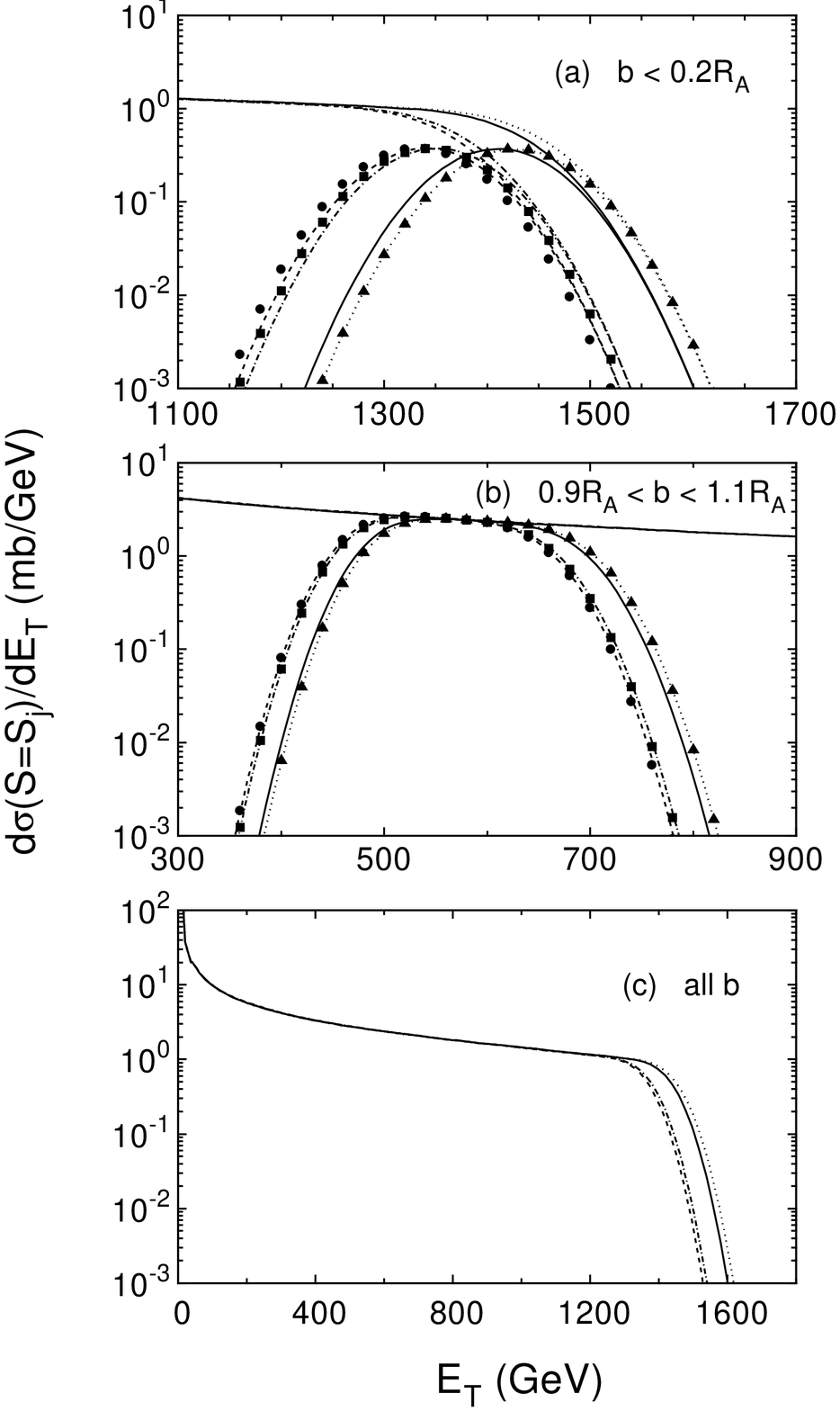}}
\caption[]{The $E_T$ distribution predicted for the STAR detector in
the interval $|y| \leq 0.9$, calculated with the MRST LO
distributions and $p_0 = 2$ GeV.  
The upper plot is for central collisions with
$b<0.2 R_A$, the middle plot shows the region $0.9R_A < b < 1.1R_A$,
and the lower plot shows the entire $E_T$ distribution.  
The lines indicate the homogeneous shadowing results: solid for no
shadowing, dashed line for $S_1$, dot-dashed for $S_2$, and dotted for
$S_3$.  The inhomogeneous shadowing results for $S_1$, circles, $S_2$,
squares, and $S_3$, diamonds are shown for $S_{\rm WS}$.}
\label{staret}
\end{figure}

\begin{figure}[htb]
\setlength{\epsfxsize=0.5\textwidth}
\setlength{\epsfysize=0.7\textheight}
\centerline{\epsffile{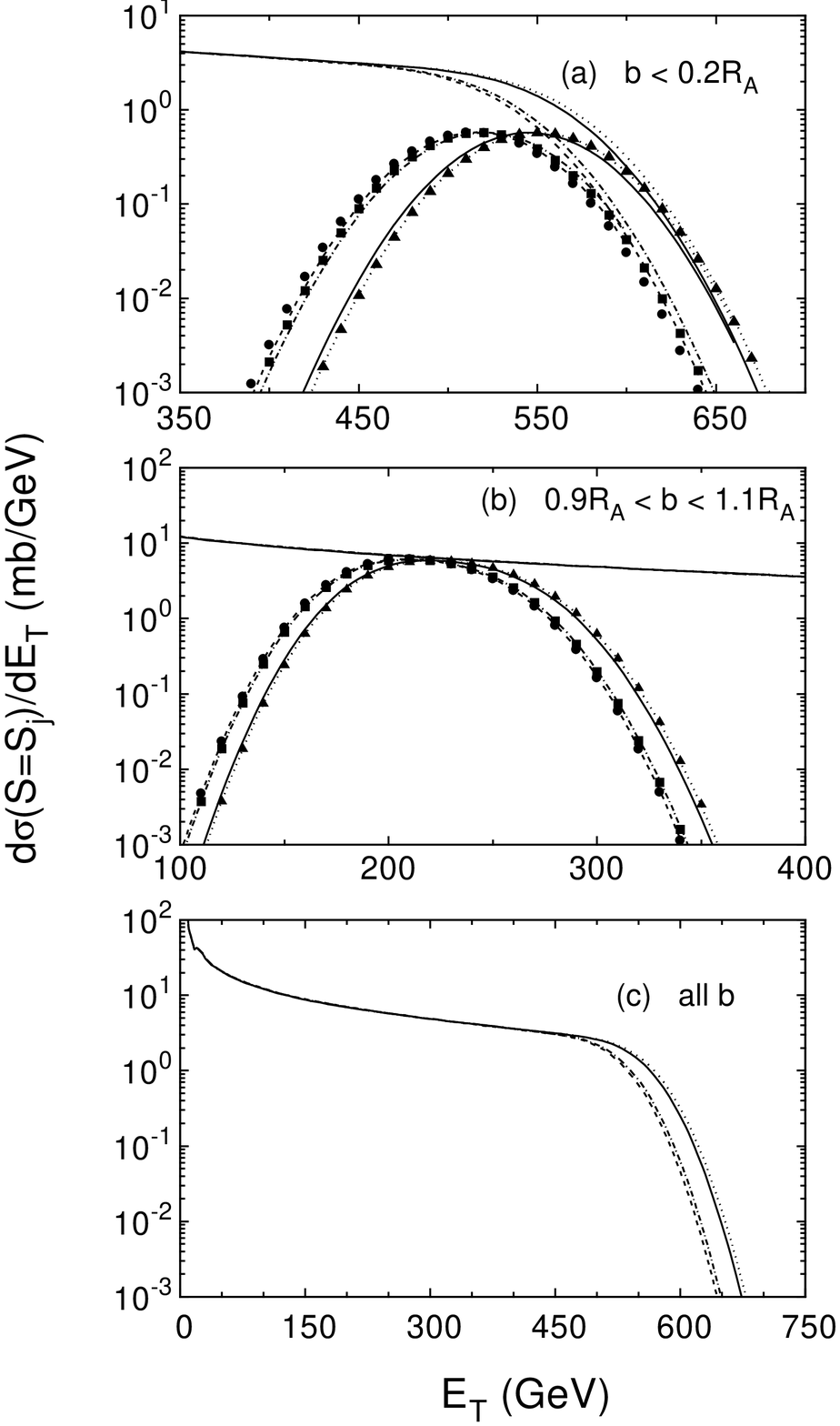}}
\caption[]{The $E_T$ distribution predicted for the PHENIX detector in
the interval $|y| \leq 0.35$, calculated with the MRST LO
distributions and $p_0 = 2$ GeV.  
The upper plot emphasizes central collisions with
$b<0.2 R_A$, the middle plot shows the region $0.9R_A < b < 1.1R_A$,
and the lower plot shows the entire $E_T$ distribution.  
The lines indicate the homogeneous shadowing results: solid for no
shadowing, dashed line for $S_1$, dot-dashed for $S_2$, and dotted for
$S_3$.  The inhomogeneous shadowing results for $S_1$, circles, $S_2$,
squares, and $S_3$, diamonds are shown for $S_{\rm WS}$.}
\label{phenixet}
\end{figure}

The $E_T$ distributions, with homogeneous and inhomogeneous shadowing
are shown for each detector in Figs.~\ref{cmset}-\ref{phenixet}.  The
hard component is calculated with the MRST LO distributions.  In each
case, we show the change in the $E_T$ distribution due to shadowing in
the most central collisions, $b<0.2R_A$, semi-central collisions,
$0.9R_A < b < 1.1R_A$, and the entire $b$ range.  The maximum $E_T$ is
reduced $30-40$\% at the LHC because the hard component,
Eq.~(\ref{sumave}), dominates the average $E_T$.  At intermediate
impact parameters, the Gaussian, Eq.~(\ref{etgauss}), is narrowed by
shadowing.  At RHIC, since the hard and soft components are
comparable, the maximum $E_T$ is shifted by only $\sim 7$\% when
shadowing is included.  Indeed, for $S_3$, since shadowing enhances
the $E_T$ moments of the hard component, the maximum $E_T$ is slightly
increased.  If the GRV 94 LO distributions are used in the calculation
of the hard part, the total $E_T$ at the LHC is nearly twice as large
and the shadowing effects are stronger.  The RHIC results are
essentially unaffected by the choice of parton distribution since the
$E_T$ moments do not depend strongly on the parton distribution, see
Tables~\ref{table3} and \ref{table4}.

These results depend on $K_{\rm jet}$ since the hard $E_T$ is
proportional to $K_{\rm jet}$.  At the LHC, $E_T$ scales nearly
linearly with $K_{\rm jet}$ since hard interactions dominate there.
At RHIC, the increase would be smaller, since only 30\%-50\% of the
$E_T$ comes from hard processes; if $K_{\rm jet}=1.5$, then the
maximum $E_T$ rises by 20\%. Similar results were found in
Ref.~\cite{LO}.

The change in the $E_T$ distribution due to shadowing is not
equivalent to scaling $E_T$ by a constant.  The shape of the
distribution is also modified because central and peripheral
collisions are affected differently.  The shape change is small at
RHIC, but clearly visible for the LHC. Figures~\ref{cmset} and
\ref{aliceet} show that the shadowed distributions are enhanced over
$S=1$ for $E_T \approx 6$ TeV and 3 TeV for CMS and ALICE
respectively.  If soft production is also affected by shadowing
\cite{capella}, the shape change may be larger for RHIC.

For semi-central through central collisions, the transverse
energy-impact parameter correlation is relatively easy to determine
but in very peripheral collisions, the entire transverse energy could
arise from a single hard collision which produces {\it e.g.} a
$J/\psi$ or a Drell-Yan pair.  Then, the simple Gaussian approximation
to Eq.~(\ref{fullet}) would break down.

\section{Drell-Yan, $J/\psi$ and $\Upsilon$ Production}

We now study the effect of inhomogeneous shadowing on the production
of hard probes.  As examples, we consider Drell-Yan and quarkonium
production.  We have previously studied the production of charm and
bottom quarks at these energies\cite{firstprl}.  We have also
considered shadowing effects on $J/\psi$ and Drell-Yan production at
the SPS, as well as their ratio as a function of $E_T$
\cite{secondprl}.  However, at the SPS, $E_T$ is dominated by the soft
component and is proportional to the number of participants
\cite{DNDY}.  We do not include final-state absorption effects on
quarkonium production.

These calculations are done at leading order to be consistent with our
calculations of minijet production.  The LO cross section for nuclei
$A$ and $B$ colliding at impact parameter $b$ and producing a vector
particle $V$ (quarkonium or $\gamma^*$) with mass $m$ at scale $Q$ is
\begin{eqnarray} \frac{d\sigma^V}{dy dm^2
d^2b d^2r} = \sum_{i,j}\int \,dz \,dz'
F_i^A(x_1,Q^2,\vec{r},z) F_j^B(x_2,Q^2,\vec{b} - \vec{r},z')
\frac{d\widehat{\sigma}_{ij}^V}{dy dm^2} \, \,  ,
\label{sigmajpsi}
\end{eqnarray}
where $\widehat{\sigma}_{ij}^V$ is the partonic $ij \rightarrow V$
cross section and the parton distributions are defined
in Eq.~(\ref{fanuc}).

The LO Drell-Yan cross section per nucleon
must include the nuclear isospin since, in general, $\sigma_{pp}^{\rm DY} \neq
\sigma_{pn}^{\rm DY} \neq \sigma_{np}^{\rm DY} \neq \sigma_{nn}^{\rm DY}$,
\begin{eqnarray}
\lefteqn{f_i^N(x_1,Q^2) f_j^N(x_2,Q^2) \frac{d\widehat{\sigma}_{ij}^{\rm DY}
}{dy dm^2} = K_{\rm exp} \, \frac{4 \pi \alpha^2}{9m^2s}} 
\label{sigmady} \\ & & \mbox{} 
\times \sum_{q=u,d,s} e_q^2 [ \left\{{Z_A \over A} f_q^p(x_1,Q^2)+ {N_A 
\over A} f_q^n(x_1,Q^2) \right\} \left\{{Z_B \over B}
f_{\overline q}^p(x_2,Q^2) + {N_B \over B} 
f_{\overline q}^n(x_2,Q^2)\right\} + q 
\leftrightarrow \overline q ] \nonumber \, \, ,
\end{eqnarray}
where $Z_A$ and $N_A$ are the number of protons and neutrons in the
nucleus.  We assume
charge symmetry, $f_u^p = f_d^n$, $f_d^p = f_u^n$ {\it etc.}, in the nuclear
environment.  In Eq.~(\ref{sigmady}), 
$x_{1,2} = Qe^{\pm y}/\sqrt{s_{NN}}$ and $Q = m$.
The factor $K_{\rm exp}$, typically $1.7-2$ for fixed-target Drell-Yan
production,  
accounts for the difference in magnitude between the calculations and the data.

\begin{figure}[htb]
\setlength{\epsfxsize=0.5\textwidth}
\setlength{\epsfysize=0.7\textheight}
\centerline{\epsffile{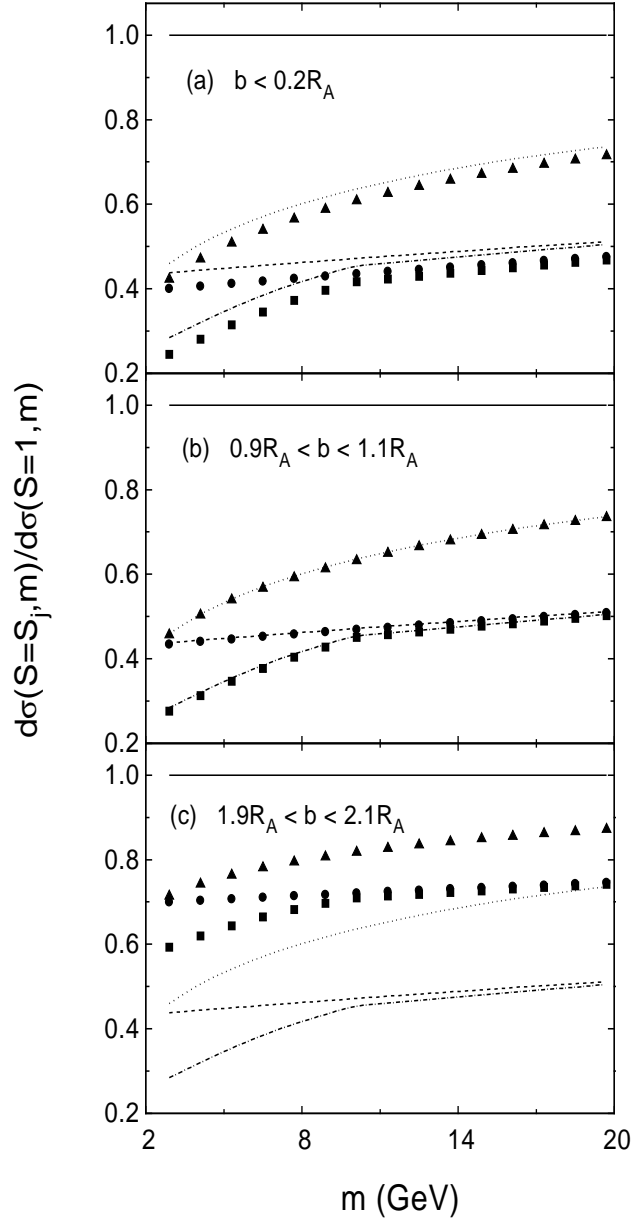}}
\caption[]{The Drell-Yan mass distribution relative to 
$S=1$ in ALICE, $|y| \leq 1$, calculated with the
MRST LO distributions.  The upper plot shows central collisions
with $b<0.2 R_A$, the middle plot shows the region $0.9R_A < b <
1.1R_A$, and the lower plot shows the peripheral region $1.9R_A < b <
2.1R_A$.  The lines are the homogeneous shadowing result.
The dashed line represents $S_1$, the dot-dashed, $S_2$, and
the dotted, $S_3$.  Equation~(\ref{wsparam})
is used to calculate the inhomogeneous shadowing ratios 
for $S_{1, {\rm WS}}$, circles, $S_{2, {\rm WS}}$,
squares, and $S_{3, {\rm WS}}$, triangles.  }
\label{dymassalice}
\end{figure}

\begin{figure}[htb]
\setlength{\epsfxsize=0.5\textwidth}
\setlength{\epsfysize=0.7\textheight}
\centerline{\epsffile{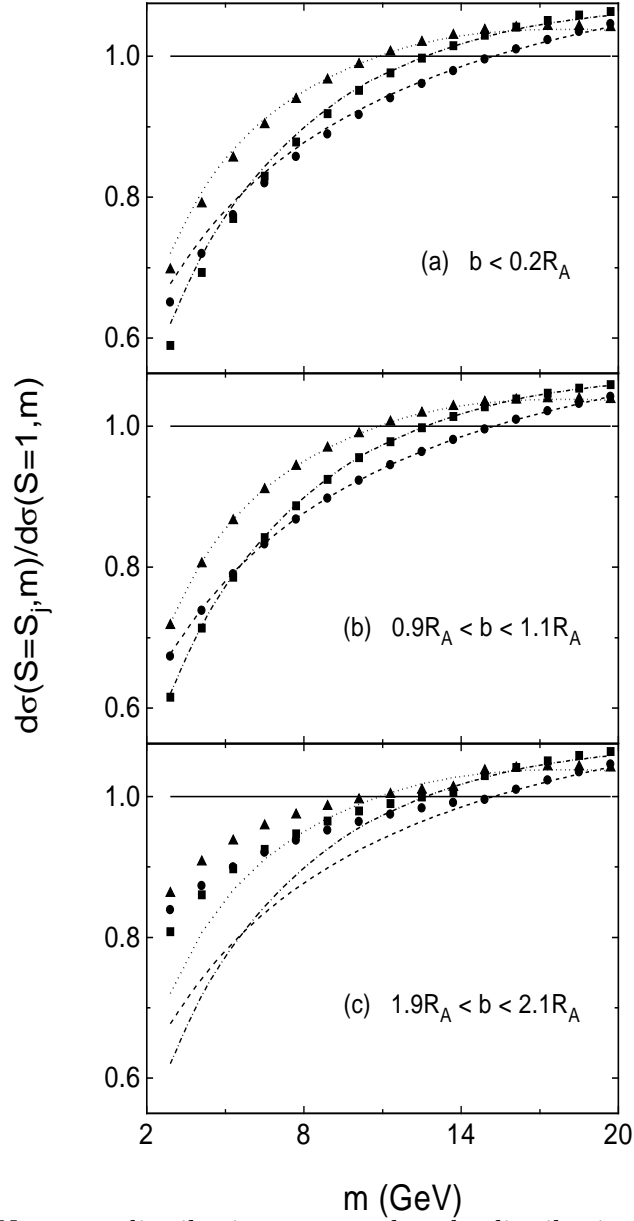}}
\caption[]{The Drell-Yan mass distribution compared to the
distribution with $S=1$ in PHENIX, $|y| \leq 0.35$, calculated with the
MRST LO distributions.  The upper plot shows central collisions
with $b<0.2 R_A$, the middle plot shows the region $0.9R_A < b <
1.1R_A$, and the lower plot shows the peripheral region $1.9R_A < b <
2.1R_A$.  The lines indicate the homogeneous shadowing
result.  The dashed line represents $S_1$, the dot-dashed, $S_2$, and
the dotted, $S_3$.  Equation~(\ref{wsparam})
is used to calculate the inhomogeneous shadowing ratios
for $S_{1, {\rm WS}}$, circles, $S_{2, {\rm WS}}$,
squares, and $S_{3, {\rm WS}}$, diamonds.  }
\label{dymassphenix}
\end{figure}

Figures~\ref{dymassalice} and \ref{dymassphenix} show the influence of
shadowing on the Drell-Yan mass distribution, calculated with the MRST
LO parton distributions. The ratio of the inhomogeneously shadowed
mass distribution to that for $S=1$ are shown in several impact
parameter bins, along with the homogeneous shadowing ratios, in the
rapidity coverage given for the ALICE and PHENIX central
detectors. The corresponding ratios for CMS and STAR are quite
similar.  Ratios are presented for the most central collisions,
$b<0.2R_A$, semi-central collisions, $0.9R_A < b < 1.1R_A$, and
peripheral collisions, $1.9R_A < b < 2.1R_A$. In the most central
collisions, the inhomogeneous shadowing, with Eq.~(\ref{wsparam}), is
somewhat stronger while in the most peripheral collisions, it is much
weaker.  In each case, the $S_3$ parameterization gives the smallest
effect.  At the LHC, evolution is also most apparent with this
parameterization.  A shortcoming of the limited $Q^2$ evolution of the
$S_2$ parameterization is obvious in Fig.~\ref{dymassalice}---the
evolution is evident up to $m = 10$ GeV after which the 10 GeV values
of the valence quark, sea quark, and gluon shadowing ratios are used
at all higher masses.  Above 10 GeV, the ratios with the $S_1$ and
$S_2$ parameterizations are then similar.  The $S_1$ results change
very slowly with mass because they lack $Q^2$ evolution.  At the lower
RHIC energy, the 10 GeV $Q^2$ cutoff in $S_2$ is less obvious because
the $x$ values are larger, in a region where shadowing is small.  At
RHIC, shadowing of the most peripheral collisions predominantly occurs for
masses below 8 GeV.  At this energy the largest mass pairs are
antishadowed.  The antishadowing is weakened in peripheral collisions,
see Fig.~\ref{dymassphenix}.

\begin{figure}[htb]
\setlength{\epsfxsize=0.7\textwidth}
\setlength{\epsfysize=0.5\textheight}
\leftline{\epsffile{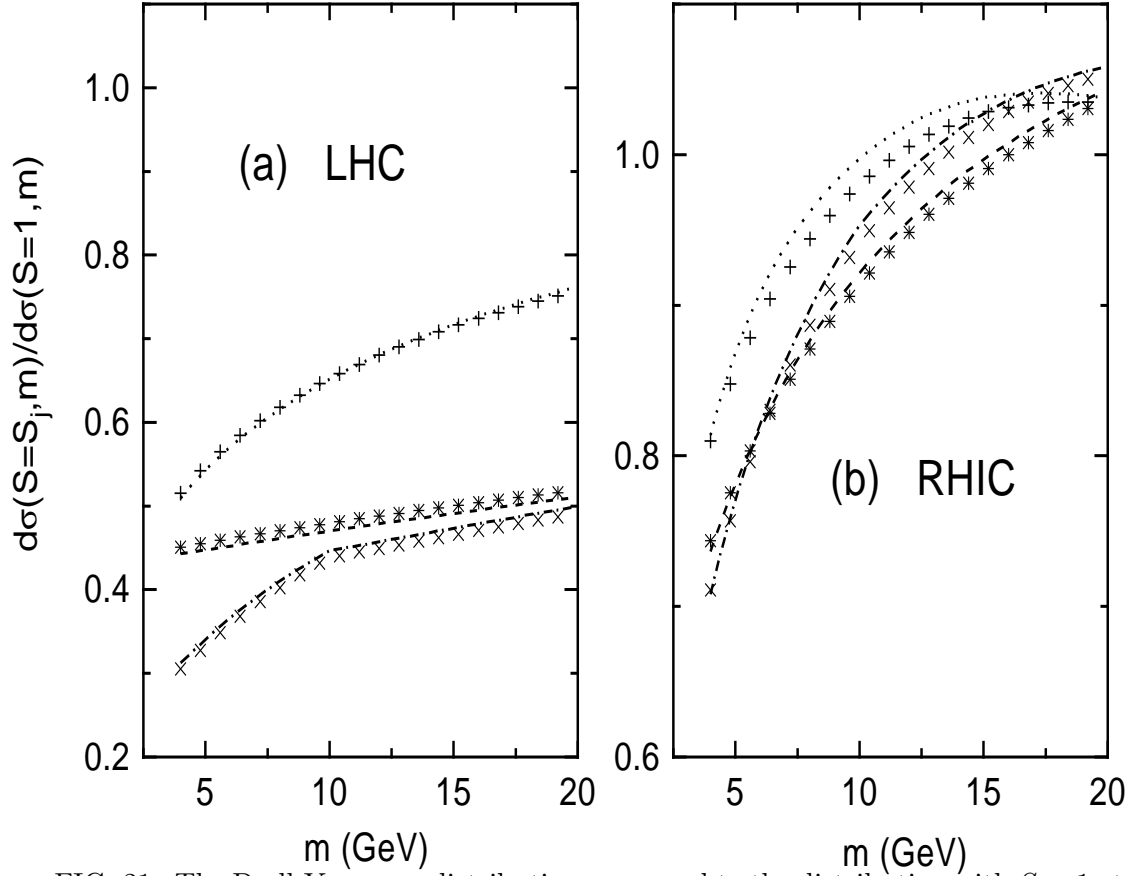}}
\caption[]{The Drell-Yan mass distribution 
compared to the distribution with $S=1$ at the LHC, $|y|
\leq 1.$, and RHIC, $|y| \leq 0.35$ calculated with the MRST 
distributions.  The ratios are shown at leading and next-to-leading
order in the Drell-Yan cross section for homogeneous shadowing.  The
dashed line represents $S_1$, the dot-dashed, $S_2$, and the dotted,
$S_3$ ratios at leading order.  The next-to-leading order ratios are
indicated by the symbols, $*$ ($S_1$), $\times$ ($S_2$), and $+$
($S_3$). }
\label{dyho}
\end{figure}

Since the next-to-leading order, NLO, Drell-Yan cross section includes
Compton scattering with an initial gluon \cite{hpdy}, it is possible
that shadowing could change significantly at NLO, especially with the
$S_2$ and $S_3$ parameterizations.  We have therefore also calculated
the Drell-Yan cross sections at NLO with all the homogeneous shadowing
parameterizations and found that the ratios do not change
significantly when the NLO terms are added.  There is a 3-4\%
difference in the ratios with shadowing at LO and NLO in Pb+Pb
collisions at 5.5 TeV and 0.5-1\% in Au+Au collisions at 200 GeV.
This should not be too surprising since the theoretical $K$ factor is
small, $K_{\rm th} = \sigma_{\rm NLO}^{\rm DY}/\sigma_{\rm LO}^{\rm
DY} \sim 1.2$ at RHIC and 1.1 at the LHC.  The effect of shadowing on
the higher order contributions must then be less than $K_{\rm th}$,
small compared to the uncertainties in the shadowing model, as
can be seen from Fig.~\ref{dyho}.

\begin{figure}[htb]
\setlength{\epsfxsize=0.5\textwidth}
\setlength{\epsfysize=0.7\textheight}
\centerline{\epsffile{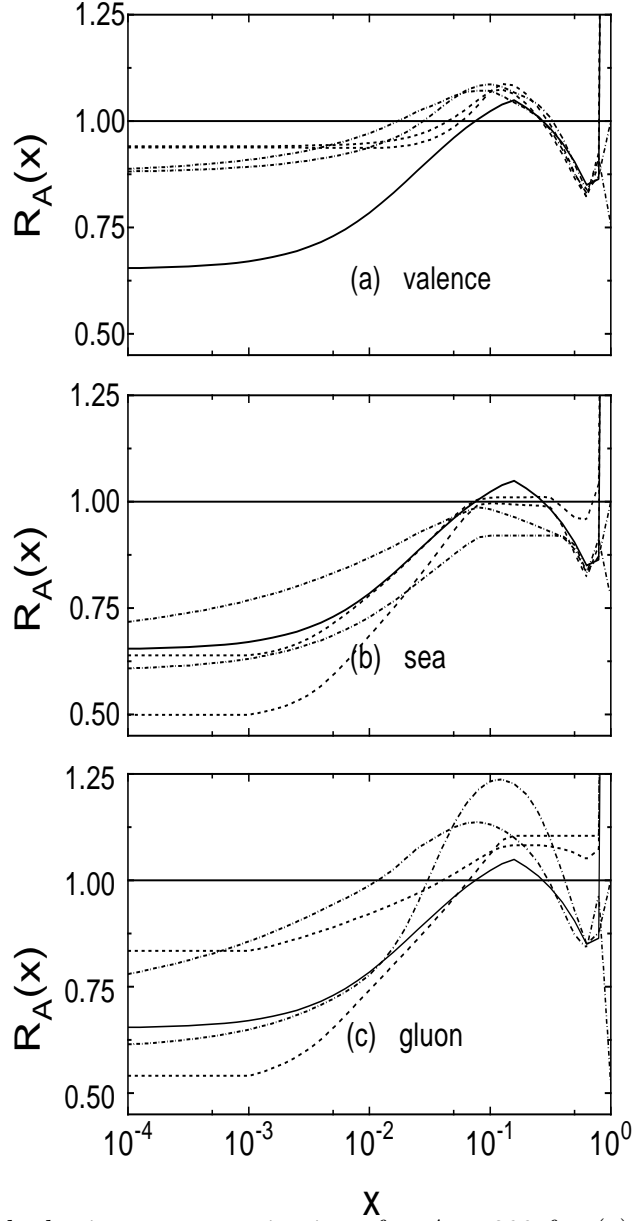}}
\caption[]{The three shadowing parameterizations for $A =
200$ for (a) valence quarks, (b) sea quarks, and (c) gluons,
relative to $S=1$.
The $S_1$ parameterization is shown in the solid curves.  The $S_2$ ratios
are given by the dashed curves.  At low $x$, the
lower curves are for $Q = 2$ GeV while the upper are for $Q=10$ GeV.  The
$S_3$ ratios, in the dot-dashed curves, are shown for the $u_V$ and $\overline
u$.  The lower curves at low $x$
are for $Q = 1.5$ GeV while the upper curves at low $x$ are for $Q=10$ GeV.}
\label{fshadow}
\end{figure}

\begin{figure}[htb]
\setlength{\epsfxsize=0.7\textwidth}
\setlength{\epsfysize=0.5\textheight}
\leftline{\epsffile{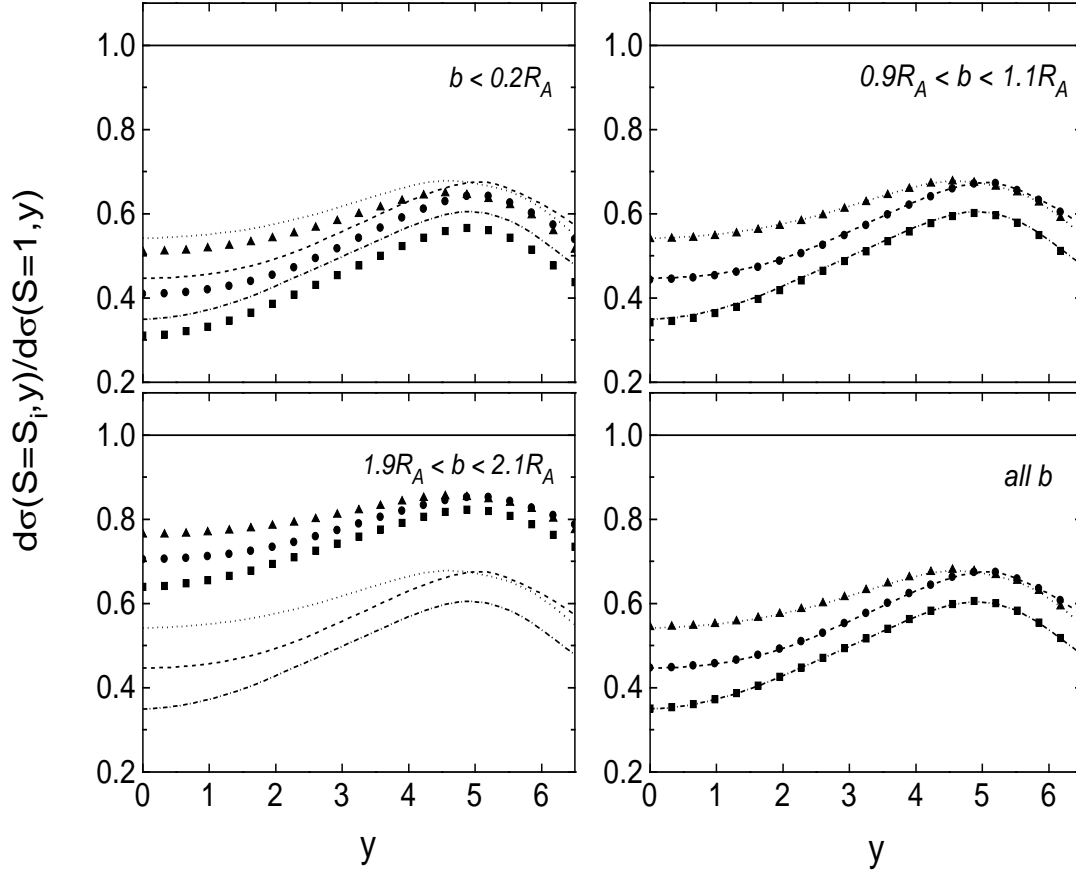}}
\caption[]{The Drell-Yan rapidity distribution in the mass interval
$4<m<9$ GeV, relative to $S=1$ for Pb+Pb collisions at the LHC,
calculated with the MRST LO distributions.  Central, $b<0.2 R_A$,
semi-central, $0.9R_A < b < 1.1R_A$, peripheral, $1.9R_A < b < 2.1R_A$
impact parameters are shown along with the integral over all $b$.  The
lines indicate the homogeneous shadowing result.  The dashed line
represents $S_1$, the dot-dashed, $S_2$, and the dotted, $S_3$.  
Equation~(\ref{wsparam})  is used to calculate the inhomogeneous
shadowing ratios for $S_1$, circles, $S_2$, squares, and
$S_3$, diamonds.}
\label{dyraplhc}
\end{figure}

\begin{figure}[htb]
\setlength{\epsfxsize=0.7\textwidth}
\setlength{\epsfysize=0.5\textheight}
\leftline{\epsffile{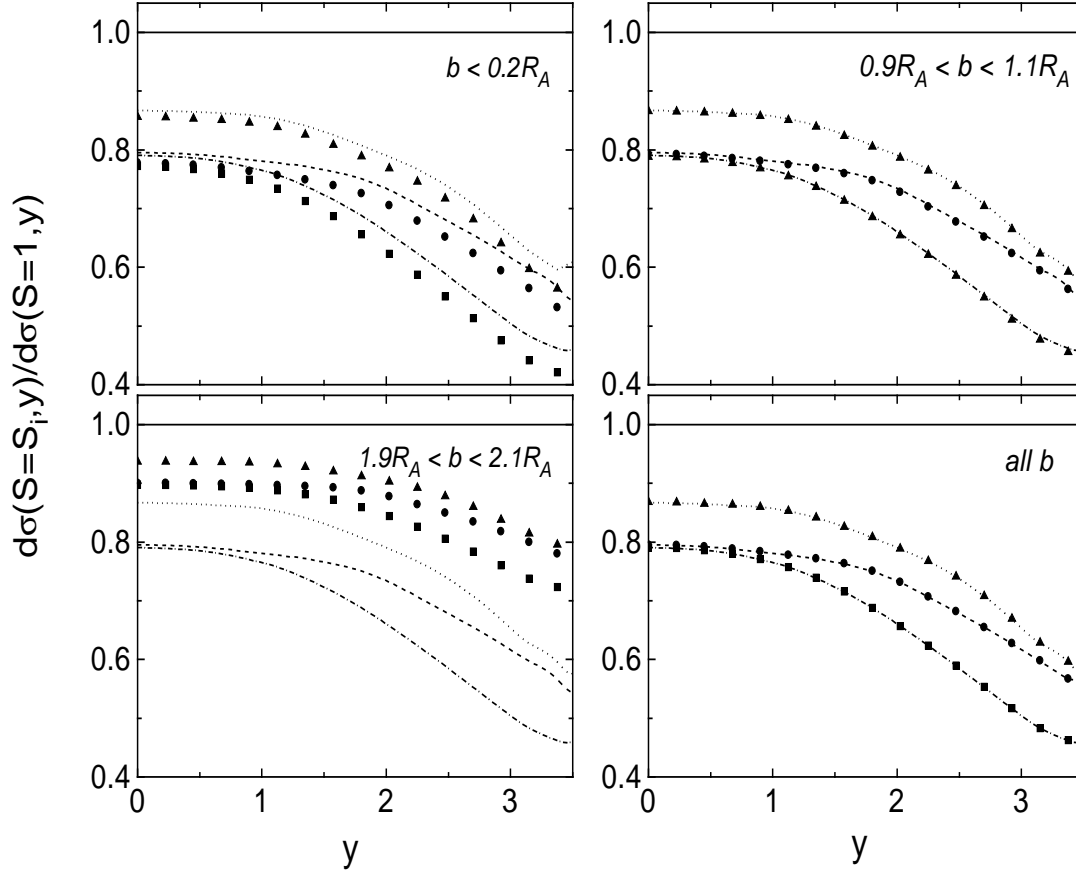}}
\caption[]{The Drell-Yan rapidity distribution in the mass interval
$4<m<9$ GeV, compared to the distribution with $S=1$ in Au+Au
collisions at RHIC, calculated with the MRST LO distributions.
Central, $b<0.2 R_A$, semi-central, $0.9R_A < b < 1.1R_A$, peripheral,
$1.9R_A < b < 2.1R_A$ impact parameters are shown along with the
integral over all $b$.  The lines indicate the homogeneous shadowing
result.  The dashed line represents $S_1$, the dot-dashed, $S_2$, and
the dotted, $S_3$.  Equation~(\ref{wsparam}) is used to
calculate the ratio for $S_1$, circles, $S_2$, squares, and $S_3$,
diamonds.  }
\label{dyraprhic}
\end{figure}

Figures~\ref{dyraplhc} and \ref{dyraprhic} show the rapidity
dependence of the shadowing for Drell-Yan production when $4<m<9$ GeV.
The homogeneous and inhomogeneous results are again compared in central,
semi-central, and peripheral collisions.  The $S_2$ parameterization
produces the strongest shadowing because the sea quark ratio is lower
at small $x$ than $S_1$ and $S_3$, see Fig.~\ref{fshadow}.  All the
LHC ratios increase with rapidity because $x_2$ remains small while
$x_1$ increases to $\sim 0.1$ at $y \sim 5$.  Recall that around $x_1
\sim 0.1$ $S_1$ shows antishadowing, $S_2\sim 1 $ for sea quarks, and
the sea quark distributions are shadowed with the $S_3$
parameterization.  Thus the change in the shadowing ratios as a
function of $y$ is smallest with $S_3$.  As $y$ and $x_1$ increase, 
the shadowing,
antishadowing, and EMC regions are traced out.  However, at forward 
rapidities, $x_2 <
10^{-4}$ so that the cross section ratios are always significantly less than
unity. 

At RHIC, the ratios decrease with rapidity.  Both $x_1$
and $x_2$ are in a region where all the parton densities are shadowed at
$y=0$ but, as the
rapidity increases, $x_2$ decreases to the low $x$ saturation region
while the $x_1$ values enter the EMC region.  The resulting convolution 
is then lower at large $y$ than at central rapidities.  
Since the Drell-Yan cross
section is calculated in the interval $4<m<9$ GeV, some influence of Fermi 
motion is apparent at the largest rapidities because 
$x_1 \sim 0.9$ when $m=9$ GeV
and $y=3$. 

\begin{figure}[htb]
\setlength{\epsfxsize=0.7\textwidth}
\setlength{\epsfysize=0.5\textheight}
\leftline{\epsffile{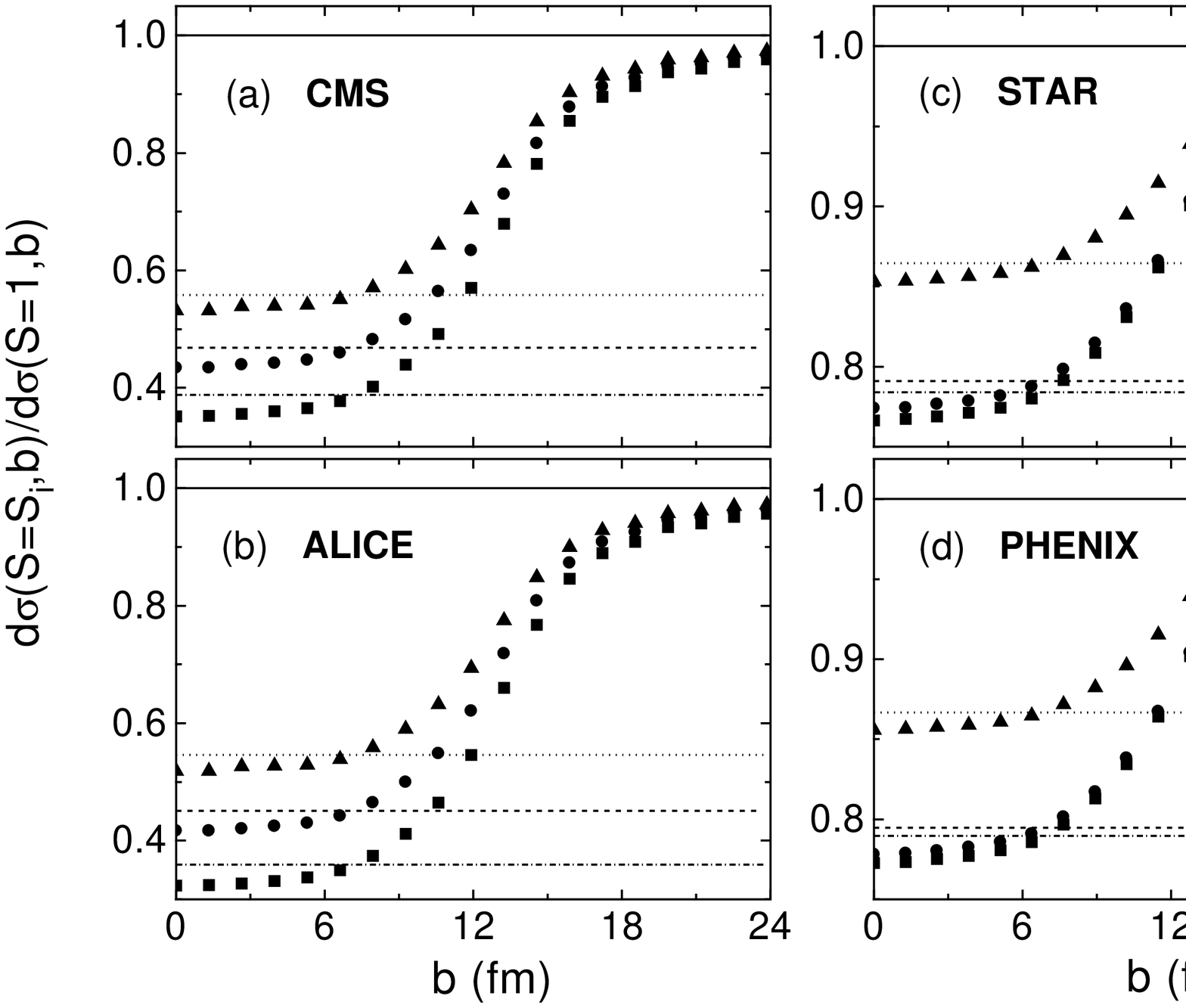}}
\caption[]{The impact parameter dependence of the Drell-Yan cross
section for $4 < m < 9$ GeV, calculated with the MRST
LO distributions.  Results are shown for the central rapidity
coverages of all four detectors.  The lines indicate the
homogeneous shadowing result: dashed for $S_1$, dot-dashed for $S_2$,
and dotted for $S_3$.  Equation~(\ref{wsparam}) is used to
calculate the ratio for $S_1$, circles, $S_2$, squares, and $S_3$,
diamonds.}
\label{dybdep49}
\end{figure}

\begin{figure}[htb]
\setlength{\epsfxsize=0.45\textwidth}
\setlength{\epsfysize=0.5\textheight}
\centerline{\epsffile{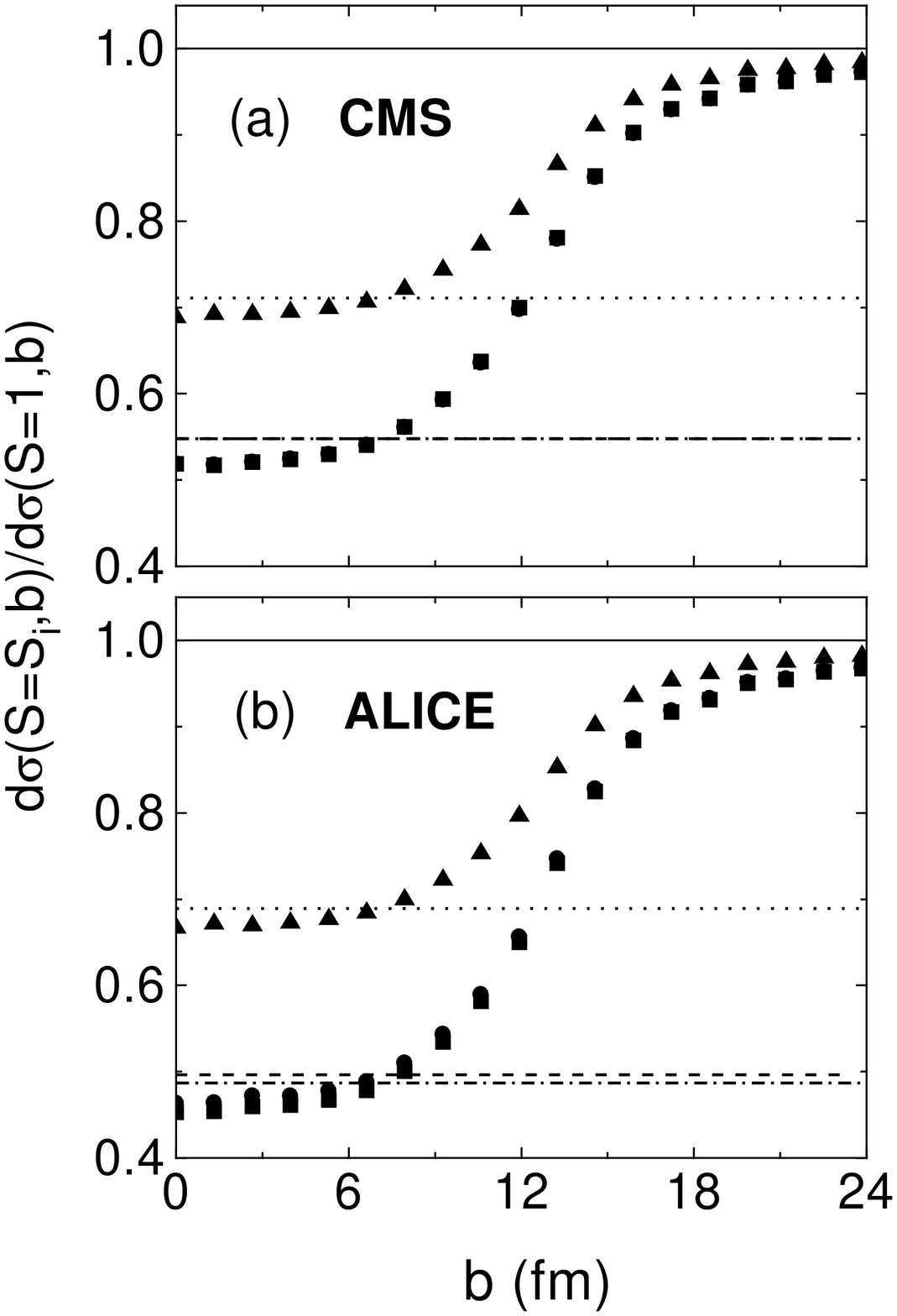}}
\caption[]{The impact parameter dependence of the Drell-Yan cross
section for $11 < m < 20$ GeV, calculated with the MRST
LO distributions.  Results are shown for the central rapidity
coverages of the two LHC detectors.  The lines indicate the
homogeneous shadowing result: dashed for $S_1$, dot-dashed for $S_2$
and dotted for $S_3$.  Equation~(\ref{wsparam})
is used to
calculate the ratio for $S_1$,
circles, $S_2$, squares, and $S_3$, diamonds.  }
\label{dybdep1120}
\end{figure}

The effect of the inhomogeneity is shown more fully in
Figs.~\ref{dybdep49} and \ref{dybdep1120}.  We have chosen two
different mass ranges, $4<m<9$ GeV and $11<m<20$ GeV, between the
$J/\psi$ and $\Upsilon$ resonances and above the $\Upsilon$ family
respectively.  The similarities between the CMS and ALICE predictions
at the LHC and the STAR and PHENIX expectations at RHIC are obvious in
these figures.  In the range $4<m<9$ GeV, shadowing is expected at all
masses.  In the larger mass region, the similarity between the $S_1$
and $S_2$ parameterizations above 10 GeV are visible in the CMS and
ALICE plots. For completeness, the LO Drell-Yan production cross
sections per nucleon pair for both mass ranges are shown in
Table~\ref{dysigtab} with and without homogeneous shadowing.  Recall
that the theoretical $K$ factor between the LO and NLO cross sections
is $\approx 1.2$.

We now consider shadowing in $J/\psi$ and $\Upsilon$ production using
two models that have been successfully employed to describe quarkonium
hadroproduction.  The first, the color evaporation model, treats all
quarkonium production identically to $Q \overline Q$ production below
the $M \overline M$ threshold, where $M$ represents the lightest meson
containing a single heavy quark $Q$, neglecting the color and spin of
the produced $Q \overline Q$ pair.  The non-relativistic QCD approach
expands quarkonium production in powers of $v$, the relative
$Q$-$\overline Q$ velocity within the bound state.  In this model, the
produced $Q \overline Q$ pair retains the information on its color,
spin and total angular momentum, requiring more parameters than the
color evaporation model.

In the color evaporation model \cite{HPC} 
\begin{eqnarray}
\lefteqn{f_i^N(x_1,Q^2) f_j^N(x_2,Q^2) \frac{d\widehat{\sigma}_{ij}^{C,B}
}{dy dm^2} = F_{C,B} K_{\rm th} \left\{
f_g^N(x_1,Q^2)f_g^N(x_2,Q^2)
\frac{\sigma_{gg}(Q^2)}{\widehat{s}} \right.} \nonumber \\ & & \mbox{} \left.
+ \sum_{q=u,d,s} [f_q^N(x_1,Q^2) f_{\overline q}^N(x_2,Q^2) + f_{\overline 
q}^N(x_1,Q^2) f_q^N(x_2,Q^2)] \frac{\sigma_{q \overline 
q}(Q^2)}{\widehat{s}} \right\}
\, \, ,
\label{psicem}
\end{eqnarray}
where $C$ and $B$ represent the produced charmonium and bottomonium
states.  The LO partonic $Q \overline Q$ cross sections are defined in
\cite{combridge} and $\widehat{s} = x_1x_2S$.  The fraction of $Q
\overline Q$ pairs below the $M \overline M$ threshold that become the
final quarkonium state, $F_{C,B}$ is fixed at NLO \cite{HPC}.  The
factor $K_{\rm th}$ matches the LO cross section to the NLO result.
Together, the multiplicative factors $F_{C,B}$ and $K_{\rm th}$
reproduce the $pp$ data in magnitude and shape. For $J/\psi$
production, we use $m_c=1.3$ GeV and $Q=m_c$ with the GRV 94 LO
distributions and $m_c=1.2$ GeV and $Q = 2m_c$ with the MRST LO
densities \cite{HPC}.  For $\Upsilon$ production, we take $m_b = Q =
4.75$ GeV with both sets of parton distributions.

\begin{figure}[htb]
\setlength{\epsfxsize=0.7\textwidth}
\setlength{\epsfysize=0.5\textheight}
\leftline{\epsffile{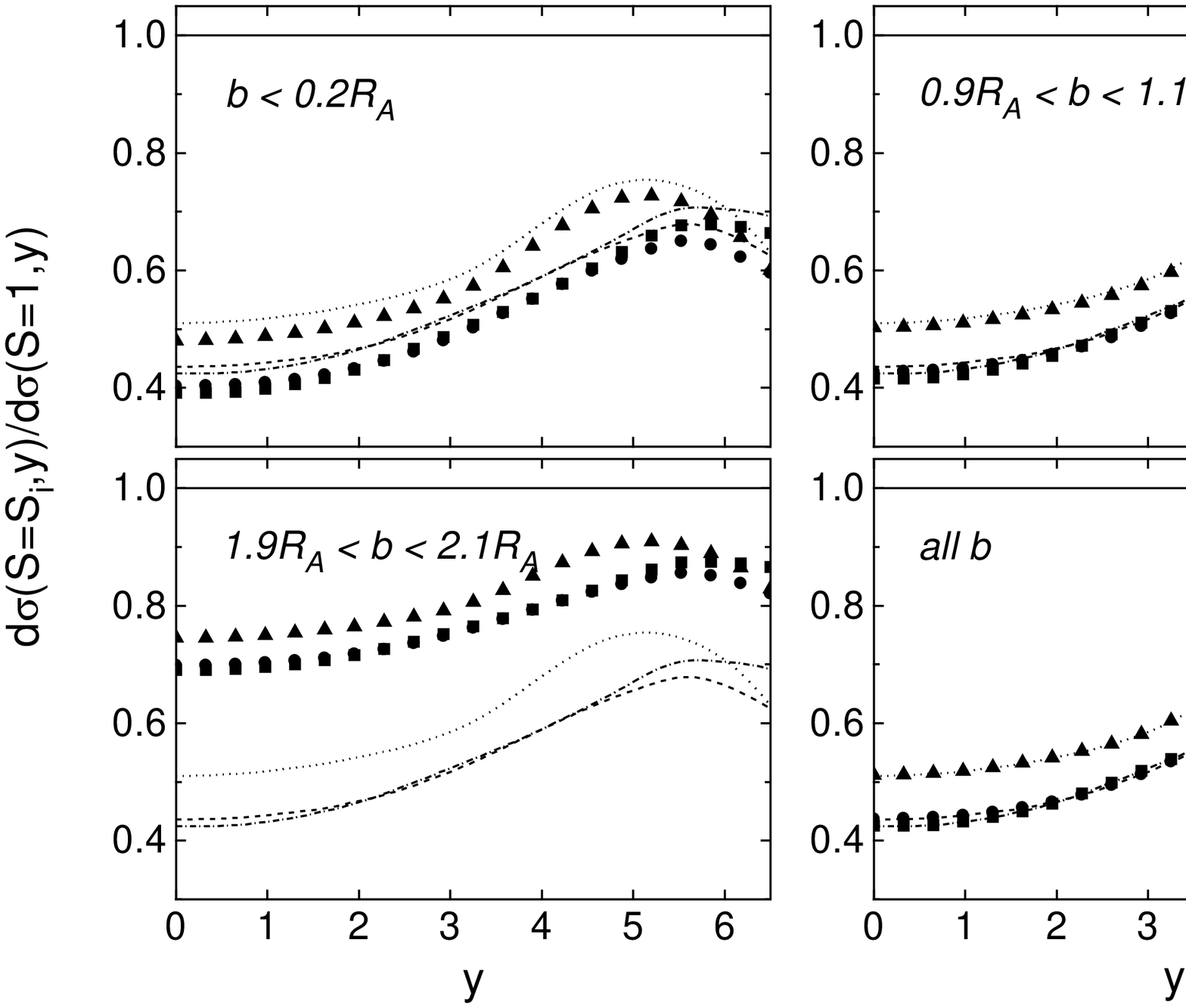}}
\caption[]{The $J/\psi$ rapidity distribution calculated in the color
evaporation model with the MRST LO distributions, compared to the
distribution with $S=1$ in Pb+Pb collisions at the LHC.  Central,
$b<0.2 R_A$, semi-central, $0.9R_A < b < 1.1R_A$, peripheral, $1.9R_A
< b < 2.1R_A$ impact parameters are shown along with the integral over
all $b$.  
The lines show the homogeneous shadowing result: dashed line
for $S_1$, dot-dashed for $S_2$, and dotted for $S_3$.  
Equation~(\ref{wsparam}) is used to
calculate the ratio for $S_1$, circles, $S_2$, squares, and $S_3$,
diamonds. }
\label{psiylhc}
\end{figure}

\begin{figure}[htb]
\setlength{\epsfxsize=0.7\textwidth}
\setlength{\epsfysize=0.5\textheight}
\leftline{\epsffile{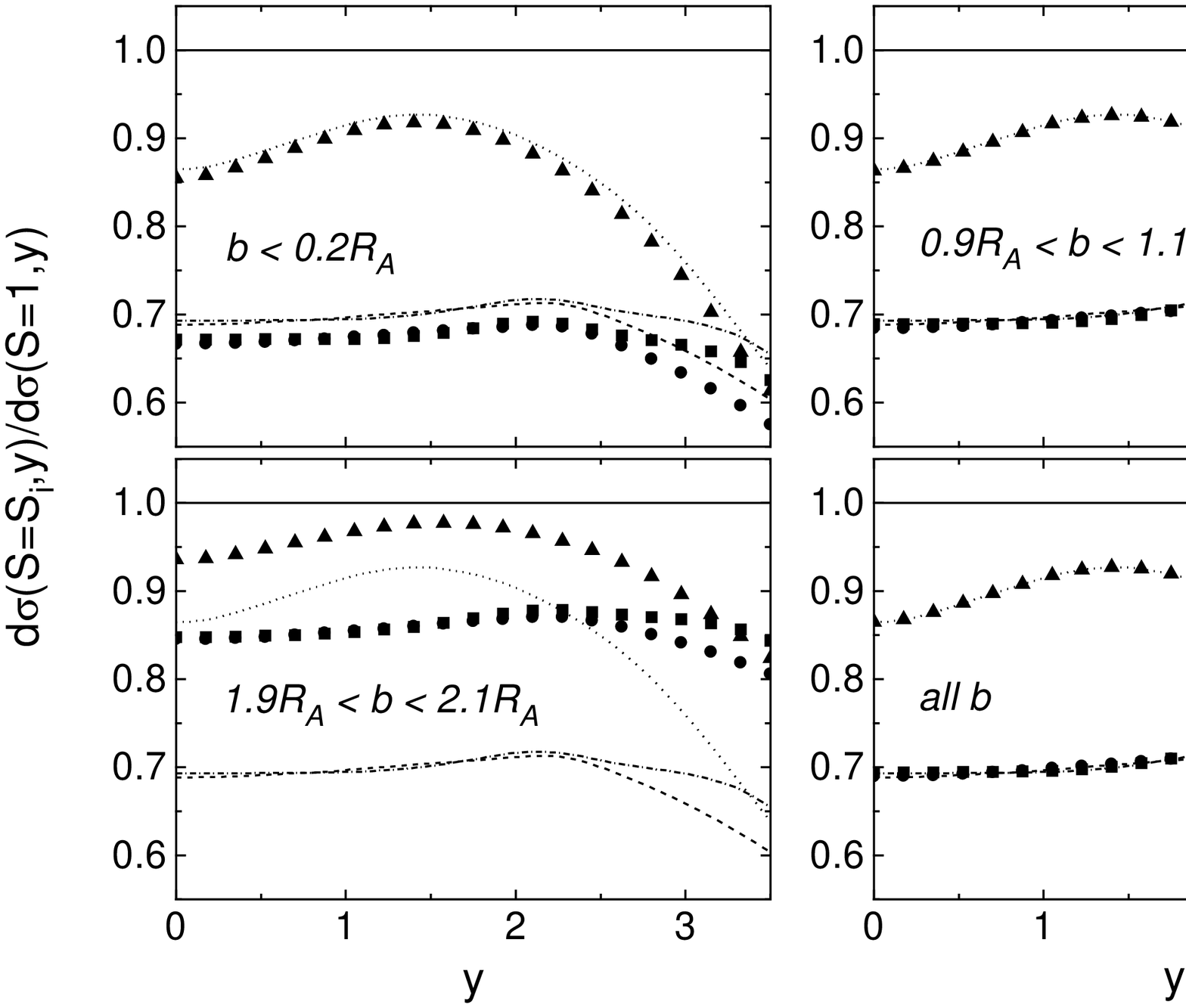}}
\caption[]{The $J/\psi$ rapidity distribution calculated in the color
evaporation model with the MRST LO distributions, compared to the
distribution with $S=1$ in Au+Au collisions at RHIC.  Central, $b<0.2
R_A$, semi-central, $0.9R_A < b < 1.1R_A$, peripheral, $1.9R_A < b <
2.1R_A$ impact parameters are shown along with the integral over all
$b$.  The lines show the homogeneous shadowing result: dashed line for
$S_1$, dot-dashed for $S_2$, and dotted for $S_3$.
Equation~(\ref{wsparam}) is used to calculate the ratio for $S_1$,
circles, $S_2$, squares, and $S_3$, diamonds. }
\label{psiyrhic}
\end{figure}

\begin{figure}[htb]
\setlength{\epsfxsize=0.7\textwidth}
\setlength{\epsfysize=0.5\textheight}
\leftline{\epsffile{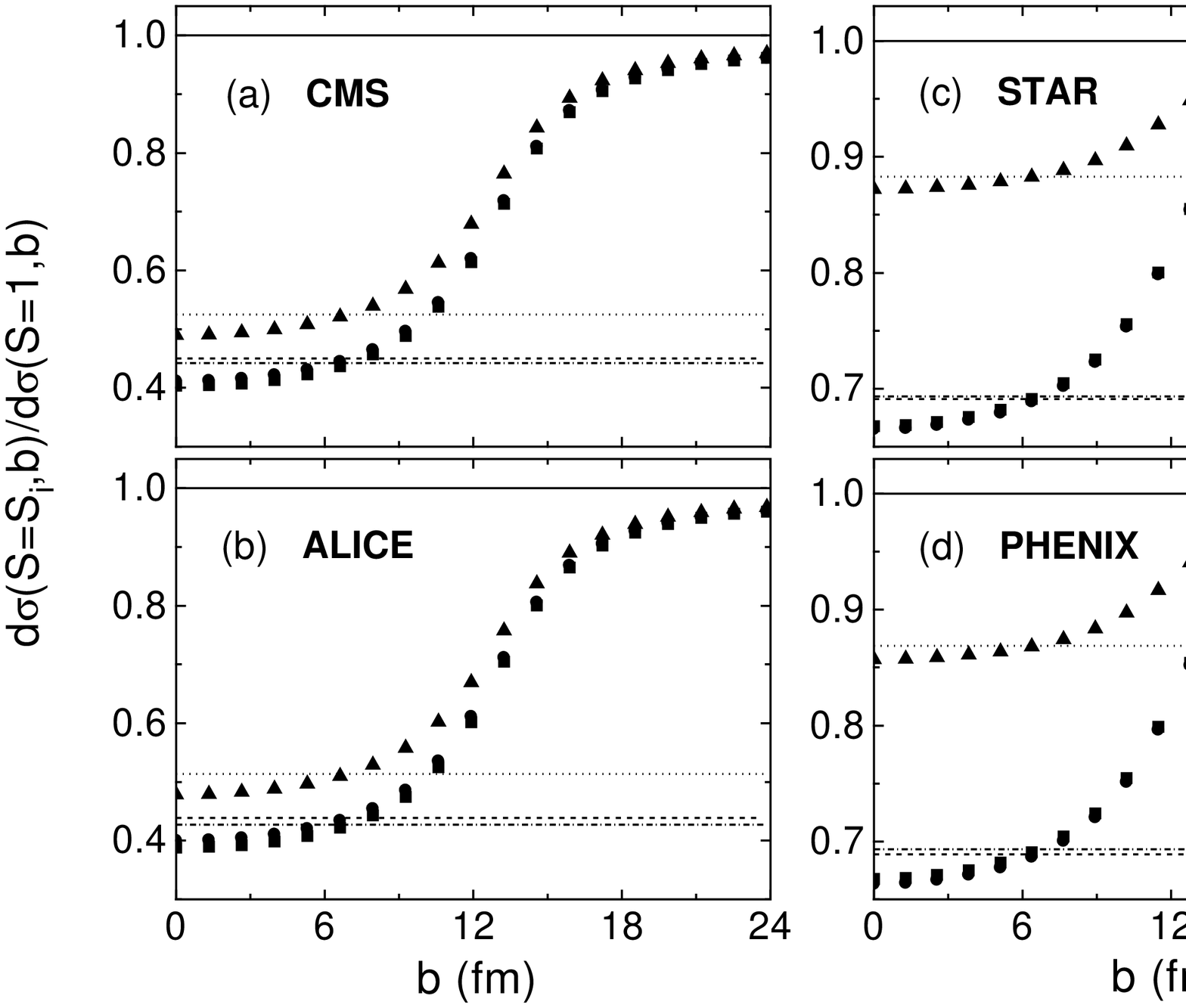}}
\caption[]{The impact parameter dependence of $J/\psi$ production calculated
in the color evaporation model with the MRST LO distributions.
Results are shown for the central rapidity coverages given for all
four detectors.  The lines show the homogeneous shadowing result:
dashed line for $S_1$, dot-dashed for $S_2$, and dotted for $S_3$.
Equation~(\ref{wsparam}) is used to calculate the ratio for $S_1$,
circles, $S_2$, squares, and $S_3$, diamonds. }
\label{jpbdepcem}
\end{figure}

The $J/\psi$ cross section ratios in the color evaporation model are
given as a function of rapidity at LHC and RHIC in Figs.~\ref{psiylhc}
and \ref{psiyrhic} respectively.  At both energies, the $S_1$ and
$S_2$ results are very similar because the product of the $S_1$
shadowing ratios and the $S_2$ gluon shadowing ratios at $Q = 2m_c =
2.4$ GeV differ by only $1-2$\% over a wide range, 5 units of rapidity at
the LHC and 2.5 units at RHIC.  The ratios with the $S_3$
parameterization are larger than with the $S_1$ and $S_2$
parameterizations.  This is due to the nature of the $S_3$ parameterization: at
low $x$ and $|y|$ there is less gluon shadowing and at large $x$ and $|y|$ the
gluon antishadowing is stronger than in $S_1$ and $S_2$.  
These effects are also obvious in the
rapidity-integrated impact parameter dependence shown in
Fig.~\ref{jpbdepcem}.  

The $J/\psi$ results in the color evaporation
model are rather sensitive to the choice of parton distributions.
This sensitivity arises from the rather low $m_c$
compared to the initial scale of many parton distributions.  The
initial scale of the MRST LO densities is $Q_0 \sim m_c \approx 1.1$ GeV
suggesting $Q = 2m_c$ is an appropriate choice.  Because
the initial scale in the GRV 94 LO densities is $Q_0 \sim m_c/2 \approx 0.63$
GeV, we use $Q = m_c$.
Choosing the scale proportional to $m_c$ is somewhat more
consistent with the calculations of the Drell-Yan and minijet
production cross sections.  However, the light charm quark mass
precludes this choice for the MRST LO densities.  We have displayed the
results with the MRST LO densities.  If the GRV 94 LO densities are used, the
shadowing is somewhat stronger at both energies and the $S_1$ and $S_2$ results
are different.

\begin{figure}[htb]
\setlength{\epsfxsize=0.7\textwidth}
\setlength{\epsfysize=0.5\textheight}
\leftline{\epsffile{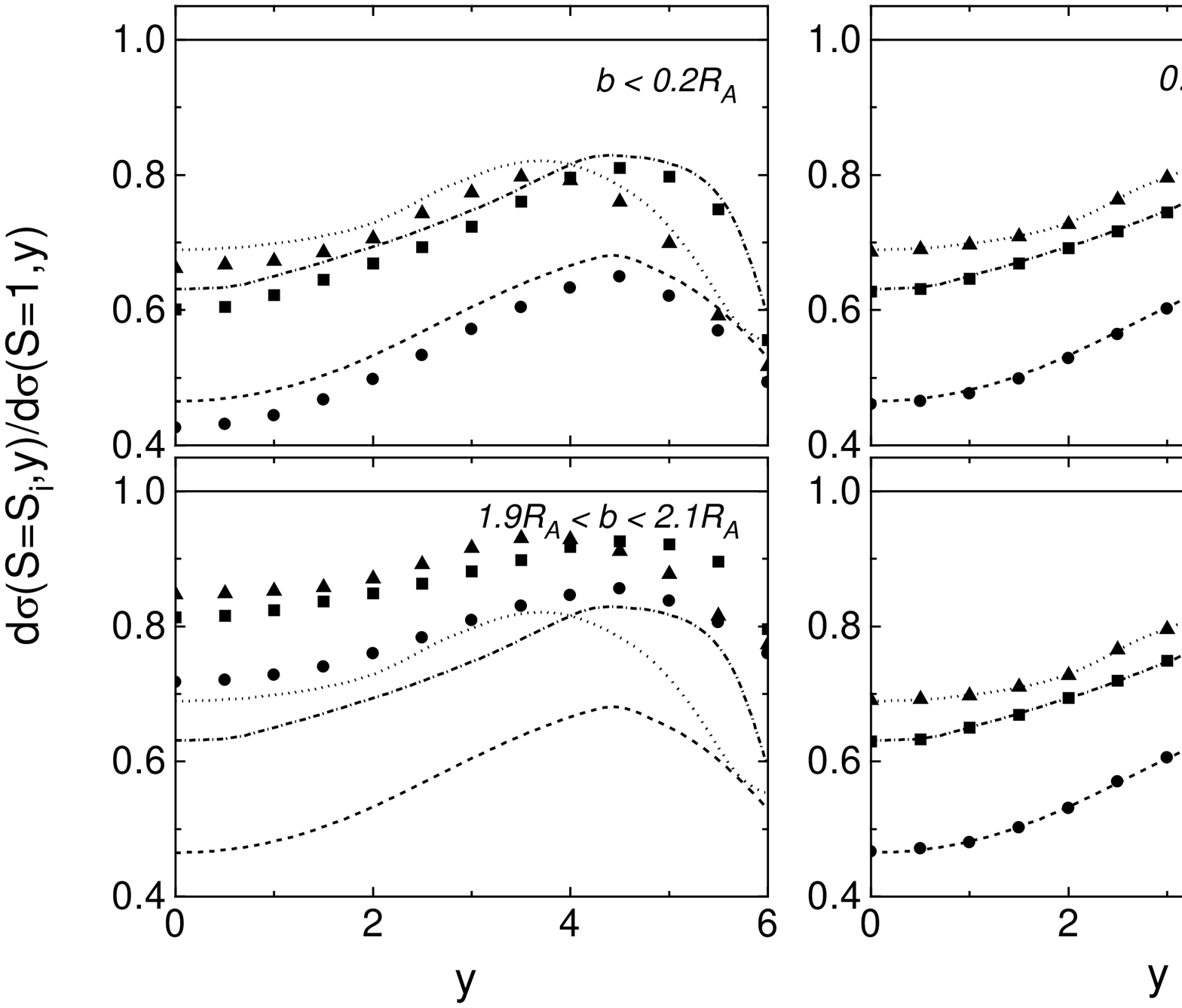}}
\caption[]{The $\Upsilon$ rapidity distribution calculated in the color
evaporation model with the MRST LO distributions, compared to the
distribution with $S=1$ in Pb+Pb collisions at the LHC.  Central,
$b<0.2 R_A$, semi-central, $0.9R_A < b < 1.1R_A$, peripheral, $1.9R_A
< b < 2.1R_A$ impact parameters are shown along with the integral over
all $b$.  
The lines show the homogeneous shadowing result: dashed line
for $S_1$, dot-dashed for $S_2$, and dotted for $S_3$.  
Equation~(\ref{wsparam}) is used to
calculate the ratio for $S_1$, circles, $S_2$, squares, and $S_3$,
diamonds. }
\label{upsylhc}
\end{figure}

\begin{figure}[htb]
\setlength{\epsfxsize=0.7\textwidth}
\setlength{\epsfysize=0.5\textheight}
\leftline{\epsffile{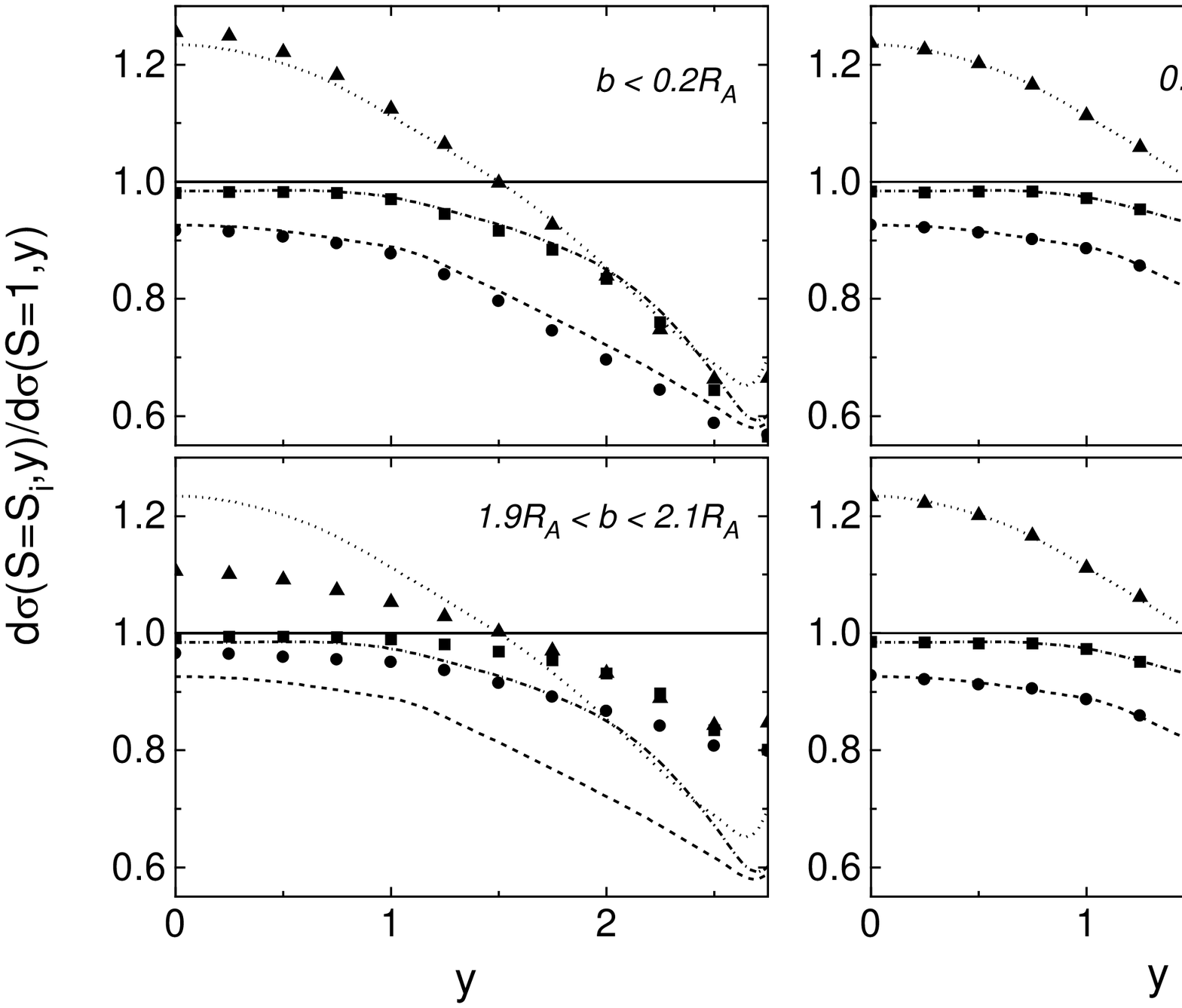}}
\caption[]{The $\Upsilon$ rapidity distribution calculated in the color
evaporation model with the MRST LO distributions, compared to the
distribution with $S=1$ in Au+Au collisions at RHIC.  Central, $b<0.2
R_A$, semi-central, $0.9R_A < b < 1.1R_A$, peripheral, $1.9R_A < b <
2.1R_A$ impact parameters are shown along with the integral over all
$b$.  The lines show the homogeneous shadowing result: dashed line for
$S_1$, dot-dashed for $S_2$, and dotted for $S_3$.
Equation~(\ref{wsparam}) is used to calculate the ratio for $S_1$,
circles, $S_2$, squares, and $S_3$, diamonds. }
\label{upsyrhic}
\end{figure}

Figures.~\ref{upsylhc} and \ref{upsyrhic} show the shadowed $\Upsilon$
cross sections, relative to $S=1$, as a function of rapidity in
several impact parameter regions.  The $S_1$ and $S_2$
parameterizations now differ due to the evolution of the $S_2$
parameterization.  The $S_1$ parameterization, without evolution,
gives an $\Upsilon$ ratio only slightly different from that of the
$J/\psi$ at $y=0$ for the LHC energy because as $x_2$ changes from
$4.4 \times 10^{-4}$ for the $J/\psi$ to $1.7 \times 10^{-3}$ for the
$\Upsilon$ at $y=0$, $S_1$ is nearly constant, see Fig.~\ref{fshadow}.
The peak at $y\sim 4.3$ with $S_1$ appears as $x_1$ goes through the
antishadowing region to the EMC region.  While the maximum in the
shadowing ratios occurs at similar rapidities in $J/\psi$ production,
$y \sim 5$ for $S_3$ and $y \sim 5.5$ for $S_1$ and $S_2$, the
$\Upsilon$ ratios peak at $y \sim 3.5$ for $S_3$, 4.5 for $S_1$ and $y
\sim 5$ for $S_2$.  In fact, now the $S_2$ and $S_3$ ratios are
similar at the LHC.  The larger gluon antishadowing associated with
$J/\psi$ production is reduced at the larger bottom mass.  At RHIC
$\Upsilon$ shadowing is further reduced relative to the $J/\psi$ than
at LHC.  In contrast to Fig.~\ref{psiyrhic}, the ratio decreases with
increasing $y$ over all rapidity.  Note also that $\Upsilon$
production is restricted to a narrower range than the $J/\psi$ because
the $\Upsilon$ is heavier.  Little $\Upsilon$ shadowing is observed
with $S_2$ while $S_3$ exhibits strong antishadowing at $y=0$ since
$x_1=x_2 = 0.048$.  The $\Upsilon$ results are less dependent on the
choice of parton distributions than the $J/\psi$.  This set of parton
distributions is weaker than that of the $J/\psi$.  This is because
$m_b > Q_0$ in both sets so that we choose $m_b = Q$, eliminating the
ambiguity in scale due to the small charm quark mass in $J/\psi$
production.

\begin{figure}[htb]
\setlength{\epsfxsize=0.7\textwidth}
\setlength{\epsfysize=0.5\textheight}
\leftline{\epsffile{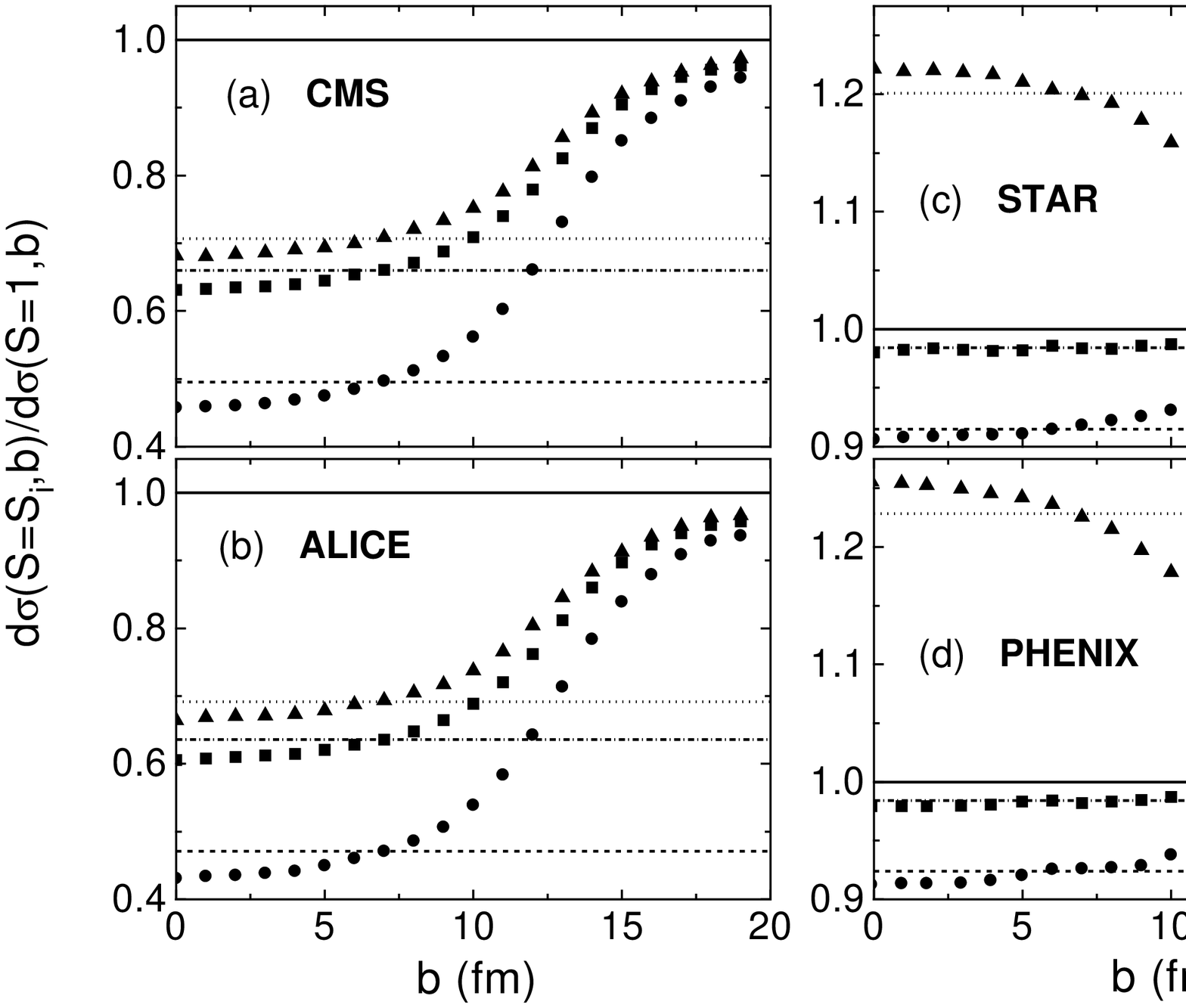}}
\caption[]{The impact parameter dependence of $\Upsilon$ production calculated
in the color evaporation model with the MRST LO distributions.
Results are shown for the central rapidity coverages given for all
four detectors.  The lines show the homogeneous shadowing result:
dashed line for $S_1$, dot-dashed for $S_2$, and dotted for $S_3$.
Equation~(\ref{wsparam}) is used to calculate the ratio for $S_1$,
circles, $S_2$, squares, and $S_3$, diamonds. }
\label{upsbdepcem}
\end{figure}

The impact parameter dependence of $\Upsilon$ production is shown for
the central rapidity coverages of the LHC and RHIC detectors in
Fig.~\ref{upsbdepcem}.  These $\Upsilon$ ratios are much more
dependent on rapidity than the corresponding $J/\psi$ ratios.  Since
the largest shadowing or antishadowing occurs in the central region, a
stronger relative $y$-integrated effect is observed in the detectors
with the narrowest rapidity acceptances.  This is particularly obvious
for the $S_3$ parameterization in PHENIX with respect to STAR.

The effects of shadowing on quarkonium production in the color
evaporation model are unchanged between LO and NLO \cite{RVe866}.
Even though at NLO quark-gluon scattering also contributes to quarkonium
production, the fraction of the total production cross section due to this
new channel is not large enough at these energies to change the shadowing
effects. 

The non-relativistic QCD, NRQCD, approach is an extension of the color
singlet model \cite{baru} which requires $J/\psi$'s to be produced
with the correct color and total angular momentum.  The color singlet
model predicts that high $p_T$ $J/\psi$ production occurs dominantly
through $\chi_{cJ}$ decays because direct $J/\psi$ production required
a hard gluon emission on a perturbative timescale.  The NRQCD model
\cite{bbl} does not restrict the angular momentum or color of the
quarkonium state to the lowest allowed color singlet state.  Then,
{\it e.g.} a $J/\psi$ may produced as a $^3P_0$ color octet which
hadronizes through the emission of nonperturbative soft gluons.

The rapidity distribution of the final-state $C$ or $B$ is
\be
\lefteqn{f_i^N(x_1,Q^2) f_j^N(x_2,Q^2) \frac{d\widehat{\sigma}_{ij}^{C,B}
}{dy}} \nonumber \\ & & = \sum_{i,j} \sum_n
\int_0^1 dx_1 dx_2 \delta (y - \frac{1}{2} \ln\left(\frac{x_1}{x_2} \right) ) 
f_i^N(x_1,Q^2)f_j^N(x_2,Q^2) C^{ij}_{Q \overline Q \, [n]} 
\langle {\cal O}_n^{C,B} \rangle \, \, . \label{psinrqcd} \ee 
The sum over $i$ and $j$ includes up, down, and strange quarks and antiquarks
as well as gluons since in NRQCD {\it e.g.}\ 
the process $ (q + \overline q)g \rightarrow
\chi_{c1} X$ also contributes to $J/\psi$ production.
The expansion coefficients $C^{ij}_{Q \overline Q \, [n]}$ are calculated
perturbatively in powers of $\alpha_s(Q^2)$ up to $\alpha_s^3$
and the nonperturbative parameters
$\langle {\cal O}_n^{C,B} \rangle$ describe the hadronization of the 
quarkonium state.  The expressions for the cross sections and the values of the
nonperturbative parameters can be found in Ref.~\cite{benrot}. Since
$\langle {\cal O}_n^{C,B} \rangle$ 
were fixed using the CTEQ 3L parton densities 
\cite{cteq3} with $m_c = 1.5$ GeV, $m_b = 4.9$ GeV, and $Q = 2m_Q$, 
we use this set with the
same $m_Q$ and $Q$ values to be consistent with
fixed target cross sections\cite{benrot}.

The total $J/\psi$ cross section includes radiative decays of the
$\chi_{cJ}$ states and hadronic decays of the $\psi'$,
\be \frac{d \sigma_{J/\psi}}{dy} =  \frac{d \sigma_{J/\psi}^{\rm dir}}{dy}
+ \sum_{J=0}^2 B(\chi_{cJ} \rightarrow J/\psi X) \frac{d 
\sigma_{\chi_{cJ}}}{dy}
+ B(\psi' \rightarrow J/\psi X) \frac{d\sigma_{\psi'}}{dy} \label{dirpsi}
\, \, . \ee 
Likewise, the total $\Upsilon$ cross section includes radiative decays from
$\chi_{bJ}(1P)$ and $\chi_{bJ}(2P)$ states and hadronic decays from the
$\Upsilon(2S)$ and $\Upsilon(3S)$ states.  We have not included radiative
decays from the proposed $\chi_{bJ}(3P)$ states since their branching ratios
to the lower bottomonium states are unknown.  Then
\be \lefteqn{\frac{d \sigma_{\Upsilon}}{dy} =  
\frac{d \sigma_{\Upsilon}^{\rm dir}}{dy}
+ \sum_{J=0}^2 B(\chi_{bJ}(1P) \rightarrow \Upsilon X) \frac{d 
\sigma_{\chi_{bJ}(1P)}}{dy}
+ B_{\rm eff}(\Upsilon(2S) 
\rightarrow \Upsilon X) \frac{d\sigma_{\Upsilon(2S)}}{dy} }
\nonumber \\ & &
+ \sum_{J=0}^2 B_{\rm eff}(\chi_{bJ}(2P) \rightarrow \Upsilon X) \frac{d 
\sigma_{\chi_{bJ}(2P)}}{dy}
+ B_{\rm eff}(\Upsilon(3S) 
\rightarrow \Upsilon X) \frac{d\sigma_{\Upsilon(3S)}}{dy} 
\label{dirups}
\, \, . \ee 
Note that as well as the direct decays of the higher bottomonium states to
the $\Upsilon$, a final-state $\Upsilon$ can be produced by a chain of hadronic
and radiative decays.  In the case of {\it e.g.}\ the $\Upsilon(3S)$, decays
to $\Upsilon(2S)$ and $\Upsilon$ are of the same order as decays to the
$\chi_{bJ}(2P)$ states.  The branching ratios above the $\chi_{bJ}(1P)$
states are labeled as $B_{\rm eff}$ to indicate that direct as well as chain
decays are included in the total branching ratio.  The perturbative part of the
production, $C^{ij}_{Q \overline Q [n]}$, is the same for the $\Upsilon$,
$\Upsilon(2S)$, and $\Upsilon(3S)$ states and for the $\chi_{bJ}(1P)$ and 
$\chi_{bJ}(2P)$ states.  Only the parameters $\langle {\cal O}_n^B \rangle$
change.  The complex feeddown of the higher bottomonium states to the
$\Upsilon$ requires more parameters than $J/\psi$ production.

In contrast, in the color evaporation model, the rapidity
distributions of all states are assumed to be the same, thus {\it
e.g.}\ $F_{J/\psi}$ in Eq.~(\ref{psicem}) includes the
$\chi_{cJ}$ and $\psi'$ decay contributions given explicitly in
Eq.~(\ref{dirpsi}).

Two differences between the NRQCD and color evaporation approaches are
relevant here.  The first concerns the $x$ values probed.  Since the
color evaporation model integrates over $Q \overline Q$ pair mass up
to the $M \overline M$ threshold, it averages over the $x$ range
$2m_Q/\sqrt{s_{NN}} < x < 2m_M/\sqrt{s_{NN}}$.  The pair mass integration 
also includes
limited $Q^2$ evolution in the parton densities and the shadowing
parameterizations.  The NRQCD formulation selects specific $x_1$ and
$x_2$ values for some of the states and only involves a convolution
over $x$ for color singlet production of {\it e.g.}\ $gg \rightarrow
J/\psi, \chi_{c1}, \chi_{c2}$ and $g( q + \overline q) \rightarrow
\chi_{c1}$. Additionally, production is at fixed $Q^2$ for all states.
The second difference is the $g (q + \overline q)$ contribution to
NRQCD production, absent in the color evaporation model.
 
\begin{figure}[htb]
\setlength{\epsfxsize=0.7\textwidth}
\setlength{\epsfysize=0.5\textheight}
\leftline{\epsffile{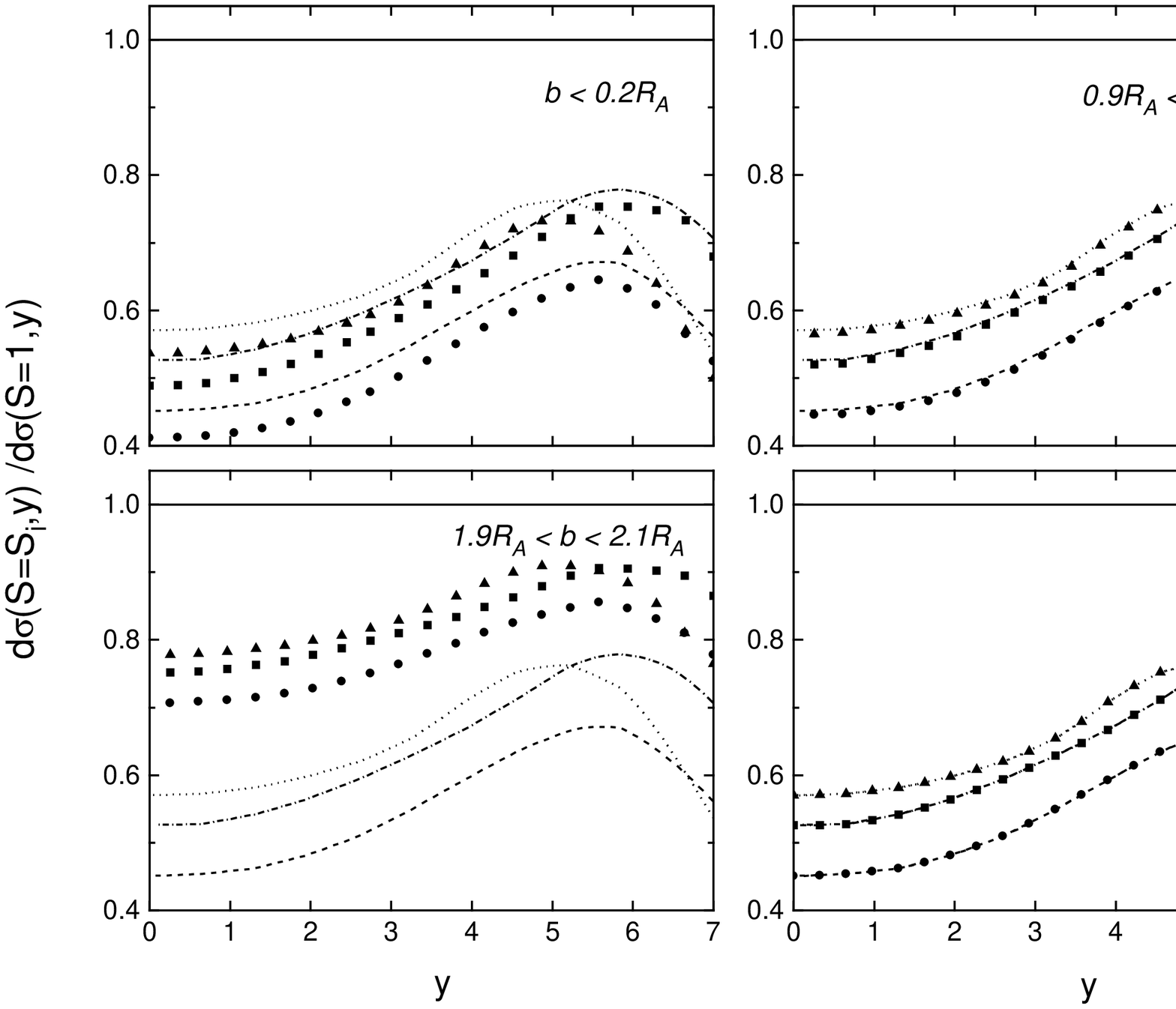}}
\caption[]{The $J/\psi$ rapidity distribution calculated in the NRQCD
model with the CTEQ 3L distributions, compared to the distribution
with $S=1$ in Pb+Pb collisions at the LHC.  Central, $b<0.2 R_A$,
semi-central, $0.9R_A < b < 1.1R_A$, peripheral, $1.9R_A < b < 2.1R_A$
impact parameters are shown along with the integral over all $b$.  The
lines show the homogeneous shadowing result: dashed line for $S_1$,
dot-dashed for $S_2$, and dotted for $S_3$.  Equation~(\ref{wsparam})
is used to calculate the ratio for $S_1$, circles, $S_2$, squares, and
$S_3$, triangles. }
\label{psiynrqcdl}
\end{figure}

\begin{figure}[htb]
\setlength{\epsfxsize=0.7\textwidth}
\setlength{\epsfysize=0.5\textheight}
\leftline{\epsffile{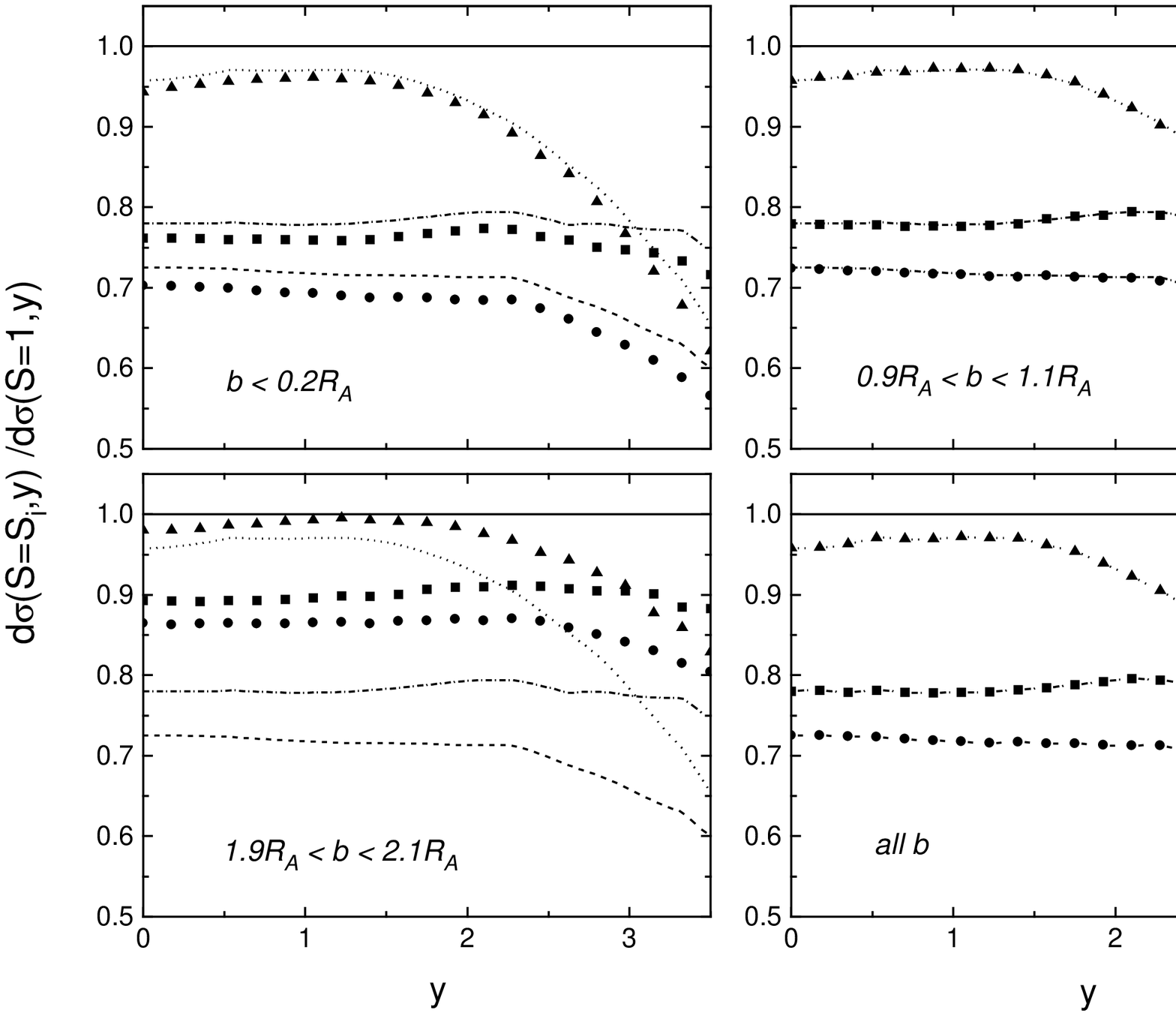}}
\caption[]{The $J/\psi$ rapidity distribution calculated in the NRQCD 
model with the CTEQ 3L distributions, compared to the
distribution with $S=1$ in Au+Au collisions at RHIC.  Central,
$b<0.2 R_A$, semi-central, $0.9R_A < b <
1.1R_A$, peripheral, $1.9R_A < b <
2.1R_A$ impact parameters are shown along with the integral over all $b$.  
The lines show the homogeneous shadowing result: dashed line
for $S_1$, dot-dashed for $S_2$, and dotted for $S_3$.  
Equation~(\ref{wsparam})
calculate the ratio for $S_1$, circles, $S_2$, squares, and $S_3$,
triangles. }
\label{psiynrqcdr}
\end{figure}

We show NRQCD results for $J/\psi$ production in
Figs.~\ref{psiynrqcdl} and \ref{psiynrqcdr}.  Since the $S_1$
parameterization is flavor and $Q^2$ independent, these results are
least influenced by the production model.  The differences between the
models are most obvious at RHIC where the $q \overline q$ contribution
is $\approx 5$\% of the color evaporation cross section and $\approx
1$\% of the NRQCD cross section.  The $g (q + \overline q)$
contribution is $\approx 3-4$\% of the NRQCD cross section.  Since the
gluon is antishadowed at RHIC, significantly less shadowing can be
expected in the NRQCD model than in the color evaporation model. The
relative reduction in shadowing is particularly obvious for the $S_3$
parameterization in Fig.~\ref{psiynrqcdr} where the cross section
ratio is $\approx 0.95$ over 1.5 units of rapidity where $x$ is
antishadowed.  At larger rapidity, $x$ is in the EMC region and the
$S_3$ gluon ratio decreases again, as shown in Fig.~\ref{fshadow}.
The $S_2$ ratio is generally flatter because the gluon ratio is not
reduced in the EMC region.  The difference between the two approaches
is significantly smaller at the LHC where the $q \overline q$
contribution is less than 1\% for both models and therefore plays
practically no role.

\begin{figure}[htb]
\setlength{\epsfxsize=0.7\textwidth}
\setlength{\epsfysize=0.5\textheight}
\leftline{\epsffile{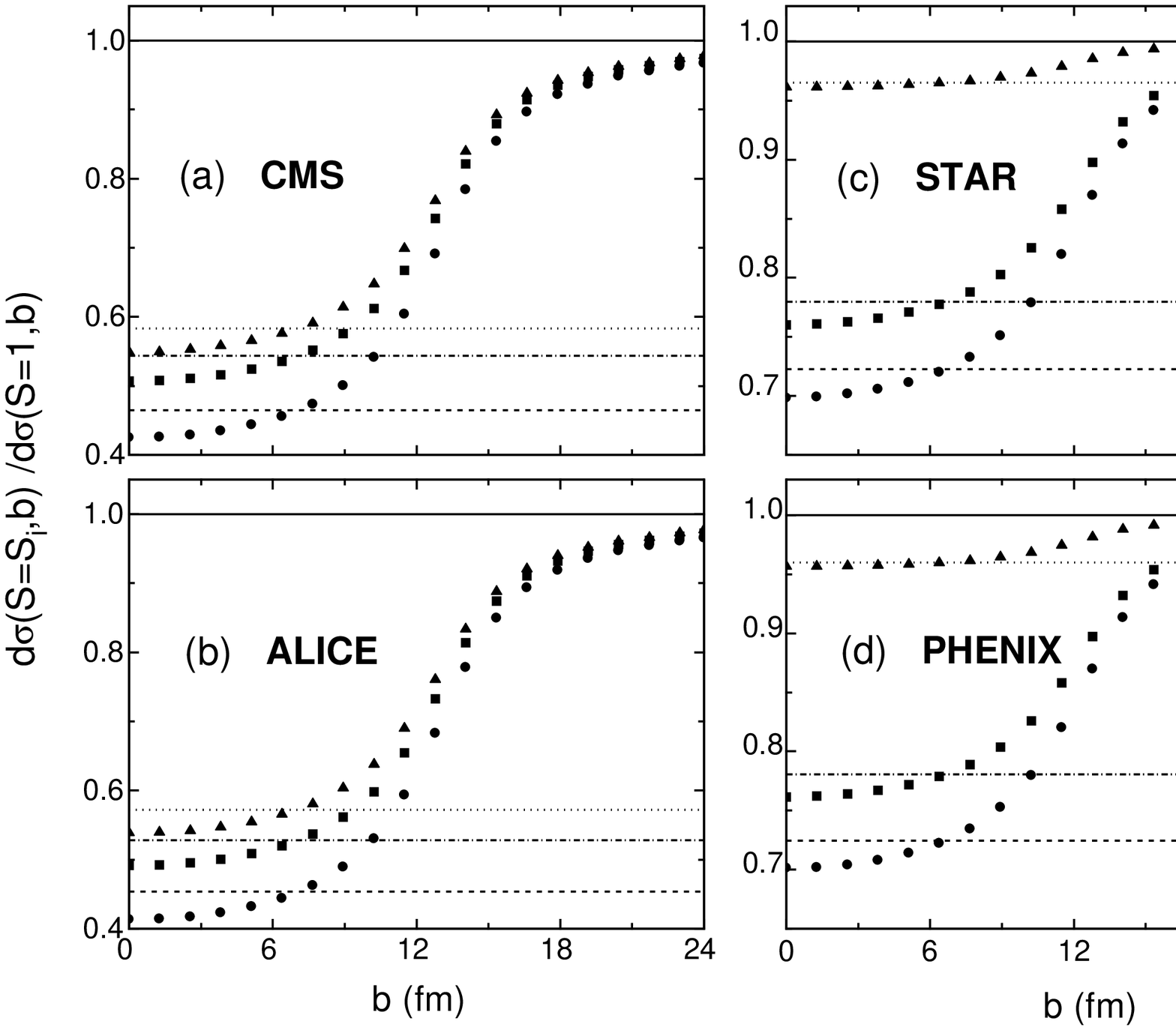}}
\caption[]{The impact parameter dependence of $J/\psi$ production calculated
in NRQCD with the CTEQ 3L distributions.  Results are shown for the
central rapidity coverages given for all four detectors.  The lines
show the homogeneous shadowing result: dashed line for $S_1$,
dot-dashed for $S_2$, and dotted for $S_3$.  Equation~(\ref{wsparam})
is used to calculate the ratio for $S_1$, circles, $S_2$, squares, and
$S_3$, triangles. }
\label{jpbdepnrqcd}
\end{figure}

The impact parameter dependence of shadowing in the NRQCD approach on
$J/\psi$ production is shown in Fig.~\ref{jpbdepnrqcd}.  The
difference between shadowing in this model and in the color
evaporation model seen in the rapidity distributions is obvious here
as well.

\begin{figure}[htb]
\setlength{\epsfxsize=0.7\textwidth}
\setlength{\epsfysize=0.5\textheight}
\leftline{\epsffile{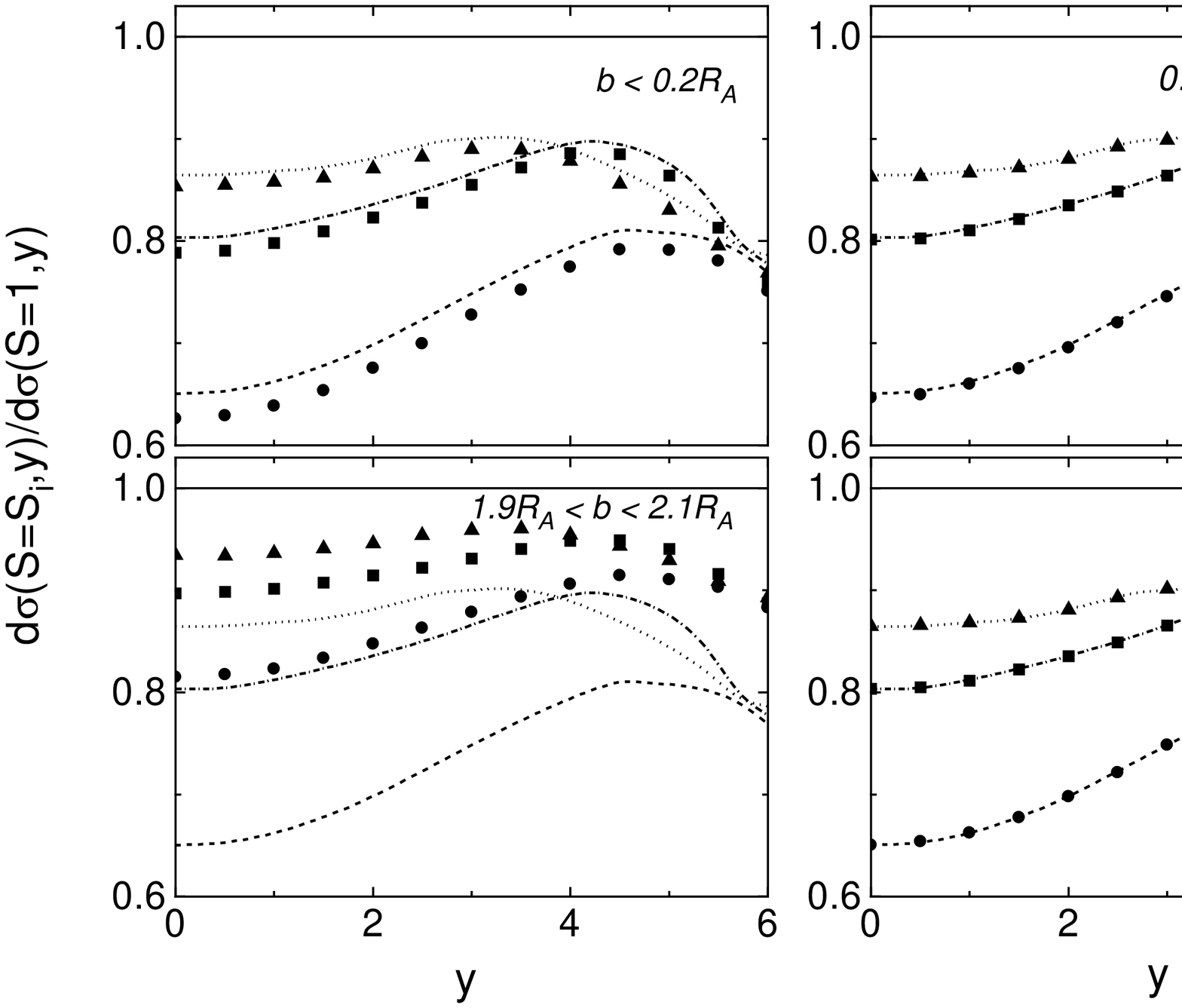}}
\caption[]{The $\Upsilon$ rapidity distribution calculated in the NRQCD
model with the CTEQ 3L distributions, compared to the distribution
with $S=1$ in Pb+Pb collisions at the LHC.  Central, $b<0.2 R_A$,
semi-central, $0.9R_A < b < 1.1R_A$, peripheral, $1.9R_A < b < 2.1R_A$
impact parameters are shown along with the integral over all $b$.  The
lines show the homogeneous shadowing result: dashed line for $S_1$,
dot-dashed for $S_2$, and dotted for $S_3$.  Equation~(\ref{wsparam})
is used to calculate the ratio for $S_1$, circles, $S_2$, squares, and
$S_3$, triangles. }
\label{upsynrqcdl}
\end{figure}

\begin{figure}[htb]
\setlength{\epsfxsize=0.7\textwidth}
\setlength{\epsfysize=0.5\textheight}
\leftline{\epsffile{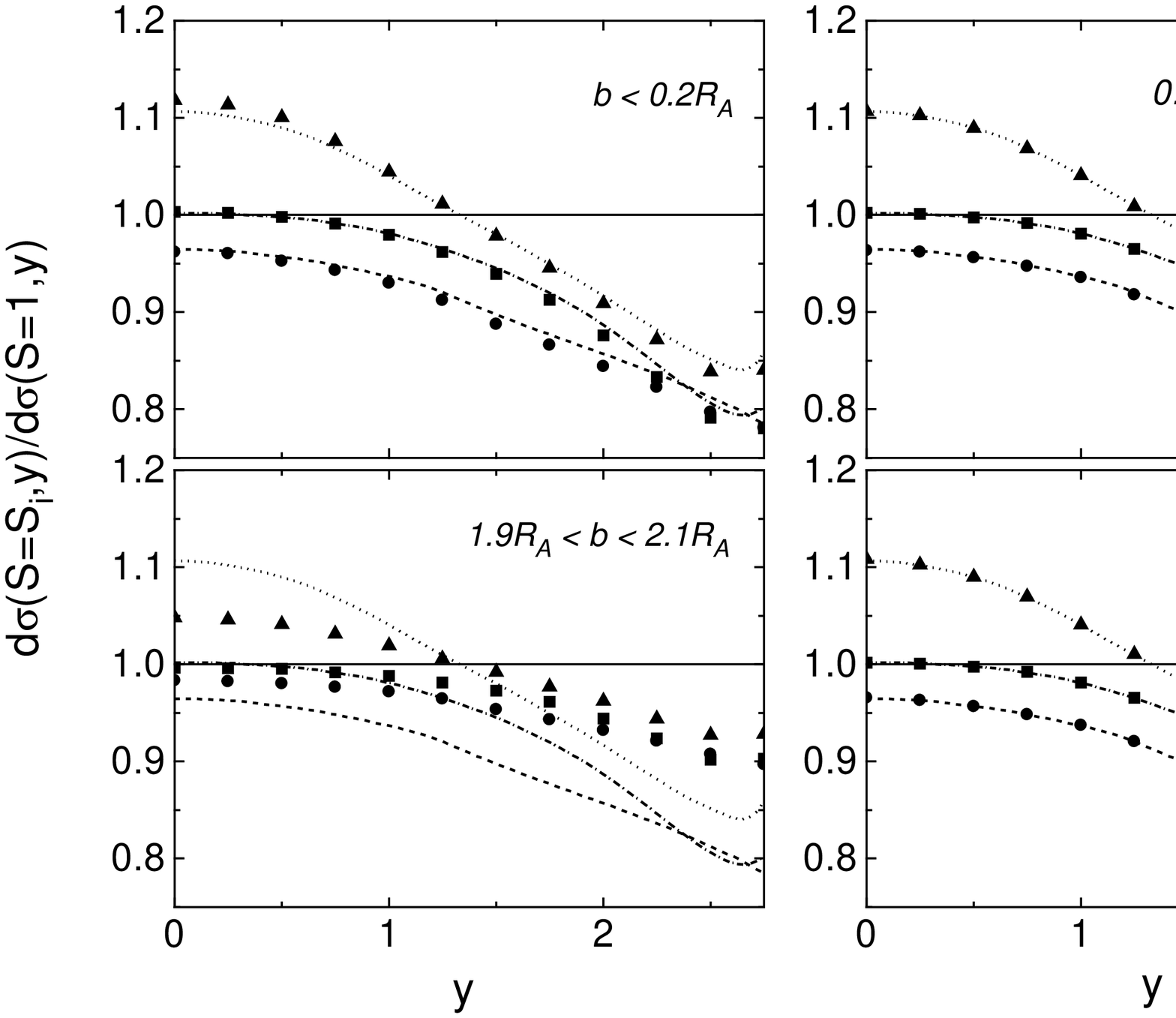}}
\caption[]{The $\Upsilon$ rapidity distribution calculated in the NRQCD 
model with the CTEQ 3L distributions, compared to the
distribution with $S=1$ in Au+Au collisions at RHIC.  Central,
$b<0.2 R_A$, semi-central, $0.9R_A < b <
1.1R_A$, peripheral, $1.9R_A < b <
2.1R_A$ impact parameters are shown along with the integral over all $b$.  
The lines show the homogeneous shadowing result: dashed line
for $S_1$, dot-dashed for $S_2$, and dotted for $S_3$.  
Equation~(\ref{wsparam})
calculate the ratio for $S_1$, circles, $S_2$, squares, and $S_3$,
triangles. }
\label{upsynrqcdr}
\end{figure}

\begin{figure}[htb]
\setlength{\epsfxsize=0.7\textwidth}
\setlength{\epsfysize=0.5\textheight}
\leftline{\epsffile{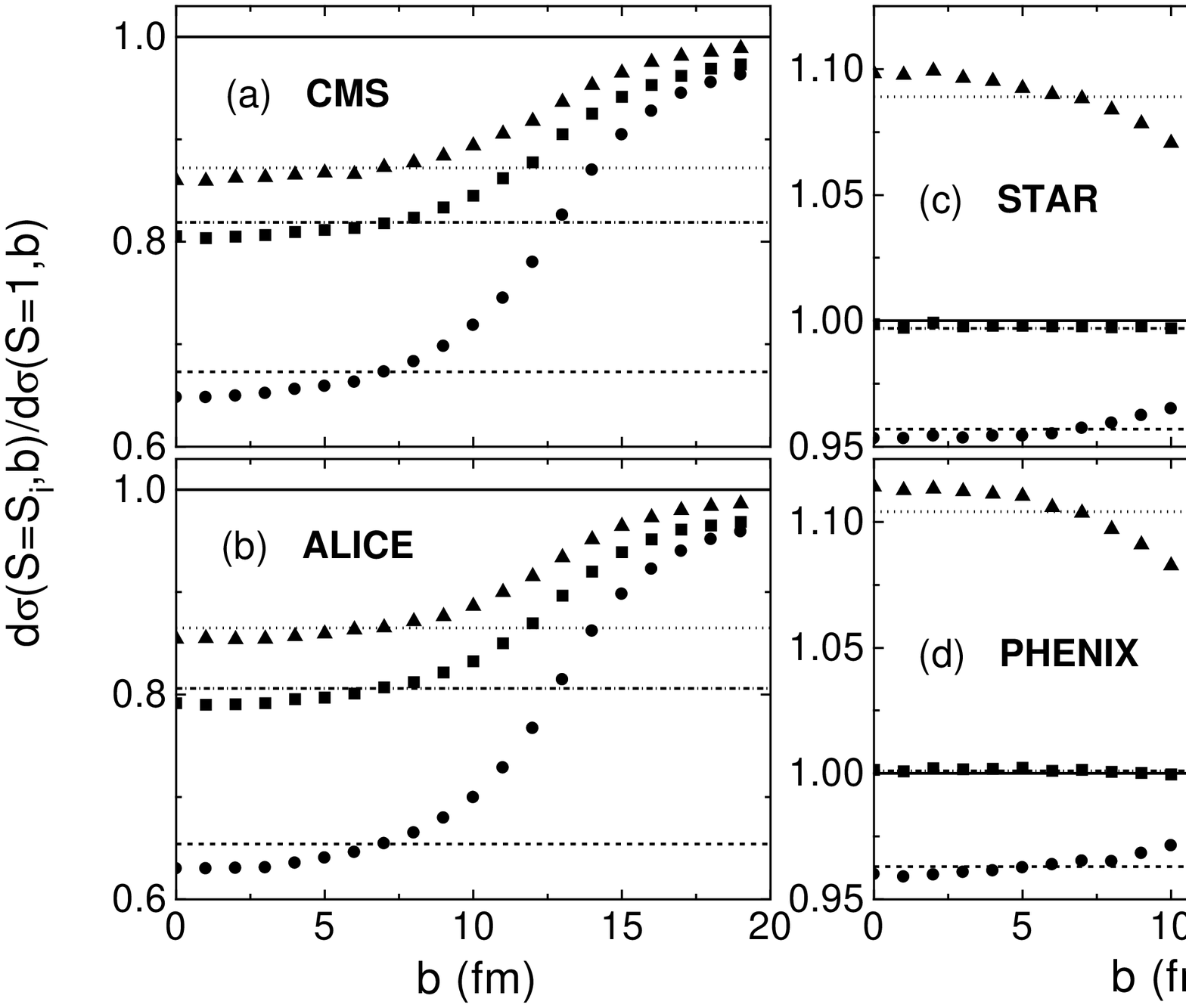}}
\caption[]{The impact parameter dependence of $\Upsilon$ production calculated
in NRQCD with the CTEQ 3L distributions.  Results are shown for the
central rapidity coverages given for all four detectors.  The lines
show the homogeneous shadowing result: dashed line for $S_1$,
dot-dashed for $S_2$, and dotted for $S_3$.  Equation~(\ref{wsparam})
is used to calculate the ratio for $S_1$, circles, $S_2$, squares, and
$S_3$, triangles. }
\label{upsbdepnrqcd}
\end{figure}

The effect of shadowing on $\Upsilon$ production in the NRQCD approach
is shown on the rapidity distributions in Figs.~\ref{upsynrqcdl} and
\ref{upsynrqcdr} and on the impact parameter dependence in
Fig.~\ref{upsbdepnrqcd}.  The same trends seen in the color
evaporation model are observed here except that shadowing or
antishadowing effects are reduced for NRQCD production.  Here, the
larger $b$ quark mass, 4.9 GeV, and scale, $Q = 2m_b$, reduce the
magnitude of the shadowing.  The importance of $q \overline q$
annihilation in the color evaporation model relative to the $q
\overline q$ and $g(q + \overline q)$ contributions in NRQCD affects
the shadowing.  The $g(q + \overline q)$ component in NRQCD is 1\% or
less of the $\Upsilon$ cross section at both RHIC and LHC.  The higher
quark mass probes larger $x$ values where the $q \overline q$
contribution is larger.  At RHIC, $q \overline q$ contributes
$13-16$\% of the total $\Upsilon$ cross section in the color
evaporation model compared to $36-56$\% of the total $\Upsilon$ cross
section in the NRQCD approach.  The larger fraction of $\Upsilon$
production by $q \overline q$ annihilation in NRQCD is due to the
large octet $\chi_{bJ}$ contribution.

The integrated $J/\psi$ cross sections per nucleon pair for both
models are shown in Table~\ref{psisigtab}.  The factor $K_{\rm th}$
is included for the color evaporation model while the NRQCD parameters
are fit to the measured cross sections at LO.  The $S=1$ cross sections agree
within $5-7$\% at RHIC and within 15\% at the LHC. The NRQCD results
are lower than the color evaporation results at RHIC but the NRQCD
cross section grows faster with energy
than the color evaporation cross section.
This behavior can be attributed to the different small $x$
behavior of the MRST LO and CTEQ 3L parton densities.  With
homogeneous shadowing, the differences are more striking, as reflected
in Figs.~\ref{psiylhc}-\ref{jpbdepnrqcd}.

Table~\ref{upssigtab} shows the integrated $\Upsilon$ production cross 
sections per nucleon pair
for both models.  The theoretical $K$ factor is
included for the color evaporation model \cite{HPC}.  The NRQCD parameters have
been fit to fixed target $\Upsilon$ production data.  The two model $\Upsilon$
cross sections do not agree as well as do those of the $J/\psi$.
Reasons for this disagreement might include the greater number of $\Upsilon$ 
parameters needed
to fit a more limited set of data or the absence of the possible
$\chi_{bJ}(3P)$ decays in this calculation.

Finally, we mention one caveat concerning quarkonium production.  Since the
initial quarkonium state is typically a color octet and obtains its
final-state identity in a later soft interaction, it is conceivable that
production and conversion occur far enough apart in position space for the
strength of the apparent shadowing to be different.  However, if shadowing is
considered to only affect quarkonium at the production point, this
separation is insignificant.  In any case, this separation is a much
bigger issue in $pp$ interactions, where the two points must be
quite close.

\section{Discussion and Conclusions}

We have studied the effect of shadowing and its position dependence on
particle production in nucleus-nucleus collisions at RHIC and LHC
energies.  Shadowing can reduce the minijet yields by up
to a factor of two at the LHC.  Assuming that hard production
dominates the determination of the initial conditions and that the
high minijet yield leads to equilibration, the initial energy density
and apparent temperature can be significantly reduced.  Fast
equilibration is unlikely, even for the gluons alone, when shadowing
is included.  The change in the initial conditions due to shadowing is
considerably smaller at RHIC, on the order of a few percent, less than
the change in the initial conditions when soft production is included.
We have compared the initial conditions in central collisions with
homogeneous and inhomogeneous shadowing and found the difference to be
small.  The inhomogeneity of the shadowing becomes more important in
peripheral collisions.  We have also showed the shadowing effects on
the $E_T$ distributions for the central rapidity acceptance of the
major detectors at the LHC and RHIC.  We note that our results at RHIC are
more stable with respect to changes in the parton densities than at the LHC
where the small $x$ behavior of the gluons can lead to unitarity violations,
the size of which depends strongly on the chosen parton densities.  Since we
assume $K_{\rm jet} = 1$, we have been very conservative in our
estimates of the initial conditions. Nonetheless, once unitarity is satisfied
at the LHC, the hard component is likely be reduced judging from the difference
between the GRV 94 LO and the MRST LO cross sections.  Thus lower number and 
energy densities may be expected for LHC collisions.

Finally we have studied the effects on the $J/\psi$, $\Upsilon$, and Drell-Yan
yields.  A careful measurement of the $J/\psi$, $\Upsilon$, 
and Drell-Yan rates as
a function of rapidity can help distinguish between shadowing models
as well as the quarkonium production mechanism since the color
evaporation and NRQCD approaches lead to quite different shadowing
patterns.  Because there is typically a larger shadowing effect on
quarkonium production in the color evaporation model than on Drell-Yan
production, {\it e.g.}\ the $J/\psi$ to
Drell-Yan ratio would be smaller than that expected for $S=1$. On the
other hand, the NRQCD approach predicts the reverse--the $J/\psi$ to
Drell-Yan ratio may be larger than expected when $S=1$.  Since the
effect of shadowing depends on the Drell-Yan pair mass, if the
Drell-Yan yield is to be used as a baseline to compare the yield of
other hard probes, the rates should be measured directly in the mass
region of interest rather than relying on calculations to extrapolate
into an unmeasured region.

One key test of the impact parameter dependence of shadowing is the
slope of the Drell-Yan mass distribution; if shadowing varies with
position, the slope of the distribution should depend on $E_T$.  If
the slopes are significantly different for central, intermediate and
peripheral collisions, this would be a clear demonstration that
shadowing depends on position.  The only complication may be due to parton
energy loss before the hard interaction.

{\bf Acknowledgements:}
V.E. and A.K. would like to thank the LBNL Relativistic Nuclear
Collisions group for their hospitality and M. Strikhanov and
V.V. Grushin for discussions and support.  We also thank K.J. Eskola
for providing the shadowing routines and for discussions.  This work
was supported in part by the Division of Nuclear Physics of the Office
of High Energy and Nuclear Physics of the U. S. Department of Energy
under Contract Number DE-AC03-76SF00098.

\begin{table}[tbp]
{\footnotesize
\begin{center}
\begin{tabular}{ccccccccc}
& \multicolumn{4}{c}{GRV 94 LO} & \multicolumn{4}{c}{MRST LO} \\ 
$S$ & $q$ & $\overline q$ & $g$ & total & $q$ & $\overline q$ & $g$ & total \\
\hline 
\multicolumn{9}{c}{$2 \sigma^f (p_0)$ (mb)} \\
$1$   & 30 & 28 & 605 & 663 & 19 & 17 & 274 & 310 \\
$S_1$ & 16 & 14 & 316 & 346 & 10 &  9 & 146 & 165 \\
$S_2$ & 14 & 13 & 329 & 356 &  9 &  8 & 156 & 173 \\
$S_3$ & 19 & 17 & 393 & 429 & 12 & 11 & 183 & 206
\\ \hline
\multicolumn{9}{c}{$\sigma^f (p_0) 
\langle E_T^f \rangle$ (mb GeV)} \\  
$1$   & 95 & 86 & 1794 & 1975 & 61 & 55 &  866 &  981 \\ 
$S_1$ & 50 & 45 &  950 & 1045 & 33 & 29 &  469 &  531 \\
$S_2$ & 48 & 42 & 1043 & 1132 & 31 & 27 &  527 &  585 \\ 
$S_3$ & 61 & 54 & 1216 & 1331 & 40 & 35 &  608 &  683
\\ \hline
\multicolumn{9}{c}{$\sigma^f(p_0) \langle E_T^{2 \, f} 
\rangle$ (mb GeV$^2$)} \\ 
$1$   & 387 & 337 & 9820 & 10544 & 262 & 228 & 5182 & 5672 \\
$S_1$ & 213 & 184 & 5206 &  5603 & 150 & 127 & 2821 & 3097 \\
$S_2$ & 222 & 182 & 6185 &  6589 & 155 & 126 & 3439 & 3720 \\
$S_3$ & 275 & 232 & 7084 &  7591 & 192 & 161 & 3915 & 4267 
\\ \hline
\end{tabular}
\end{center}
}
\caption[]{The minijet cross section, Eq.~(\ref{dsdy}),
first and second moments of the transverse energy distribution, 
Eqs.~(\ref{etmom}) and (\ref{etmom2}) respectively with $p_0 = 2$ GeV, 
integrated over $b$ and $r$ and
divided by $AB$, within CMS, $|y| \leq 2.4$.  Results for both sets of 
parton distributions used are separated into 
contributions from quarks,
antiquarks and gluons as well as the total.  The calculations are done without
shadowing, $S=1$, and with shadowing parameterizations $S_1$, $S_2$, 
and $S_3$.}
\label{table1}
\end{table}

\begin{table}[tbp]
{\footnotesize
\begin{center}
\begin{tabular}{ccccccccc}
& \multicolumn{4}{c}{GRV 94 LO} & \multicolumn{4}{c}{MRST LO} \\ 
$S$ & $q$ & $\overline q$ & $g$ & total & $q$ & $\overline q$ & $g$ & total \\
\hline 
\multicolumn{9}{c}{$2 \sigma^f (p_0)$ (mb)} \\
$1$   & 14 & 13 & 296 & 323 &  8 &  7 & 120 & 135 \\
$S_1$ &  7 &  7 & 152 & 166 &  4 &  4 &  63 &  71 \\
$S_2$ &  6 &  6 & 158 & 170 &  4 &  4 &  67 &  75 \\
$S_3$ &  8 &  8 & 190 & 206 &  5 &  4 &  80 &  89 \\ \hline
\multicolumn{9}{c}{$\sigma^f(p_0) 
\langle E_T^f \rangle$ (mb GeV)} \\  
$1$   & 43 & 40 & 882 & 965 & 25 & 24 & 381 & 430 \\ 
$S_1$ & 22 & 21 & 459 & 502 & 13 & 12 & 202 & 227 \\
$S_2$ & 21 & 19 & 504 & 544 & 12 & 11 & 228 & 251 \\ 
$S_3$ & 27 & 25 & 592 & 644 & 16 & 15 & 265 & 296 \\ \hline
\multicolumn{9}{c}{$\sigma^f(p_0) \langle E_T^{2 \, f} 
\rangle$ (mb GeV$^2$)} \\ 
$1$   & 172 & 159 & 4006 & 4337 & 105 &  97 & 1858 & 2060 \\
$S_1$ &  92 &  85 & 2100 & 2277 &  58 &  53 & 1001 & 1112 \\
$S_2$ &  94 &  83 & 2498 & 2675 &  59 &  52 & 1222 & 1333 \\
$S_3$ & 120 & 108 & 2876 & 3104 &  75 &  68 & 1401 & 1544
\\ \hline
\end{tabular}
\end{center}
}
\caption[]{The minijet cross section, Eq.~(\ref{dsdy}),
first and second moments of the transverse energy distribution, 
Eqs.~(\ref{etmom}) and (\ref{etmom2}) respectively with $p_0 =2$ GeV, 
integrated over $b$ and $r$ and
divided by $AB$, within ALICE, $|y| \leq 1$.  Results 
for both sets of parton distributions are separated into 
contributions from quarks,
antiquarks and gluons as well as the total.  The calculations are done without
shadowing, $S=1$, and with shadowing parameterizations $S_1$, $S_2$, 
and $S_3$.}
\label{table2}
\end{table}

\begin{table}[tbp]
{\footnotesize
\begin{center}
\begin{tabular}{ccccccccc}
& \multicolumn{4}{c}{GRV 94 LO} & \multicolumn{4}{c}{MRST LO} \\ 
$S$ & $q$ & $\overline q$ & $g$ & total & $q$ & $\overline q$ & $g$ & total \\
\hline 
\multicolumn{9}{c}{$2 \sigma^f(p_0)$ (mb)} \\
$1$   & 0.76 & 0.45 & 5.55 & 6.77 & 0.66 & 0.41 & 4.53 & 5.60 \\
$S_1$ & 0.63 & 0.38 & 4.54 & 5.55 & 0.55 & 0.34 & 3.74 & 4.63 \\
$S_2$ & 0.64 & 0.36 & 4.64 & 5.64 & 0.55 & 0.32 & 3.84 & 4.72 \\
$S_3$ & 0.74 & 0.42 & 5.69 & 6.85 & 0.64 & 0.38 & 4.68 & 5.70 \\ \hline
\multicolumn{9}{c}{$\sigma^f(p_0) 
\langle E_T^f \rangle$ (mb GeV)} \\  
$1$   & 2.14 & 1.24 & 14.62 & 18.00 & 1.85 & 1.11 & 12.13 & 15.09 \\ 
$S_1$ & 1.80 & 1.03 & 12.11 & 14.94 & 1.57 & 0.94 & 10.15 & 12.66 \\
$S_2$ & 1.84 & 1.00 & 12.49 & 15.33 & 1.60 & 0.90 & 10.52 & 13.02 \\ 
$S_3$ & 2.10 & 1.15 & 15.23 & 18.48 & 1.83 & 1.04 & 12.77 & 15.64 \\ \hline
\multicolumn{9}{c}{$\sigma^f(p_0) \langle E_T^{2 \, f} 
\rangle$ (mb GeV$^2$)} \\ 
$1$   & 6.92 & 3.71 & 52.38 & 63.00 & 6.10 & 3.35 & 44.32 & 53.77 \\
$S_1$ & 6.02 & 3.17 & 44.17 & 53.36 & 5.33 & 2.89 & 37.84 & 46.05 \\
$S_2$ & 6.19 & 3.09 & 45.97 & 55.25 & 5.48 & 2.83 & 39.54 & 47.85 \\
$S_3$ & 6.96 & 3.51 & 56.24 & 66.71 & 6.15 & 3.21 & 48.08 & 57.44 \\ \hline
\end{tabular}
\end{center}
}
\caption[]{The minijet cross section, Eq.~(\ref{dsdy}),
first and second moments of the transverse energy distribution, 
Eqs.~(\ref{etmom}) and (\ref{etmom2}) respectively for $p_0 = 2$ GeV, 
integrated over $b$ and $r$ and
divided by $AB$, within STAR, $|y| \leq 0.9$.  Results 
for both sets of parton distributions are separated into 
contributions from quarks,
antiquarks and gluons as well as the total.  The calculations are done without
shadowing, $S=1$, and with shadowing parameterizations $S_1$, $S_2$, 
and $S_3$.}
\label{table3}
\end{table}

\begin{table}[tbp]
{\footnotesize
\begin{center}
\begin{tabular}{ccccccccc}
& \multicolumn{4}{c}{GRV 94 LO} & \multicolumn{4}{c}{MRST LO} \\ 
$S$ & $q$ & $\overline q$ & $g$ & total & $q$ & $\overline q$ & $g$ & total \\
\hline 
\multicolumn{9}{c}{$2 \sigma^f(p_0)$ (mb)} \\
$1$   & 0.29 & 0.18 & 2.28 & 2.75 & 0.25 & 0.16 & 1.81 & 2.22 \\
$S_1$ & 0.24 & 0.15 & 1.86 & 2.25 & 0.20 & 0.13 & 1.48 & 1.81 \\
$S_2$ & 0.24 & 0.14 & 1.91 & 2.29 & 0.21 & 0.13 & 1.53 & 1.87 \\
$S_3$ & 0.28 & 0.17 & 2.32 & 2.77 & 0.24 & 0.15 & 1.85 & 2.24 \\ \hline
\multicolumn{9}{c}{$\sigma^f(p_0) 
\langle E_T^f \rangle$ (mb GeV)} \\  
$1$   & 0.81 & 0.50 & 6.01 & 7.32 & 0.69 & 0.44 & 4.85 & 5.98 \\ 
$S_1$ & 0.68 & 0.41 & 4.98 & 6.07 & 0.58 & 0.37 & 4.04 & 4.99 \\
$S_2$ & 0.70 & 0.40 & 5.14 & 6.24 & 0.59 & 0.35 & 4.20 & 5.14 \\ 
$S_3$ & 0.80 & 0.46 & 6.23 & 7.49 & 0.68 & 0.41 & 5.06 & 6.15 \\ \hline
\multicolumn{9}{c}{$\sigma^f(p_0) \langle E_T^{2 \, f} 
\rangle$ (mb GeV$^2$)} \\ 
$1$   & 2.60 & 1.49 & 19.01 & 23.10 & 2.23 & 1.33 & 15.64 & 19.20 \\
$S_1$ & 2.25 & 1.27 & 16.03 & 19.55 & 1.94 & 1.13 & 13.30 & 16.37 \\
$S_2$ & 2.31 & 1.24 & 16.67 & 20.22 & 1.99 & 1.11 & 13.93 & 17.03 \\
$S_3$ & 2.61 & 1.41 & 20.22 & 24.24 & 2.24 & 1.27 & 16.78 & 20.29 \\ \hline
\end{tabular}
\end{center}
}
\caption[]{The minijet cross section, Eq.~(\ref{dsdy}), first and
second moments of the transverse energy distribution,
Eqs.~(\ref{etmom}) and (\ref{etmom2}) respectively with $p_0=2$ GeV,
integrated over $b$ and $r$ and divided by $AB$, within PHENIX, $|y| \leq
0.35$.  Note that the cross sections and moments are given over all azimuth.
Results for both sets of parton distributions are separated
into contributions from quarks, antiquarks and gluons as well as the
total.  The calculations are done without shadowing, $S=1$, and with
shadowing parameterizations $S_1$, $S_2$, and $S_3$.}
\label{table4}
\end{table}

\begin{table}[tbp]
{\footnotesize
\begin{center}
\begin{tabular}{ccccccccc} 
& \multicolumn{4}{c}{GRV 94 LO} & \multicolumn{4}{c}{MRST LO} \\ 
& \multicolumn{2}{c}{gluon} & \multicolumn{2}{c}{total} &
\multicolumn{2}{c}{gluon} & \multicolumn{2}{c}{total} \\
$S$ & HS & IHS & HS & IHS & HS & IHS & HS & IHS \\ \hline 
\multicolumn{9}{c}{$\epsilon_i^f(b=0, p_0)$ (GeV/fm$^3$)} \\ 
$1$   & 835 &  -  & 920 &  -  & 404 &  -  & 457 &  -  \\
$S_1$ & 443 & 413 & 487 & 454 & 219 & 204 & 248 & 231 \\
$S_2$ & 486 & 457 & 527 & 496 & 246 & 233 & 273 & 257 \\
$S_3$ & 566 & 543 & 620 & 594 & 283 & 272 & 318 & 306 \\
\hline
\multicolumn{9}{c}{$n_i^f(b=0, p_0)$ (1/fm$^3$)} \\
$1$   & 282 &  -  & 309 &  -  & 128 &  -  & 144 &  -  \\
$S_1$ & 148 & 138 & 162 & 151 &  68 & 63  &  77 & 72  \\
$S_2$ & 154 & 145 & 166 & 155 &  73 & 68  &  81 & 76  \\
$S_3$ & 183 & 174 & 200 & 191 &  85 & 82  &  96 & 92  \\
\end{tabular}
\end{center}
}
\caption[]{The energy density, Eq.~(\ref{edens}), and number density,
Eq.~(\ref{ndens}), at $b=0$ from minijet
production alone with $p_0 = 2$ GeV within CMS, $|y| \leq 2.4$ are given
for both sets of parton distributions.
Results are shown for homogeneous (HS) and inhomogeneous (IHS)
shadowing, with the latter based on $S_{\rm WS}$. Both the gluon contribution
alone and the total for gluons with three light quark flavors are
presented.  The calculations are done without shadowing, $S=1$, and
with shadowing parameterizations $S_1$, $S_2$, and $S_3$.}
\label{table5}
\end{table}

\begin{table}[tbp]
{\footnotesize
\begin{center}
\begin{tabular}{ccccccccc} 
& \multicolumn{4}{c}{GRV 94 LO} & \multicolumn{4}{c}{MRST LO} \\ 
& \multicolumn{2}{c}{gluon} & \multicolumn{2}{c}{total} &
\multicolumn{2}{c}{gluon} & \multicolumn{2}{c}{total} \\
$S$ & HS & IHS & HS & IHS & HS & IHS & HS & IHS \\ \hline 
\multicolumn{9}{c}{$\epsilon_i^f(b=0, p_0)$ (GeV/fm$^3$)} \\ 
$1$   & 986 &   -  & 1079 &  -  & 426 &   - & 481 &  -  \\
$S_1$ & 513 & 478  & 561  & 522 & 226 & 210 & 256 & 237 \\
$S_2$ & 563 & 531  & 607  & 572 & 255 & 240 & 281 & 264 \\
$S_3$ & 661 & 634  & 719  & 689 & 296 & 285 & 331 & 317 \\ 
\hline
\multicolumn{9}{c}{$n_i^f(b=0, p_0)$ (1/fm$^3$)} \\
$1$   & 332 &  -  & 362 &  -  & 134 &  -  & 151 &  -   \\
$S_1$ & 170 & 158 & 186 & 173 &  70 & 65  &  79 & 73   \\
$S_2$ & 177 & 167 & 190 & 178 &  75 & 70  &  84 & 77   \\
$S_3$ & 213 & 203 & 231 & 220 &  89 & 85  &  99 & 95   \\ 
\end{tabular}
\end{center}
}
\caption[]{The energy density, Eq.~(\ref{edens}), and number density,
Eq.~(\ref{ndens}), at $b=0$ from minijet
production alone with $p_0 = 2$ GeV within ALICE, $|y| \leq 1$ are given
for both sets of parton distributions.
Results are shown for homogeneous (HS) and inhomogeneous (IHS)
shadowing, with the latter based on $S_{\rm WS}$. Both the gluon contribution
alone and the total for gluons with three light quark flavors are
presented.  The calculations are done without shadowing, $S=1$, and
with shadowing parameterizations $S_1$, $S_2$, and $S_3$.}
\label{table6}
\end{table}

\begin{table}[tbp]
{\footnotesize
\begin{center}
\begin{tabular}{ccccccccc} 
& \multicolumn{4}{c}{GRV 94 LO} & \multicolumn{4}{c}{MRST LO} \\ 
& \multicolumn{2}{c}{gluon} & \multicolumn{2}{c}{total} &
\multicolumn{2}{c}{gluon} & \multicolumn{2}{c}{total} \\
$S$ & HS & IHS & HS & IHS & HS & IHS & HS & IHS \\ \hline 
\multicolumn{9}{c}{$\epsilon_i^f(b=0, p_0)$ (GeV/fm$^3$)} \\ 
$1$   & 18.9 &   -  & 23.2 &   -  & 15.7 &   -  & 19.5 &   -   \\
$S_1$ & 15.6 & 15.3 & 19.4 & 19.0 & 13.1 & 12.8 & 16.3 & 16.0 \\
$S_2$ & 16.1 & 15.9 & 19.8 & 19.5 & 13.6 & 13.3 & 16.8 & 16.5 \\
$S_3$ & 19.6 & 19.7 & 23.8 & 23.9 & 16.5 & 16.5 & 20.2 & 20.2 \\ \hline
\multicolumn{9}{c}{$n_i^f(b=0, p_0)$ (1/fm$^3$)} \\
$1$   & 7.2 &  -  & 8.7 &  -  & 5.8 &  -  & 7.2 &  -  \\
$S_1$ & 5.9 & 5.8 & 7.2 & 7.1 & 4.8 & 4.7 & 6.0 & 5.8 \\
$S_2$ & 6.0 & 5.9 & 7.3 & 7.2 & 5.0 & 4.8 & 6.1 & 6.0 \\
$S_3$ & 7.3 & 7.3 & 8.8 & 8,8 & 6.0 & 6.1 & 7.4 & 7.4 \\ 
\end{tabular}
\end{center}
}
\caption[]{The energy density, Eq.~(\ref{edens}), and number density,
Eq.~(\ref{ndens}), at $b=0$ from minijet
production alone with $p_0 = 2$ GeV within the STAR calorimeter, $|y|
\leq 0.9$ are
given for both sets of parton distributions.  
Results are shown for homogeneous (HS) and inhomogeneous
(IHS) shadowing, with the latter based on $S_{\rm WS}$.  Both the gluon
contribution alone and the total for gluons with three light quark
flavors are presented.  The calculations are done without shadowing,
$S=1$, and with shadowing parameterizations $S_1$, $S_2$, and $S_3$.}
\label{table7}
\end{table}

\begin{table}[tbp]
{\footnotesize
\begin{center}
\begin{tabular}{ccccccccc} 
& \multicolumn{4}{c}{GRV 94 LO} & \multicolumn{4}{c}{MRST LO} \\ 
& \multicolumn{2}{c}{gluon} & \multicolumn{2}{c}{total} &
\multicolumn{2}{c}{gluon} & \multicolumn{2}{c}{total} \\
$S$ & HS & IHS & HS & IHS & HS & IHS & HS & IHS \\ \hline
\multicolumn{9}{c}{$\epsilon_i^f(b=0, p_0)$ (GeV/fm$^3$)} \\ 
$1$   & 19.9 &   -  & 24.3 &   -  & 16.1 &   -  & 19.9 &   -  \\
$S_1$ & 16.5 & 16.2 & 20.2 & 19.8 & 13.5 & 13.2 & 16.6 & 16.2 \\
$S_2$ & 17.0 & 16.8 & 20.7 & 20.3 & 14.0 & 13.7 & 17.1 & 16.8 \\
$S_3$ & 20.7 & 20.7 & 24.9 & 24.6 & 16.8 & 16.9 & 20.5 & 20.6 \\ \hline
\multicolumn{9}{c}{$n_i^f(b=0, p_0)$ (1/fm$^3$)} \\
$1$   & 7.6 &  -  & 9.1 &  -  & 6.0 &  -  & 7.4 &  -  \\
$S_1$ & 6.2 & 6.1 & 7.5 & 7.3 & 4.9 & 4.8 & 6.0 & 5.9 \\
$S_2$ & 6.3 & 6.2 & 7.6 & 7.4 & 5.1 & 5.0 & 6.2 & 6.1 \\
$S_3$ & 7.7 & 7.7 & 9.2 & 9.2 & 6.2 & 6.2 & 7.5 & 7.5 \\ 
\end{tabular}
\end{center}
}
\caption[]{The energy density, Eq.~(\ref{edens}), and number density,
Eq.~(\ref{ndens}), at $b=0$ from minijet
production alone with $p_0 = 2$ GeV within PHENIX, $|y| \leq 0.35$ are given
for both sets of parton distributions.
Results are shown for homogeneous (HS) and inhomogeneous (IHS)
shadowing, with the latter based on $S_{\rm WS}$.  Both the gluon contribution
alone and the total for gluons with three light quark flavors are
presented.  The calculations are done without shadowing, $S=1$, and
with shadowing parameterizations $S_1$, $S_2$, and $S_3$.}
\label{table8}
\end{table}

\begin{table}[tbp]
{\footnotesize
\begin{center}
\begin{tabular}{cccc} 
Detector & Rapidity & 
$\overline E_T^S$ (mb GeV) & $\overline{E_T^2}^S$ (mb GeV$^2$) \\ \hline
CMS     & $|y| \leq 2.4$  & 135  & 450 \\
ALICE   & $|y| \leq 1$    & 56   & 188 \\
STAR    & $|y| \leq 0.9$  & 34   & 112 \\
PHENIX  & $|y| \leq 0.35$ & 13 & 44 \\ \hline
\end{tabular}
\end{center}
}
\caption[]{The first and second $E_T$ moments of the soft contribution
adjusted to the acceptance of the experiments at the LHC and RHIC.  We assume 
$\sigma_S^{pp} = 40$ mb at RHIC and $\sigma_S^{pp} = 60$ mb at the LHC.}
\label{softet}
\end{table}

\begin{table}
\begin{tabular}{ccccc}
Detector & $\sigma(S=1)$ (nb) &  $\sigma(S=S_1)$ (nb) &  $\sigma(S=S_2)$ (nb)
& $\sigma(S=S_3)$ (nb) \\ \hline
\multicolumn{5}{c}{$4<m<9$ GeV} \\
CMS    & 4.05  & 1.90  & 1.57  & 2.26  \\
ALICE  & 1.89  & 0.86  & 0.68  & 1.04  \\
STAR   & 0.32  & 0.26  & 0.26 & 0.28  \\
PHENIX & 0.13  & 0.10  & 0.10  & 0.11  \\
\multicolumn{5}{c}{$11<m<20$ GeV} \\
CMS    & 0.48 & 0.25 & 0.24 & 0.33  \\
ALICE  & 0.23 & 0.10 & 0.10 & 0.15 \\
\end{tabular}
\caption{Leading order Drell-Yan cross section, in units of nb per 
nucleon pair, integrated over all
impact parameters, for the MRST LO parton densities.  Full azimuthal coverage
is assumed.}
\label{dysigtab}
\end{table}

\begin{table}
\begin{tabular}{ccccc}
Detector & $\sigma(S=1)$ ($\mu$b) &  $\sigma(S=S_1)$ ($\mu$b) &
$\sigma(S=S_2)$ ($\mu$b) & $\sigma(S=S_3)$ ($\mu$b) \\ \hline
\multicolumn{5}{c}{Color Evaporation Model} \\
CMS    & 43.5 & 19.6 & 19.2 & 22.8 \\
ALICE  & 18.8 & 8.23 & 8.00 & 9.63 \\
STAR   & 1.62 & 1.12 & 1.12 & 1.43 \\
PHENIX & 0.65 & 0.44 & 0.40 & 0.56 \\
\multicolumn{5}{c}{NRQCD} \\
CMS    & 51.0 & 23.8 & 27.1 & 29.8 \\
ALICE  & 21.5 & 9.75 & 11.1 & 12.3 \\
STAR   & 1.54 & 1.11 & 1.16 & 1.49 \\
PHENIX & 0.60 & 0.44 & 0.45 & 0.58 \\
\end{tabular}
\caption{$J/\psi$ production cross sections in the color evaporation and
NRQCD approach in units of $\mu$b per nucleon pair.  No nuclear absorption of
the $J/\psi$ in the final-state is included.  
The color evaporation cross sections were calculated with 
the MRST LO parton
densities and the NRQCD results were obtained with the CTEQ 3L distributions.
Both are normalized so as to agree with results from charmonium 
hadroproduction.  Full azimuthal coverage
is assumed.}
\label{psisigtab}
\end{table}

\begin{table}
\begin{tabular}{ccccc}
Detector & $\sigma(S=1)$ (nb) &  $\sigma(S=S_1)$ (nb) &
$\sigma(S=S_2)$ (nb) & $\sigma(S=S_3)$ (nb) \\ \hline
\multicolumn{5}{c}{Color Evaporation Model} \\
CMS    & 377  &  187 &  249 &  267 \\
ALICE  & 169  &   80 &  107 &  117 \\
STAR   & 4.80 & 4.38 & 4.72 & 5.76 \\
PHENIX & 1.92 & 1.77 & 1.89 & 2.36 \\
\multicolumn{5}{c}{NRQCD} \\
CMS    &  419 &  282 &  343 &  365 \\
ALICE  &  181 &  119 &  146 &  157 \\
STAR   & 6.19 & 5.92 & 6.17 & 6.74 \\
PHENIX & 2.52 & 2.43 & 2.52 & 2.78 \\
\end{tabular}
\caption{$\Upsilon$ production cross sections in the color evaporation and
NRQCD approach in units of $\mu$b per nucleon pair.  No nuclear absorption of
the $\Upsilon$ in the final-state is included.  
The color evaporation cross sections were calculated with 
the MRST LO parton
densities and the NRQCD results were obtained with the CTEQ 3L distributions.
Both are normalized so as to agree with results from bottomonium
hadroproduction.  Full azimuthal coverage
is assumed.}
\label{upssigtab}
\end{table}

\end{document}